%% file: main.tex
\documentclass[reprint,superscriptaddress,amsmath,amssymb,aps,prl,longbibliography]{revtex4-2}

\usepackage{graphicx,tabularx,makecell}
\usepackage{amsmath,amsfonts,amssymb}
\usepackage{mathtools}
\usepackage{bm,dsfont}
\usepackage[dvipsnames]{xcolor}
\usepackage[bookmarks=true,colorlinks=true,urlcolor=blue,linkcolor=blue,citecolor=blue,breaklinks]{hyperref}
\usepackage[all]{hypcap}
\usepackage{xspace}
\usepackage{ulem}

\graphicspath{{figures/}}

\definecolor{pu}{RGB}{200,50,200}
\definecolor{gr}{RGB}{0,187,0}
\definecolor{bl}{RGB}{68,34,200}
\definecolor{re}{RGB}{200,34,68}
\definecolor{ye}{RGB}{255,165,0}
\definecolor{oran}{RGB}{255,170,0}

\newcommand{\moire}[0]{moir\'{e}\xspace}

\newcommand{\RNum}[1]{\uppercase\expandafter{\romannumeral #1\relax}}

\begin{document}

\title{Itinerant Magnetism in Twisted Bilayer WSe$_2$ and MoTe$_2$}

\author{Liangtao Peng}
\affiliation{Department of Physics, Washington University in St. Louis, St. Louis, Missouri 63130, United States}
\author{Christophe De Beule}
\affiliation{Department of Physics and Astronomy, University of Pennsylvania, Philadelphia, Pennsylvania 19104, USA}
\author{Yiyang Lai}
\affiliation{Department of Physics, Washington University in St. Louis, St. Louis, Missouri 63130, United States}
\author{Du Li}
\affiliation{Department of Physics, Washington University in St. Louis, St. Louis, Missouri 63130, United States}
\author{Li Yang}
\affiliation{Department of Physics, Washington University in St. Louis, St. Louis, Missouri 63130, United States}
\affiliation{Institute of Materials Science and Engineering, Washington University in St. Louis, St. Louis, Missouri 63130, USA}
\author{E. J. Mele}
\affiliation{Department of Physics and Astronomy, University of Pennsylvania, Philadelphia, Pennsylvania 19104, USA}
\author{Shaffique Adam}
\affiliation{Department of Physics, Washington University in St. Louis, St. Louis, Missouri 63130, United States}
\affiliation{Department of Physics and Astronomy, University of Pennsylvania, Philadelphia, Pennsylvania 19104, USA}
\affiliation{Department of Materials Science and Engineering, 
National University of Singapore, 9 Engineering Drive 1, 
Singapore 117575}


\date{\today}

\begin{abstract}
Using a self-consistent Hartree-Fock theory, we show that the recently observed ferromagnetism in twisted bilayer WSe$_2$ [Nat.\ Commun.\ \textbf{16}, 1959 (2025)] can be understood as a Stoner-like instability of interaction-renormalized moir\'e bands. We quantitatively reproduce the observed Lifshitz transition as function of hole filling and applied electric field that marks the boundary between layer-hybridized and layer-polarized regimes.  The former supports a ferromagnetic valley-polarized ground state below half-filling, developing a topological charge gap at half-filling for smaller twist angles. At larger twist angles, the system hosts a gapped triangular N\'eel antiferromagnet. On the other hand, the layer-polarized regime supports a stripe antiferromagnet below half-filling and a wing-shaped multiferroic ground state above half-filling. We map the evolution of these states as a function of filling factor, electric field, twist angle, and interaction strength. 
Our results demonstrate that long-range exchange in a symmetry-unbroken parent state with strongly renormalized moir\'e bands provides a broadly applicable framework to understand itinerant magnetism in moir\'e TMDs.
\end{abstract}

\maketitle

Moir\'e superlattice engineering of small rotational mismatches between atomically thin layers has opened up new avenues for studying strongly-correlated electronic phases in two-dimensional systems. In graphene-based moir\'es, such as magic-angle twisted bilayer graphene (TBG), electron-electron interactions in the nearly flat \moire bands near charge neutrality give rise to a rich phenomenology including correlated insulators \cite{cao_correlated_2018}, superconductivity \cite{cao_unconventional_2018}, among others. More recently, there has been a lot of experimental progress on twisted transition metal dichalcogenides (TMDs), where strong spin-orbit coupling, topology, and in-situ moir\'e band engineering provide new pathways for realizing correlated quantum phases \cite{anderson_programming_2023,cai_signatures_2023,kang_evidence_2024}. Here we focus on twisted bilayer WSe$_2$ (tWSe$_2$) where recent experiments have reported correlated magnetic states \cite{ghiotto_stoner_2024,knuppel_correlated_2025} and superconductivity \cite{xia_superconductivity_2025,guo_superconductivity_2025}.
\begin{figure}[ht!]
    \centering
    \includegraphics[trim={1cm 0.5cm 1.3cm 0.5cm},clip,width=\linewidth]{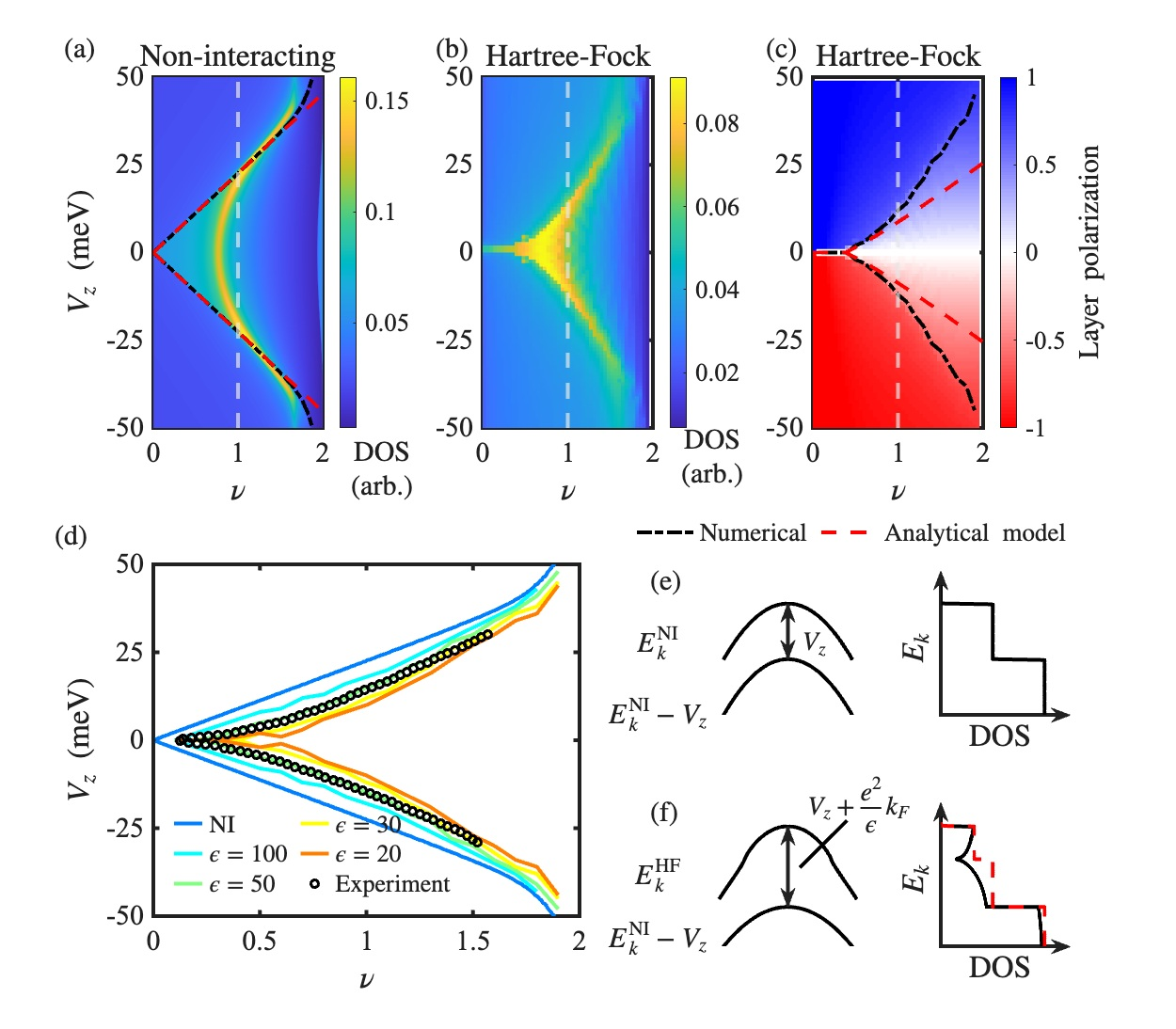}
    \caption{Interaction-renormalized moir\'e bands for the {\it symmetry unbroken} (``parent") Hartree-Fock state. Compared to the noninteracting theory (a), interactions shift the Van Hove singularity and soften the Lifshitz transition (b). This coincides with the change from layer-hybridized to layer-polarized regimes shown by the dotted line in (c). The red dashed line is an analytical result obtained in the absence of moir\'e potentials. (d) By matching the Lifshitz transition to experiment \cite{knuppel_correlated_2025} we can reproduce its shape with $\theta = 3.65^\circ$ and $\epsilon \approx 30$. Panels (e) and (f) illustrate how the linear boundary in (a) follows from the constant density of states of a 2DEG while the quadratic shape arises from the interplay between Fock interactions and \moire tunneling. }
    \label{fig:fig1}
    \vspace{-0.1in}
\end{figure}

Critical observations in tWSe$_2$ remain unresolved within a free-particle framework. First, experiments at twist angle $\theta = 3.65^\circ$ reveal a quadratic boundary \cite{xia_superconductivity_2025,knuppel_correlated_2025} 
between the layer-hybridized and layer-polarized phases in the $\nu \mbox{--} V_z$ plane (i.e. $V_z\propto\nu^2$). Here $\nu = n/n_M$ is the filling of the valence moir\'e bands (with $n_M$ the density per moir\'e cell) and $V_z$ is the interlayer bias due to an applied electric field perpendicular to the layers.  The single-particle theory predicts a linear boundary $V_z\propto\nu$ \cite{devakul_magic_2021}. Second, reflective magnetic circular dichroism (MCD), which probes the out-of-plane magnetization, identifies a pronounced ferromagnetic state in a triangular region of the phase diagram near zero electric field and $\nu < 1$, which develops ``wings" for $\nu > 1$ in finite electric fields and small magnetic fields \cite{knuppel_correlated_2025}. By contrast, the single-particle theory predicts a U-shaped feature that tracks the Van Hove singularity (VHS). Third, a superconducting phase emerges at half-filling ($\nu = 1$) for small electric fields, transitioning into a correlated insulator with increasing field strength before ultimately becoming metallic. These experimental findings motivate the need for a framework beyond the noninteracting theory.

In this Letter, we address the first two experimental observations by incorporating long-range Coulomb interactions within a Hartree-Fock (HF) scheme. The third feature, namely the observation of superconductivity at both small \cite{xia_superconductivity_2025} and large \cite{guo_superconductivity_2025} electric fields, has been discussed in some very recent theoretical works \cite{kim_theory_2024,zhu_theory_2024,guerci_topological_2024,schrade_nematic_2024,christos_approximate_2024,tuo_theory_2024,qin_kohn-luttinger_2024,xie_superconductivity_2024,fischer_theory_2024}. Our main results are summarized in Fig.~\ref{fig:fig1}.  We demonstrate that both the experimentally observed phase boundary and triangular ferromagnetic region can be understood within a self-consistent mean-field theory. We interpret our findings in terms of a Stoner instability of interaction-renormalized moir\'e bands of the symmetry-unbroken HF state. The success of our approach suggests a more broadly applicable hierarchy of approximations, where the theory of symmetry-broken phases should be based on an unbroken parent state with strongly renormalized moir\'e bands. To guide future experiments, we further investigate the phase diagram and orbital magnetization as a function of hole filling and electric field, interaction-driven Lifshitz transitions, as well as the nature of the HF ground state at half-filling.

\emph{Model} --- Our starting point is the Hamiltonian $\mathcal{H} = \mathcal{H}_0 + \mathcal{H}_\text{int}$ where $\mathcal{H}_0$ is the noninteracting low-energy moir\'e theory for the valence band near valley $K/K'$. Due to strong Ising spin-orbit coupling, spin $S_z$ is locked to the valley and $\mathcal{H}_0 = \sum_{\sigma = \uparrow,\downarrow} \int d^2 \bm r \, \psi_\sigma^\dag(\bm r) \mathcal H_0^\sigma \psi_\sigma(\bm r)$ with
\begin{equation} \label{eq:H0}
\mathcal H_0^\sigma =
\begin{bmatrix}
    \tfrac{\hbar^2 \nabla^2}{2m_*} + \varepsilon(\bm r) + \frac{V_z}{2} & t(-\sigma \bm r) \\ t(\sigma \bm r) & \tfrac{\hbar^2 \nabla^2}{2m_*} + \varepsilon(-\bm r) - \frac{V_z}{2}
\end{bmatrix},
\end{equation}
in the layer basis within the local-stacking approximation (LSA) \cite{jung_ab_2014,jung_origin_2015}.
Here we take $m_* = 0.45m_e$ for the effective mass, $V_z$ is an interlayer bias from the applied electric field, and $\varepsilon(\bm r)$ [$t(\bm r)$] are intralayer [interlayer] moir\'e potentials. These can be written as \textcolor{red}{}
\begin{align}
    \varepsilon(\bm r) & = \sum_j 2V_j \sum_{n=1}^3 \cos\left[ \bm b_j^{(n)} \cdot \bm \phi(\bm r) + \psi_j \right], \label{eq:intra} \\
    t(\bm r) & = \sum_j w_j \sum_{|\bm K + \bm b| = \text{const.}(j)} e^{i ( \bm K + \bm b ) \cdot \bm \phi(\bm r)}, \label{eq:inter}
\end{align}
where the first sum runs over monolayer reciprocal vectors $\bm b$. Here $\bm b_j^{(2,3)} = R(\pm2\pi/3) \bm b_j^{(1)}$ and $\bm K = 4\pi \hat x/3a$ with $a = 3.317$~\r A. For example, for the first star: $\bm b_1^{(1)} = 4\pi \hat y/\sqrt{3}a$ and $|\bm K + \bm b| = 4\pi/3a$. Here $\bm \phi(\bm r) = (a/a_M) \hat z \times \bm r + \bm u(\bm r)$ is the local stacking
with moir\'e length $a_M = a/[2\sin(\theta/2)]$ and $\bm u(\bm r)$ is the displacement field due to lattice relaxation \cite{carr_relaxation_2018,ezzi_analytical_2024}. The latter effectively modifies the rigid lattice moir\'e potentials in Eqs.\ \eqref{eq:intra} and \eqref{eq:inter}.
We find that lattice relaxation becomes significant for twist angles $\theta < 3^\circ$. 
In this work, we consider up to three stars of reciprocal lattice vectors in the expansion in Eqs.\ \eqref{eq:intra} and \eqref{eq:inter} using parameters obtained from density-functional theory (DFT) for untwisted bilayers. Computational details are in the Supplemental Material (SM) \footnote{See Supplemental Material [url] for \ldots} and nonzero values are shown in Table \ref{tab:parameters}.

\begin{table}[t!]
    \centering
    \begin{tabular}{c | c | c | c | c }
        \Xhline{1pt}
        $(V_1,\psi_1)$ & $(V_3,\psi_3)$ & $w_1$ & $w_2$ & $w_3$ \\
        \Xhline{1pt}
        $(6.61,89^\circ)$ & $(0.21,-94^\circ)$ & $12.96$ & $-1.95$ & $-0.58$ \\
        \hline
        $\sqrt{3}$ & $2\sqrt{3}$ & $1$ & $2$ & $\sqrt{7}$ \\
        \Xhline{1pt}
    \end{tabular}
    \caption{(top row) Parameters (in meV) of $H_0$ calculated from DFT for untwisted WSe$_2$ bilayers. (bottom row) Values for $|\bm b|$ (intralayer) and $|\bm K+\bm b|$ (interlayer) in units $4\pi/3a$.}
    \label{tab:parameters}
\end{table}

For the twist angles we consider, the long-wavelength moir\'e leaves valleys decoupled, and the system has U(1) valley charge conservation, i.e., $[\mathcal{H},S_z]=0$. In a single valley, the remaining symmetries are the magnetic point group $D_3(C_3) = \left< \mathcal C_{3z}, \mathcal C_{2y} \mathcal T \right>$ where $\mathcal T = i\sigma_y K$ is time reversal, and moir\'e translations. In addition, for $V_z = 0$ the moir\'e bands are spin degenerate: $E_n^\sigma(\bm k) \overset{\mathcal P}{=} E_n^\sigma(-\bm k) \overset{\mathcal T}{=} E_n^{-\sigma}(\bm k)$. Here $\mathcal P: (\bm r \mapsto -\bm r, \tau_x)$ where $\tau_x$ flips the layers, is an ``intravalley inversion" symmetry built into Eq.\ \eqref{eq:H0} by the LSA.
It is only exact in the first star of the moir\'e potentials and is broken when either $V_z \neq 0$ (which also breaks $\mathcal C_{2y}$) or we go beyond LSA \footnote{In this case $t^*(\bm r) \neq t(-\bm r)$ and we only have $\varepsilon_2(x,y) = \varepsilon_1(-x,y)$ by $\mathcal C_{2y}\mathcal T$ symmetry. For example, $\mathcal P$ symmetry is broken by $\arg(w_1/w_2) \neq 0,\pi$ and $\psi_2 \neq 0,\pi$ which is the phase of the second star that is equal for both layers.}. The latter lifts the spin degeneracy of the moir\'e bands, which can be seen along the $\gamma-m$ line in DFT calculations \cite{zhang_polarization-driven_2024,jia_moire_2024}.

We treat electron-electron interactions in the self-consistent HF approximation \cite{peng_many-body_2025} (see SM \cite{Note1} for details). In particular, we consider a dual-gate screened Coulomb interaction: $V_{\bm q} = 2\pi e^2\tanh{(d q)}/(\epsilon q)$ where $d$ is the gate-to-sample distance. Throughout this work, we set $d = 20\,\mathrm{nm}$ and treat the relative dielectric constant $\epsilon$ as a phenomenological parameter to study how physical properties depend on the interaction strength. The bandwidth of the topmost moir\'e band for $\theta = 3.65^\circ$ is about $40$ meV \cite{Note1}, similar to the Coulomb energy $e^2 / \epsilon a_M \approx 28\,\mathrm{meV}$ for $\epsilon = 10$. Hence the system is in the intermediate to strongly interacting regime.

\emph{Hartree-Fock parent state} --- We first consider the symmetry-unbroken HF state. While this is not necessarily the ground state, it serves as a {\it{parent state}} for symmetry-broken phases. In contrast to twisted bilayer graphene \cite{ezzi_self-consistent_2024,lewandowski_does_2021} and graphene on h-BN \cite{klein_imaging_2024} where away from charge neutrality band-renormalization is dominated by Hartree term, we find that for tWSe$_2$ the self-energy is always dominated by Fock term. Hence the same exchange interactions responsible for magnetic order also dominate band renormalization. In Fig.\ \ref{fig:fig1} we show the density of states (DOS) of the single-particle moir\'e bands (a) and the parent unbroken HF state (b) for $\theta = 3.65^\circ$ as function of hole doping $\nu$ and interlayer bias $V_z$. We find that the DOS is significantly modified by Coulomb interactions. This reshapes the VHS
and softens the Lifshitz transition between layer-hybridized and layer-polarized regimes [Fig.\ \ref{fig:fig1}(c)] as experimentally observed \cite{knuppel_correlated_2025}. 
In Fig.\ \ref{fig:fig1}(d) we compare the Lifshitz transition of the parent state to the experimental data of Ref.\ \cite{knuppel_correlated_2025}. We find that our results can reproduce the data for $\epsilon \approx 30$ where we assume that the interlayer bias is linear in the applied electric field.  From the experimental geometry, one would expect $\epsilon \sim 4$; however, on very general grounds, we can anticipate a strong suppression of the Fock interactions in \moire materials that is captured phenomenologically using a larger effective dielectric constant. 
We note that other theory works on TMDs also use larger values for the dielectric constant \cite{pan_topological_2022,qiu_interaction-driven_2023,li_electrically_2024}.

\begin{figure}
    \centering
    \includegraphics[trim={0cm 0 0cm 0},clip,width=1\columnwidth]{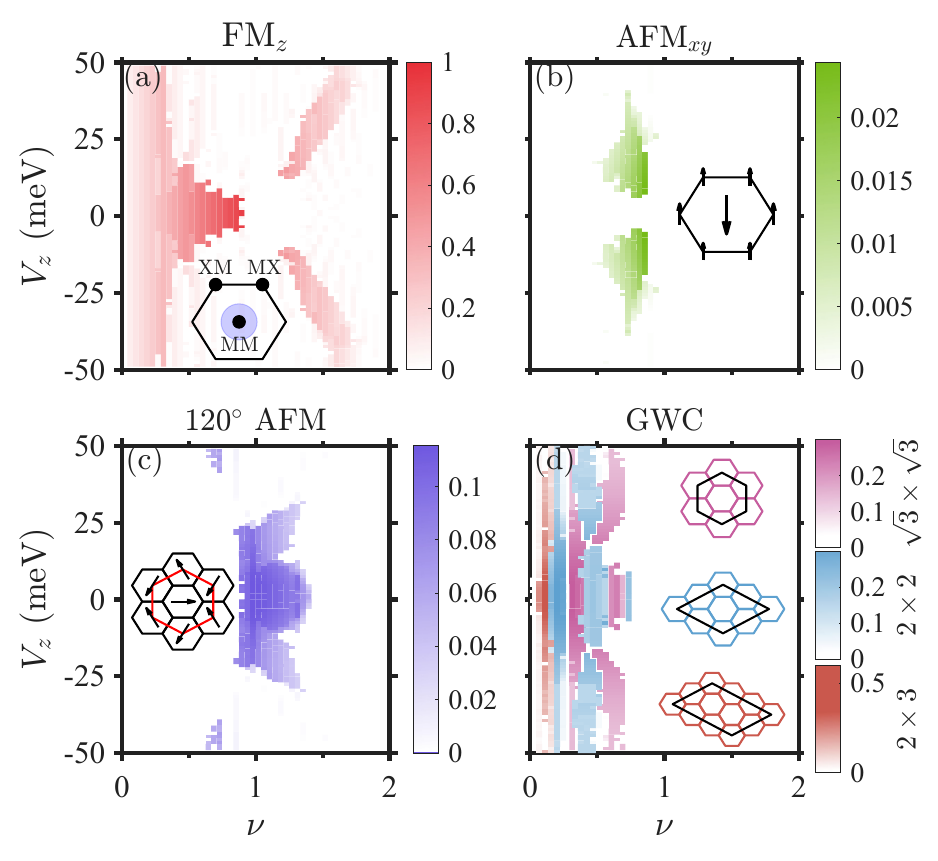}
    \caption{Ground state magnetic order and layer polarization for $\theta = 3.65^\circ$ and $\epsilon = 25$. (a) Out-of-plane spin polarization corresponding to the FM$_z$ phase, which exists both near charge neutrality and in regions where Fock interactions enhance the DOS [Fig.\ \ref{fig:fig1}(b)].
    The experimentally observed ferromagnetic region at small $V_z$ below half-filling ($\nu = 1$) is clearly visible. (b) In-plane spin polarization of the IVC stripe antiferromagnet, which exists only in a small $\nu$ window in layer-polarized regions. (c) Order parameter of the $120^\circ$ AFM: $\sum_{\bm k} f_{\bm k} \langle c_{\bm k,\uparrow}^\dag  c_{\bm k+\bm q,\downarrow} \rangle$. (d) Order parameter of the generalized Wigner crystal state with $\sqrt{3}\times\sqrt{3}$, $2\times2$, and $2\times3$ reconstructions of the moir\'e unit cell.}
    \label{fig:fig2}
\end{figure}

Some features of our numerical results can be understood from a toy model that neglects the moir\'e potentials. In this scenario, we have two parabolic bands $E^{\mathrm{NI}}_{\bm k}$ and $E^{\mathrm{NI}}_{\bm k}-V_z$ with $E^{\mathrm{NI}}_{\bm k} = -k^2/2m_*$ and $V_z > 0$, as illustrated in Fig.\ \ref{fig:fig1}(e). A Lifshitz transition occurs when the Fermi energy reaches the top of the second band, i.e, for $\nu =2 \Omega_M \int_{-V_{z,c}}^{0} \mathrm{DOS}(E) \, dE$, with $\Omega_M$ the \moire cell area, and the factor 2 accounts for spin. Consequently, we obtain $V_{z,c} = \pi \nu/\Omega_Mm^*$ which yields the linear boundary in the single-particle limit as shown in Fig.\ \ref{fig:fig1}(a).
Adding Coulomb interactions renormalizes the top band: $E^\text{HF}_{\bm k} = E^\text{NI}_{\bm k} + \Sigma^\text{F}_{\bm k}$, where $\Sigma^\text{F}_{\bm k}=  \int d^2\bm q V_{\bm k + \bm q} f_{\bm q}/(2\pi)^2$ \footnote{For a uniform electron gas the Hartree term is canceled by the jellium background} is the Fock self-energy with $f_{\bm q}$ the occupation. The integral can be carried out analytically \cite{Note1} and $E^\text{HF}_{\bm k} \simeq (e^2k_\text{F}/\epsilon) \left[ 1 - (k/2k_\text{F})^2 \right]$ where $k_\text{F} = \sqrt{2\pi \nu/\Omega_M}$. Hence the Fock correction shifts the upper band upward and reduces the mass $1/m_*^\text{F} = 1/m_*+e^2/2\epsilon k_{\mathrm{F}}$. 
For a given filling, the critical bias is now determined by
\begin{equation}
  \nu = 2\Omega_M \left( \int_{-V_{z,c}}^{E_F} \frac{dE m_*}{2\pi} + \int_{E_F}^{e^2 k_\text{F}/\epsilon} \frac{dE m_*^\text{F}}{2\pi} \right),
\end{equation}
where we approximated the DOS as illustrated by the red dashed line in Fig.\ \ref{fig:fig1}(f), yielding $V_{z,c} = \pi \nu/m_* - 3e^2 k_\text{F}/4\epsilon$. Hence interactions lead to two effects: (i) Below filling factor $\nu_0 = 9 e^2 m_*^2/(8\pi \epsilon^2 \Omega_m)$ the Fock term shifts the top band upwards such that it accommodates all holes and the layer-hybridized regime is never reached, and (ii) the critical $V_z$ gains a nonlinear correction given by $3e^2 k_\text{F}/4\epsilon \sim \sqrt{\nu}$. In Fig.\ \ref{fig:fig1}(c), we compare these results to the full numerics using an effective $\epsilon_*=40$.
We find that this simple model captures the trend that the layer-hybridized regime is reduced. The difference is attributed to the interlayer moir\'e coupling, which enhances the layer-hybridized regime. Hence, we conclude that the interplay between moir\'e tunneling and Fock interactions give rise to the parabolic-like transition observed both in our numerics and in experiment \cite{knuppel_correlated_2025}. Since these arguments are generic, we expect similar behavior in other tTMDs such as tMoTe$_2$ \cite{anderson_programming_2023}.

\emph{Magnetic orders} --- Next, we consider three symmetry-breaking magnetic orders: (i) a spin and valley polarized ferromagnetic state (FM$_z$) with finite $S_z$ polarization; (ii) an intervalley coherent (IVC) antiferromagnetic state that conserves moir\'e translations (AFM$_{xy}$); (iii) an IVC triangular N\'eel AFM that has a $\sqrt{3}\times\sqrt{3}$ reconstruction of the moir\'e cell (120$^\circ$ AFM), and (iv) a generalized Wigner crystal (GWC) with $\sqrt{3}\times\sqrt{3}$, $2\times2$, and $2\times3$ reconstructions of the moir\'e unit cell, which realizes a valley-polarized spin-density wave. The AFM$_{xy}$ state is an uncompensated stripe antiferromaget while the 120$^\circ$ AFM hosts clockwise rotating spins for $V_z >0$, and anticlockwise for $V_z <0$.  While all the IVC orders break $\mathcal T$ and valley U(1), they conserve an effective time-reversal symmetry $\mathcal T' = e^{i\pi\sigma_z/2} \mathcal T$ with $(\mathcal T')^2 = 1$ \cite{bultinck_ground_2020,wang_topology_2023}.

In Fig.\ \ref{fig:fig2} we show the three magnetic order parameters in the ground state, for $\theta = 3.65^\circ$ as a function of filling and interlayer bias. These are $S_z$, $S_{x,y}$, and $\langle c_{\bm k,\uparrow}^\dag  c_{\bm k+\bm q,\downarrow} \rangle$ with $\bm q = \pm 4\pi \hat y/(3L)$. We find the FM$_z$ phase [Fig.\ \ref{fig:fig2}(a)] exhibits a fully-polarized phase near $V_z=0$ below half-filling, and two ``wings" with partial polarization at finite $V_z$. This can be understood in terms of Stoner-like ferromagnetism \cite{stoner_collective_1997} of the HF parent state, where band renormalization from Fock interactions warps the VHS and induces a triangular region with enhanced DOS below half-filling near zero interlayer bias. These features are consistent with recent MCD measurements on tWSe$_2$ \cite{knuppel_correlated_2025} and similar observations in tMoTe$_2$ \cite{anderson_programming_2023}. 
We also find FM$_z$ at small filling, which may be invisible in MCD due to the low carrier density. On the other hand, the uncompensated stripe AFM$_{xy}$ has a small in-plane magnetization [Fig.\ \ref{fig:fig2}(b)] and appears at finite interlayer bias below half-filling, while the $120^\circ$ AFM state [Fig.\ \ref{fig:fig2}(c)] emerges at near half-filling in a finite $V_z$ window. 
Moreover, we find this state has a finite charge gap at $\nu=1$ as observed in experiment \cite{xia_superconductivity_2025,knuppel_correlated_2025}.
In addition, we find generalized Wigner crystal states as the ground states near $1/3$, $1/4$, and $1/6$ filling (one hole per reconstructed cell), and they remain stable under finite interlayer bias. This trend is consistent with experimental observations near $3.5^\circ$ twist \cite{knuppel_correlated_2025}. Allowing translational symmetry breaking also produces insulating states at $\nu=1/2$, $2/3$, and $3/4$, although these are much weaker with only a slight energy gain relative to the competing unreconstructed ferromagnet \cite{Note1}. For completeness, we also compute the phase diagram for tMoTe$_2$ at $\theta = 3.5^\circ$~\cite{Note1}, where the main difference is that the FM$_z$ state becomes the ground state at half filling matching recent experiments~\cite{li_universal_2025}.

\begin{figure}
    \includegraphics[trim={0.3cm 0.4cm 0.4cm 0.25cm},clip,width=0.95\columnwidth]{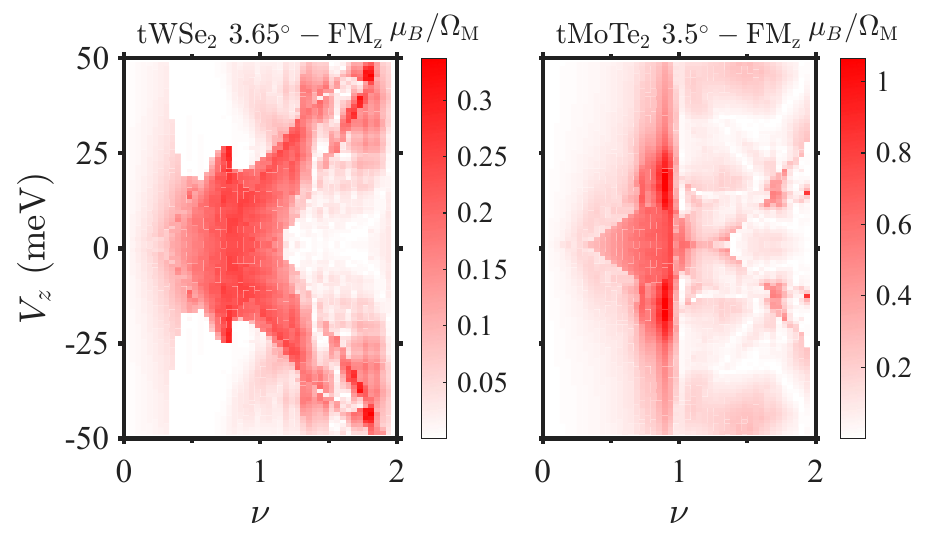}
    \caption{Orbital magnetization per \moire unit cell.  Results are shown as a function of filling $\nu$ and interlayer potential $V_z$ for the FM$_z$ phase for tWSe$_2$ at $\theta = 3.65^\circ$ (left) and tMoTe$_2$ at $\theta = 3.5^\circ$ (right). Twisted WSe$_2$ peaks at about $0.4~\mu_B$ near half filling, while twisted MoTe$_2$ shows a stronger response of about $1.2~\mu_B$ that is further enhanced by finite $V_z$, reflecting the larger Berry curvature and narrower bands.}
    \label{fig:fig3}
\end{figure}

\begin{figure}
    \centering
    \includegraphics[trim={0.1cm 0 1.6cm 0},clip,width=1\columnwidth]{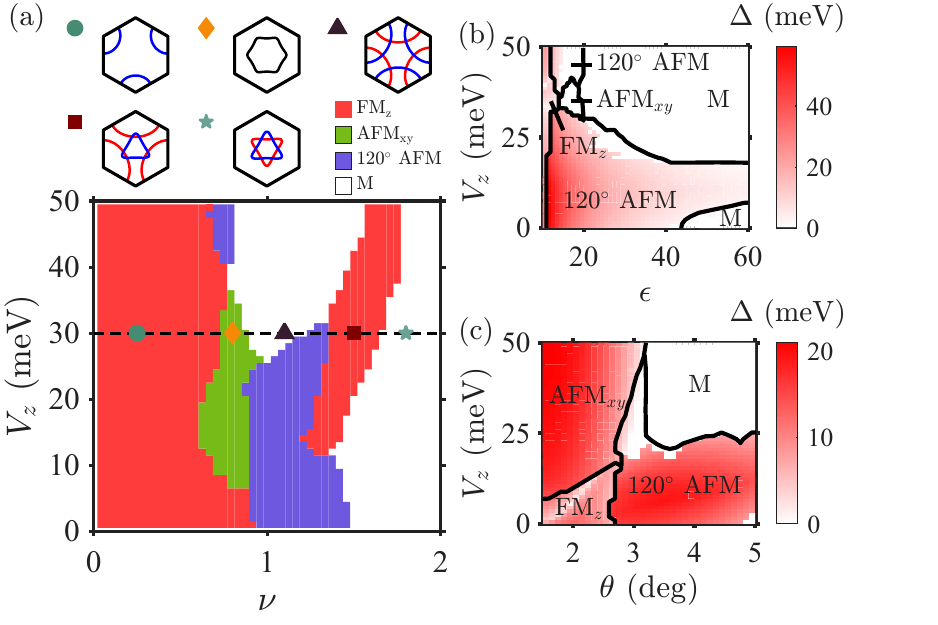}
    \caption{Phase diagram, Fermi surface topology and ground state at half filling versus twist angle and interaction strength. (a) Phase diagram for $\theta=3.65^\circ$ and $\epsilon = 25$, shown together with a representative Fermi surface of the gapless phases. Red and blue indicate the Fermi surfaces of the two valleys. The $\nu=1$ phase diagram versus (b) interaction strength ($1/\epsilon$) and (c) twist angle, where the color gives the charge gap $\Delta$. The gapped IVC $120^\circ$ AFM persist even for weak interactions in regions where the Fermi energy is close to the VHS on the unbroken parent state (M).}
    \label{fig:fig4}
\end{figure}

The complete $\nu$–$V_z$ phase diagram for $\theta = 3.65^\circ$ and $\epsilon = 25$ is shown in Fig.~\ref{fig:fig4}(a). A key feature is an interaction-driven Lifshitz transition that is absent in the non-interacting model. The parent state has a single Lifshitz boundary separating layer-hybridized (cyan pentagram) and layer-polarized (black triangle) phases. Interactions shift this boundary, and each symmetry-broken phase develops its own Fermi surface. Below half filling, the FM$_z$ phase hosts a single valley-polarized pocket. As the filling increases, the AFM$_{xy}$ stripe phase with IVC quasiparticles, becomes the ground state and exhibits a single connected pocket. With further filling, the FM$_z$ phase returns and the pocket within one valley merges again. These results show that long-range exchange sets both the symmetry and the Fermi-surface topology across the phase diagram.

\emph{Spin and orbital magnetization} --- To investigate the origin of the observed magnetization, we compare spin and orbital contributions. The spin contribution is given by the out-of-plane polarization shown in Fig.~\ref{fig:fig2}(a), and reaches about $1\ \mu_B$ per moir\'e cell near half filling. The orbital part \cite{Note1,xiao_berry_2005,shi_quantum_2007} is shown in Fig.~\ref{fig:fig3}(a). We find that tMoTe$_2$ has a much larger orbital response than tWSe$_2$, with a peak of about $1.2\ \mu_B$ per moir\'e cell slightly below half filling, which is further enhanced for finite $V_z$. This reflects larger Berry curvature and narrower valley Chern bands. Unlike TBG at odd fillings—where spin polarization is weak \cite{tschirhart_imaging_2021}, the magnetism in tWSe$_2$ and tMoTe$_2$ has comparable spin and orbital parts due to spin–valley locking. The moir\'e-band contributions match DFT-based estimates \cite{song_collective_2024} but are a factor of $\sim$2 lower than experiments~\cite{redekop_direct_2024}. As discussed in Refs.~\cite{deilmann_ab_2020, wozniak_exciton_2020}, this discrepancy may stem from the fact that our continuum model is constructed from $d_{x^2-y^2} \pm i d_{xy}$ orbitals rather than the full Bloch wave functions.

\emph{Correlated insulator at half-filling} --- At half-filling, transitions between a superconductor, a correlated insulator, and a metal were recently reported \cite{xia_superconductivity_2025}.  
To understand the origin of these phases in WSe$_2$, we examine the ground state in the self-consistent HF approximation. Our results at half-filing are shown here for completeness and are in agreement with previous reports in the literature focusing mostly on tMoTe$_2$~\cite{pan_topological_2022,qiu_interaction-driven_2023, xie_nematic_2023,li_electrically_2024,wang_diverse_2024}.
We emphasize that the filling dependence of these states has not been considered previously.  In Fig.\ \ref{fig:fig4}(b) and (c), we show the phase diagram and charge gap as a function of the interaction strength, interlayer bias, and twist angle.  For a small interlayer bias $V_z$, as we increase the interaction strength, the metallic symmetry-unbroken phase (M) gives way to a $120^\circ$ AFM with a topologically trivial gap. The experimentally observed superconductivity \cite{xia_superconductivity_2025,guo_superconductivity_2025} is believed to be mediated by magnons of the $120^\circ$ AFM. At larger $V_z$ the single-particle bandwidth increases yielding a metallic phase except for strong interactions. However, for intermediate $V_z$, the $120^\circ$ AFM persists even for weak interactions. This is due to the proximity of the Fermi energy to the VHS of the unbroken state. The FM$_z$ phase only becomes the ground state for very strong interactions yielding a quantum anomalous Hall insulator with unit Chern number. For fixed $\epsilon=25$ the FM$_z$ state appears for small twists close to $V_z = 0$ meV transitioning to a trivially gapped stripe AFM as $V_z$ increases. For larger twists, a gapped $120^\circ$ AFM dominates for small interlayer bias until $|V_z| = 25~\text{meV}$ where the gap closes to a the unbroken phase due to the increased single-particle bandwidth. For small twist angles the FM$_z$ and AFM$_{xy}$ states differ in energy by less than $0.1$ meV per moir\'e cell~\cite{Note1}, making the ground state sensitive to strain and other small perturbations. In this regime the FM–AFM balance at half filling is less reliable, and a weak out-of-plane magnetic field ($B \gtrsim 0.3$ T) is needed to stabilize the FM$_z$ state in experiments.

\emph{Conclusions}--- Our self-consistent Hartree-Fock calculations offer a useful semi-quantitative guide of the ferromagnetic behavior observed experimentally. Our results emphasize the importance of first having the interaction-renormalized band structure before incorporating the effects of strong correlations.  In such \moire systems, the interaction driven modification of the band structures can be significant, even at the mean-field level.  As illustrated here for the case of tWSe$_2$, experimental observations of correlated states can be understood as a simple Stoner-like ferromagnetism once the interaction-modification of the bands are accounted for.  These results highlight the importance of long-range Coulomb interactions in capturing the quantum phases in twisted 2D materials and the methods we employ are generally applicable to a wide class of \moire systems. Several future directions can be pursued within this framework. For instance, the superconducting phase at half-filling could be explored by incorporating superconducting diagrams within the Nambu formalism \cite{nambu_quasi-particles_1960}.

\let\oldaddcontentsline\addcontentsline 
\renewcommand{\addcontentsline}[3]{} 
\begin{acknowledgments}
\emph{Acknowledgments} --- We thank Valentin Cr\'epel, Charles Kane, Daniel Mu\~noz-Segovia, and Wenjin Zhao for helpful discussions. Computation was done on Stampede3 at the Texas Advanced Computing Center through allocation PHY240263 from the ACCESS program supported by the U.S. National Science Foundation. L.P.\ and S.A.\ are supported by a start-up grant at Washington University in St. Louis.  C.D.B.\ and E.J.M.\ are supported by the U.S.\ Department of Energy under Grant No.\ DE-FG02-84ER45118.  D.L.\ and L.Y.\ acknowledge support from U.S. National Science Foundation  DMR-2124934.  
\end{acknowledgments}

\emph{Note Added} --- A few days after our work was posted to arXiv, a similar Hartree-Fock study on magnetism in tWSe$_2$ appeared with similar conclusions but a different focus~\cite{munoz-segovia_twist-angle_2025}.

\bibliography{references}
\let\addcontentsline\oldaddcontentsline 

\include{SM.tex}

\end{document}

%% file: SM.tex
\clearpage
\onecolumngrid
\begin{center}
\textbf{\Large Supplemental Material for ``Magnetism in Twisted Bilayer WSe$_2$''}
\end{center}
 
\setcounter{equation}{0}
\setcounter{figure}{0}
\setcounter{table}{0}
\setcounter{page}{1}
\setcounter{secnumdepth}{2}
\makeatletter
\renewcommand{\thepage}{S\arabic{page}}
\renewcommand{\thesection}{S\arabic{section}}
\renewcommand{\theequation}{S\arabic{equation}}
\renewcommand{\thefigure}{S\arabic{figure}}
\renewcommand{\thetable}{S\arabic{table}}

\tableofcontents
\vspace{1cm}

\twocolumngrid

\section{Continuum moir\'e theory} \label{sm:moire}

We construct the moir\'e theory using the local-stacking approximation starting from untwisted bilayers. In this way, we can easily account for lattice relaxation. 

\subsection{Untwisted bilayers}

We consider the valence band of parallel stacked bilayer 2H WSe$_2$, and we construct an effective model near the valence band edge at $K/K'$ as a function of the stacking. Because the 2H monolayer breaks inversion symmetry, Ising spin-orbit coupling is allowed which leads to a substantial splitting of the topmost valence band at $K/K'$. Hence we only need to consider spin-up (spin-down) states at $K$ ($K'$). The effective Hamiltonian for the bilayer near the valence band maximum can then be written as 
\begin{equation}
    H(\bm k,\bm \phi) = \begin{bmatrix}
        -\tfrac{\hbar^2 k^2}{2m_*} + \varepsilon(\bm \phi) & t(-\bm \phi) \\ t(\bm \phi) & -\tfrac{\hbar^2 k^2}{2m_*} + \varepsilon(-\bm \phi)
    \end{bmatrix},
\end{equation}
where $m_* = 0.45m_e$ is the effective mass of monolayer WSe$_2$ at the $K/K'$ point and $\bm \phi = (\phi_x, \phi_y)$ is the stacking vector giving the relative lateral shift between the two layers.
For example, $\bm \phi = (0,0)$ corresponds to MM stacking while $\bm \phi = (0, \pm a /\sqrt{3})$ is MX/XM stacking with $a$ the lattice constant. Here we already used that,
\begin{equation} \label{eq:homobilayer}
    H(\bm k,\bm \phi) = \tau_x H(\bm k,-\bm \phi) \tau_x,
\end{equation}
which follows from the mirror symmetry of the 2H monolayer \cite{wu_topological_2019}. This turns out to only be an approximate symmetry of the moir\'e theory. Moreover, periodicity implies $H(\bm k,\bm \phi + \bm a) = H(\bm k,\bm \phi)$ where $\bm a$ is a monolayer lattice vector. We further defined the intralayer potential $\varepsilon(\bm \phi)$ and the tunneling amplitude $t(\bm \phi)$ from layer $1$ to $2$.

\subsubsection{Symmetry}

The point group of the 2H monolayer is given by $D_{3h} = D_3 \times \sigma_h$ with $D_3 = \left< \mathcal C_{3z}, \mathcal C_{2y} \right>$. Here $\mathcal C_{3z}$ is a threefold rotation about the $z$ axis and $\mathcal C_{2y}$ is twofold rotation about the $y$ axis. This implies
\begin{align}
    H(\bm k,\bm \phi) & = H(\bm k, \mathcal C_{3z} \bm \phi), \\
    & = \tau_x H^*(\bm k,\phi_x,-\phi_y) \tau_x \\
    & = H^*(\bm k,-\phi_x,\phi_y).
\end{align}
Here complex conjugation is due to exchanging $K$ and $K'$ under $\mathcal C_{2y}$ since they are related by time-reversal symmetry. And the last line follows from Eq.\ \eqref{eq:homobilayer}. We obtain the following conditions
\begin{align}
    \varepsilon(\bm \phi) & = \varepsilon(-\phi_x,\phi_y) = \varepsilon(\mathcal C_{3z} \bm \phi), \\
    t(\bm \phi) & = t^*(-\bm \phi) = t(\phi_x,-\phi_y) = t(\mathcal C_{3z} \bm \phi).
\end{align}
Furthermore, since $\epsilon$ is real, the symmetry-allowed Fourier expansion is given by
\begin{align}
    \varepsilon(\bm \phi) & = 2V_1 \sum_{i=1}^3 \cos(\bm b_i \cdot \bm \phi + \psi_1) \\
    & + 2V_2 \sum_{i=1}^3 \cos(\bm b_i' \cdot \bm \phi) \\
    & + 2V_3 \sum_{i=1}^3 \cos(2\bm b_i \cdot \bm \phi + \psi_3) \\ 
    & + \cdots,
\end{align}
where $\bm b_i$ ($\bm b_i'$) are reciprocal lattice vectors of the first (second) star that are related by $\mathcal C_{3z}$ rotations. For example $\bm b_1' = \bm b_1 - \bm b_2$. The constant term is omitted as it only gives an overall energy shift.
\begin{figure}
    \centering
    \includegraphics[width=.8\linewidth]{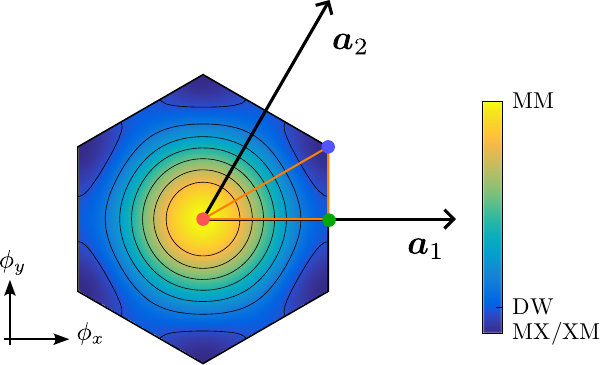}
    \caption{Illustration of the energy landscape of parallel-stacked bilayers of 2H TMDs as a function of the stacking configuration $\bm \phi$. The orange triangles gives all configurations not related by symmetry. The dots correspond to MM (red), MX/XM (blue), and DW (green) stacking.}
    \label{fig:stacking}
\end{figure}

\subsubsection{Interlayer tunneling}

Noting that the states at $K/K'$ are comprised mostly of $d$ orbitals of the metal atoms, we require that the maximal amplitude of the interlayer coupling corresponds to MM stacking centers, while $t(\text{MX})=t(\text{XM})=0$ in lowest order. Taking into account the symmetry constraints, we find
\begin{align}
    t(\bm \phi) & = w_1 e^{i \bm K \cdot \bm \phi} \left[ 1 + e^{i \bm b_2 \cdot \bm \phi} + e^{i \left( \bm b_1 + \bm b_2 \right) \cdot \bm \phi} \right] \\
    & + w_2 e^{i \bm K \cdot \bm \phi} \left[ e^{i \bm b_1 \cdot \bm \phi} + e^{-i \bm b_1 \cdot \bm \phi} + e^{i \left( \bm b_1 + 2 \bm b_2 \right) \cdot \bm \phi} \right] \\
    & + \cdots,
\end{align}
with $\bm K = -\left( \bm b_1 + 2 \bm b_2 \right) / 3$. Here $w_1$ and $w_2$ are real, $\bm b_1 = 4\pi \hat y/\sqrt{3}a$, and $\bm b_2 = \mathcal C_{3z} \bm b_1$. 

\subsubsection{Fitting to DFT}

\begin{figure}
    \centering
    \includegraphics[width=1\columnwidth]{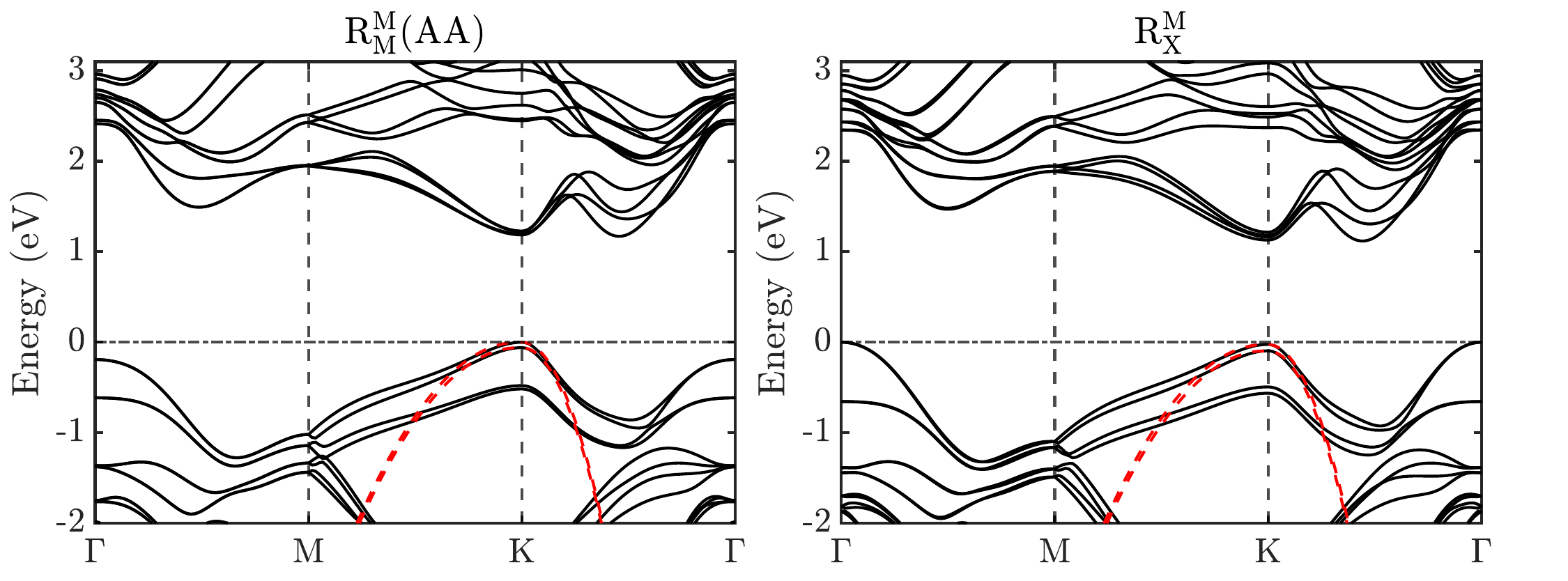}
    \caption{DFT band structures for MM and MX stacking (black solid lines) including spin-orbit coupling. The red dashed line corresponds to the fitted bands near the $K$ and $K'$ valleys. We find the following fitting parameters for the first star: $V_1=6.6$~meV, $\psi_1=89^\circ$, and $w_1=13$~meV.}
    \label{fig:BandsFittingWSe2_K}
\end{figure}

In general, we can write the Hamiltonian as
\begin{equation}
    H(\bm k, \bm \phi) = \left[ -\frac{\hbar^2k^2}{2m_*} + d_0(\bm \phi) \right] \tau_0 + \bm d(\bm \phi) \cdot \bm \tau,
\end{equation}
with energies
\begin{equation}
    E_\pm(\bm k,\bm \phi) = -\frac{\hbar^2k^2}{2m_*} + d_0(\bm \phi) \pm |\bm d(\bm \phi)|.
\end{equation}
We fit these expressions to the valence band maximum and the splitting at the $K$ point taking into account up to three stars. The results are shown in Table \ref{tab:parameters} of the main text and illustrated and Fig.\ \ref{fig:BandsFittingWSe2_K}.

The DFT calculation used to obtain Fig.\ \ref{fig:BandsFittingWSe2_K} was performed using the local density approximation exchange-correlation functional with spin orbital coupling following Ref.~\cite{troullier_efficient_1991}, and implemented using the Quantum ESPRESSO package \cite{giannozzi_advanced_2017} with a wavefunction cutoff of 60 Ry, a charge density cutoff of 400 Ry and an $18 \times 18 \times 1$ momentum grid. Moreover, a vacuum of 20 \r A between adjacent layers is used to avoid spurious interactions between periodic images along the out-of-plane direction of two-dimensional bilayer structures.

\subsection{Local-stacking approximation}

In the long-wavelength limit, a twist moir\'e is defined by the local stacking configuration 
\begin{equation} \label{eq:lsa}
    \bm \phi(\bm r) = \frac{a}{L} \hat z \times \bm r + \bm u(\bm r),
\end{equation}
with $L = a/2\sin(\theta/2)$ the moir\'e lattice constant and where $\bm u(\bm r)$ is the acoustic displacement field due to lattice relaxation. The latter reconstructs the rigid moir\'e pattern leading to domain-wall formation at sufficiently small twist angles. The moir\'e lattice is then defined by
\begin{equation}
    \bm \phi(\bm r+\bm L) = \bm \phi(\bm r) + \bm a,
\end{equation}
with $\bm L$ moir\'e lattice vectors. 

In the local-stacking approximation, we replace the constant configuration $\bm \phi$ in the Hamiltonian by $\bm \phi(\bm r)$ from Eq.\ \eqref{eq:lsa} \cite{jung_origin_2015}. The moir\'e continuum theory then becomes $H_0 = \sum_{\sigma = \uparrow,\downarrow} \int d^2 \bm r \, \psi_\sigma^\dag(\bm r) \mathcal H_0^\sigma \psi_\sigma(\bm r)$ with
\begin{equation} \label{eq:continuum1}
    \mathcal H_0^\sigma =
    \begin{bmatrix}
        \tfrac{\hbar^2 \nabla^2}{2m_*} + \varepsilon(\bm r) & t(-\sigma\bm r) \\ t(\sigma\bm r) & \tfrac{\hbar^2 \nabla^2}{2m_*} + \varepsilon(-\bm r)
    \end{bmatrix},
\end{equation}
where $\psi_\sigma = ( \psi_{\sigma1}, \psi_{\sigma2} )^T$ fermion field operators in layer basis. Here we defined $\varepsilon(\bm r) = \varepsilon[\bm \phi(\bm r)]$ and $t(\bm r) = t[\bm \phi(\bm r)]$ where we note that
\begin{equation}
    \bm b \cdot \bm \phi(\bm r) = \bm g \cdot \bm r + \bm b \cdot \bm u(\bm r),
\end{equation}
with moir\'e reciprocal vectors $\bm g = 2 \sin \left( \tfrac{\theta}{2} \right) \bm b \times \hat z$.

\subsection{Symmetries of the moir\'e}

The moir\'e magnetic point group at $K/K'$ is given by $D_3(C_3) = C_3 + (D_3 / C_3) \mathcal T = \left< \mathcal C_{3z}, \mathcal C_{2y} \mathcal T \right>$. The action of the symmetries on the field operators is chosen as
\begin{align}
    \mathcal T \psi_\sigma(\bm r) \mathcal T^{-1} & = \psi_{-\sigma}(\bm r), \\
    \mathcal C_{2y} \psi_\sigma(x,y) \mathcal C_{2y}^{-1} & = \psi_{-\sigma}(-x,y), \\
    \left( \mathcal C_{2y} \mathcal T \right) \psi_\sigma(x,y) \left( \mathcal C_{2y} \mathcal T \right)^{-1} & = \sigma_x \psi_\sigma(-x,y),
\end{align}
with $\mathcal T i \mathcal T^{-1} = -i$ and
\begin{equation}
    \mathcal C_{3z} \psi_\sigma(\bm r) \mathcal C_{3z}^{-1} = \psi_\sigma(\mathcal C_{3z} \bm r),
\end{equation}
Given a symmetry $\mathcal S$ we require $[H_0,\mathcal S] = 0$. The local-stacking moir\'e continuum Hamiltonian obeys all these symmetries. However, in general we have
\begin{equation}
    \mathcal H_0^\sigma =
    \begin{bmatrix}
        \tfrac{\hbar^2 \nabla^2}{2m_*} + \varepsilon_1(\bm r) & t^*(\bm r) \\ t(\bm r) & \tfrac{\hbar^2 \nabla^2}{2m_*} + \varepsilon_2(\bm r)
    \end{bmatrix}.
\end{equation}
After a change of integration variables, we find that $\mathcal C_{2y} \mathcal T$ symmetry implies
\begin{align}
    \varepsilon_1(x,y) & = \varepsilon_2(-x,y) = \varepsilon(x,y), \\
    t(x,y) & = t(-x,y),
\end{align}
for the intralayer moir\'e potentials and moir\'e tunneling. Similarly, threefold rotation symmetry yields
\begin{equation}
    \varepsilon(\mathcal C_{3z}^{-1} \bm r) = \varepsilon(\bm r), \quad t(\mathcal C_{3z}^{-1} \bm r) = t(\bm r).
\end{equation}
These symmetries impose the following conditions on the moir\'e bands: $E_n^\sigma(\bm k) = E_n^\sigma(k_x,-k_y) = E_n^\sigma (\mathcal C_{3z}\bm k)$, and $E_n^{-\sigma}(\bm k) = E_n^\sigma(-\bm k)$ where $n$ is the band index.

In the local-stacking approximation, the theory has an additional model ``intravalley inversion" symmetry
\begin{equation}
    \mathcal P \psi_\sigma(\bm r) \mathcal P^{-1} = \sigma_x \psi_s(-\bm r),
\end{equation}
which is incorporated in Eq.\ \eqref{eq:continuum1}. This symmetry is inherited from the mirror symmetry about the $xy$ plane of the 2H monolayer. One finds that $\mathcal P$ symmetry implies
\begin{equation}
    \varepsilon_1(\bm r) = \varepsilon_2(-\bm r), \qquad t^*(\bm r) = t(-\bm r).
\end{equation}
Together with the symmetries of the moir\'e we find
\begin{equation}
    \varepsilon(x,y) = \varepsilon(x,-y), \qquad t^*(x,y) = t(x,-y),
\end{equation}
as before. One consequence of $\mathcal P$ is that the moir\'e bands become spin degenerate.

Importantly, if we restrict the moir\'e potentials to the first moir\'e star then $\mathcal P$ symmetry is satisfied for real $w_1$. If $w_1$ is not real, the action of $\mathcal P$ can be modified since $w_1$ can always be made real by introducing a relative phase between layers: $\psi_\sigma(\bm r) \mapsto e^{i \arg(w) \tau_z / 2} \psi_\sigma(\bm r)$. In this case
\begin{align}
    \varepsilon(\bm r) & = 2V_1 \sum_{i=1}^3 \cos \left( \bm g_i \cdot \bm r + \psi_1 \right), \\
    t(\bm r) & = w_1 \sum_{i=1}^3 e^{-i \bm q_i \cdot \bm r},
\end{align}
where $\bm q_1 = k_\theta \left( 0, 1 \right)$, $\bm q_2 = k_\theta ( -\sqrt{3}/2, -1/2 )$, and $\bm q_3 = -\bm q_1 -\bm q_2$ with $k_\theta = 4\pi/3L$. One needs to go to the second moir\'e star to break $\mathcal P$. For example
\begin{align}
    \varepsilon_1(\bm r) & = 2V_1 \sum_{i=1}^3 \cos \left( \bm g_i \cdot \bm r + \psi_1 \right) \\
    & + 2V_2 \sum_{i=1}^3 \cos \left( \bm g_i' \cdot \bm r + \psi_2 \right), \nonumber \\
    \varepsilon_2(\bm r) & = 2V_1 \sum_{i=1}^3 \cos \left( \bm g_i \cdot \bm r - \psi_1 \right) \\
    & + 2V_2 \sum_{i=1}^3 \cos \left( \bm g_i' \cdot \bm r + \psi_2 \right), \nonumber
\end{align}
where $\psi_2 \neq 0$ breaks $\mathcal P$. Here $\bm g_1' = \bm g_1 - \bm g_2$ and the rest are related by threefold rotations. Another way to break $\mathcal P$ is through a nonzero relative phase $\arg(w_1/w_2)$ of the first and second star of the interlayer tunneling
\begin{equation}
    t(\bm r) = w_1 \sum_{i=1}^3 e^{-i \bm q_i \cdot \bm r} + w_2 \sum_{i=1}^3 e^{-i \bm q_i' \cdot \bm r},
\end{equation}
where $\arg(w_1/w_2)$ is invariant under a relative phase change between layers. Here $\bm q_1' = -2\bm q_1$ and the rest are related by $\mathcal C_{3z}$ rotations. These terms will lift the spin degeneracy between the moir\'e bands along the $\gamma-m$ line of the moir\'e Brillouin zone, for example.

\subsection{Lattice relaxation}

The acoustic displacement field $\bm u(\bm r)$ due to lattice relaxation inherits the $C_{6v}$ symmetry of the stacking-fault energy of bilayer WSe$_2$ with parallel stacking. This implies $\bm \phi(-\bm r) = -\bm \phi(\bm r)$ such that Eq.\ \eqref{eq:continuum1} still holds when we account for lattice relaxation. For ``large" twist angles $\theta > 3^\circ$, the displacement field is well approximated by \cite{ezzi_analytical_2024}
\begin{equation}
    \bm u(\bm r) = 2u_1 \sum_{i=1}^3 \hat z \times \hat g_i \sin(\bm g_i \cdot \bm r),
\end{equation}
with $u_1 = \sqrt{3} a c_1 / 2\pi \theta^2$ and where the sum runs over the three moir\'e reciprocal vectors of the first star related by $\mathcal C_{3z}$ rotations. Here $c_1 = U_1 / \mu \sim 4 \times 10^{-4}$ for tWSe$_2$ near $0^\circ$ with $U_1$ the first-star Fourier coefficient of the stacking-fault energy and $\mu$ the shear modulus. In the local-stacking approximation, the first star becomes
\begin{align}
    \varepsilon(\bm r) & = 2V_1 \sum_{n=1}^3 \cos \left[ \bm b_n \cdot \bm \phi(\bm r) + \psi_1 \right], \\
    t(\bm r) & = w_1 \sum_{n=1}^3 e^{i R(2\pi n/3) \bm K \cdot \bm \phi(\bm r)}.
\end{align}
Note that lattice relaxation does not break $\mathcal C_{3z}$ and $\mathcal C_{2y} \mathcal T$ symmetry because
\begin{align}
    \bm u(x,y) & = \text{diag} \left( 1, -1 \right) \bm u(-x,y), \\
    \bm u(\mathcal C_{3z} \bm r) & = \mathcal C_{3z} \bm u(\bm r),
\end{align}
respectively, and where in the first line an additional minus sign appears due to layer reversal.

In lowest order, we find
\begin{align}
    \frac{\varepsilon(\bm r)}{2V} & = \sum_{i=1}^3 \cos \left( \bm g_i \cdot \bm r + \psi \right) - \frac{c_1}{\theta^2} \sum_{i=1}^3 \cos \left( \bm g_i \cdot \bm r - \psi \right) \\
    & + \frac{2 c_1 \cos \psi}{\theta^2} \sum_{i=1}^3 \cos \left( \bm g_i' \cdot \bm r \right) \\
    & + \frac{2 c_1}{\theta^2} \sum_{i=1}^3 \cos \left( 2 \bm g_i \cdot \bm r + \psi \right),
\end{align}
which obeys all the symmetries of the moir\'e. For example, the first star is renormalized as $V e^{i \psi} \mapsto V \left( e^{i\psi} - c_1 e^{-i\psi} / \theta^2 \right)$. Here we did not include an overall constant $V_0 - 12V_1c_1 \cos(\psi) / \theta^2$. Similarly, for the interlayer tunneling in lowest order,
\begin{align}
    \frac{t(\bm r)}{w} & = \sum_{i=1}^3 \left( 1 - \frac{4\pi u_1}{\sqrt{3}a} \right) e^{-i\bm q_i \cdot \bm r} \\
    & + \frac{2 \pi u_1}{\sqrt{3}a} \sum_{i=1}^6 e^{-i\bm q_i' \cdot \bm r} \\
    & = \sum_{i=1}^3 \left( 1 - \frac{2 c_1}{\theta^2} \right) e^{-i\bm q_i \cdot \bm r} + \frac{c_1}{\theta^2} \sum_{i=1}^6 e^{-i\bm q_i'' \cdot \bm r},
\end{align}
where $\bm q_1'' = 3 \bm q_1 + \bm q_2$ and $\bm q_4'' = 2 \bm q_1 - \bm q_2$ and the rest are related by $\mathcal C_{3z}$. The latter correspond to the third interlayer moir\'e shell. Note that lattice relaxation effectively reduces the interlayer coupling because it leads to an increase of the MX/XM stacking regions. This explains why the bands in DFT calculations \cite{zhang_polarization-driven_2024} are more dispersive than those from moir\'e continuum theories that are fitted to large twist angles \cite{devakul_magic_2021}. We further note that the renormalization of the moir\'e tunneling is identical to the change in the AA interlayer tunneling in twisted bilayer graphene under lattice relaxation  \cite{ezzi_analytical_2024}. In general, we perform the Fourier transform of the relaxed moir\'e potential numerically to extract the new values for generated moir\'e reciprocal stars.

\section{Toy model}

The Fock integral for a 2DEG is given by
\begin{equation}
    \Sigma_F(\bm k) = \int \frac{d^2\bm q}{(2\pi)^2} V_{\bm k+\bm q} f_{\bm q}.
\end{equation}
At zero temperature for a Coulomb potential, we find
\begin{widetext} 
\begin{equation}
    \Sigma_F(\bm k) = \frac{e^2k_\text{F}}{8\pi^2\epsilon} \int_0^1 dq \int_0^{2\pi} d\theta \, \frac{q}{\sqrt{q^2 + 2kq\cos\theta + k^2}},
\end{equation}
with $k$ in units of $k_\text{F}$. Substituting $u = \cos \theta$ we find
\begin{align}
    \Sigma_F(\bm k) & = \frac{e^2}{4\pi^2\epsilon} \int_{-1}^1 \frac{du}{\sqrt{1-u^2}} \left( \sqrt{1 + k^2 + 2ku} - k - ku \ln \left[ \frac{\sqrt{1 + k^2 + 2ku} - ku -1}{k(1-u)} \right] \right) \\
    & = \frac{e^2}{4\pi\epsilon} F(k),
\end{align}
with
\begin{equation}
    F(k) = \frac{1}{2\pi} \left\{ ( 1 - k ) K \left[ \frac{4k}{(1+k)^2} \right]
    + ( 1 + k ) E \left[ \frac{4k}{(1+k)^2} \right]
    + |1-k| E \left[ -\frac{4k}{(1-k)^2} \right]
    + \frac{1-k^2}{|1-k|} K \left[ -\frac{4k}{(1-k)^2} \right] \right\},
\end{equation}
\end{widetext}
where
\begin{align}
    K(z) & = \int_0^{\pi/2} \frac{d\theta}{\sqrt{1 - z \sin^2 \theta}}, \\
    E(z) & = \int_0^{\pi/2} d\theta \sqrt{1 - z \sin^2 \theta}, 
\end{align}
are complete elliptic integrals. Moreover, we find
\begin{equation}
    F(k) \simeq 1 - \frac{k^2}{4} - \frac{3k^4}{64} + \mathcal O(k^6).
\end{equation}

\section{Hartree-Fock Theory}

In this section, we introduce the Coulomb interaction and develop the formalism required for Hartree-Fock calculations. Our theoretical framework is based primarily on the many-body perturbation approach for moir\'e systems from Ref.~\cite{peng_many-body_2025}. Here we generalize this method to handle translation symmetry-breaking orders by introducing enlarged unit cells, effectively folding the original moir\'e Brillouin zone (mBZ). 

We begin with the full Hamiltonian $H = H_{0} + H_{\mathrm{I}}$. The non-interacting Hamiltonian is given by
\begin{equation}
H_{0}
=\sum_{\tilde{\mathbf{k}},\tilde{\mathbf{k}}^{\prime}}\sum_{\sigma}  H^{\sigma}(\tilde{\mathbf{k}},\tilde{\mathbf{k}}^{\prime}) \hat{c}^{\dagger}_{\tilde{\mathbf{k}},\sigma} \hat{c}_{\tilde{\mathbf{k}}^{\prime},\sigma},
\end{equation}
with electron creation and annihilation operators $\hat{c}^{\dagger}_{\tilde{\mathbf{k}},\sigma}$ and $\hat{c}_{\tilde{\mathbf{k}},\sigma}$ for momentum $\tilde{\mathbf{k}}$ and spin $\sigma$. The interacting Hamiltonian takes the form
\begin{equation}
\begin{aligned}
H_{\mathrm{I}}&=
\frac{1}{2}
\sum_{\tilde{\mathbf{k}},\tilde{\mathbf{k}}^{\prime},\tilde{\mathbf{q}}}
\sum_{\sigma,\sigma^{\prime}}
V_{\tilde{\mathbf{q}}}
\hat{c}_{\tilde{\mathbf{k}}+\tilde{\mathbf{q}},\sigma}^{\dagger}
\hat{c}_{\tilde{\mathbf{k}}^{\prime}-\tilde{\mathbf{q}},\sigma^{\prime}}^{\dagger}
\hat{c}_{\tilde{\mathbf{k}}^{\prime},\sigma^{\prime}}
\hat{c}_{\tilde{\mathbf{k}},\sigma},\\
\end{aligned}
\end{equation}
where momenta $\tilde{\mathbf{k}},\tilde{\mathbf{k}}^{\prime},\tilde{\mathbf{q}}$ lie within the Brillouin zone, and the screened Coulomb potential is $V_{\tilde{\mathbf{q}}}=2\pi e^2\tanh(d q)/(\epsilon q)$, with gate-to-sample distance $d$.

To handle translation symmetry-breaking orders, we introduce enlarged unit cells, or equivalently fold the mBZ \cite{xie_phase_2023}. Each order corresponds to a specific reciprocal lattice vector pair $\mathbf{Q}_{1,2}$. The folding number is defined as
\begin{equation} N_F=\frac{|\mathbf{b}_1\times \mathbf{b}_2|}{|\mathbf{Q}_1\times \mathbf{Q}_2|}, \end{equation} 
which represents the number of times the Brillouin zone is folded, where $\mathbf{b}_i$ are reciprocal lattice vectors of the mBZ. Any momentum $\tilde{\mathbf{k}}$ in the BZ can then be decomposite as
\begin{equation}
\tilde{\mathbf{k}}=\mathbf{k}+\mathbf{Q}+\mathbf{G},
\end{equation}
where $\mathbf{Q}=l_1\mathbf{Q}_1+l_2\mathbf{Q}_2$ with $l_1$ and $l_2$ are integer number stand for all the $N_F$ reciprocal vectors of the folded moir\'e Brillouin zones in the unfolded moir\'e Brillouin zones. Specifically, for FM$_z$ and AFM$_{xy}$, we take $\mathbf{Q}_1=\mathbf{b}_1$, $\mathbf{Q}_2=\mathbf{b}_2$ so that $N_F=1$. For the $120^\circ$ AFM and the $\sqrt{3}\times\sqrt{3}$ GWC state, we choose $\mathbf{Q}_1=(\mathbf{b}_2-\mathbf{b}_1)/3$, $\mathbf{Q}_2=(\mathbf{b}_1+2\mathbf{b}_2)/3$, giving $N_F=3$. For the $2\times2$ GWC state, we use $\mathbf{Q}_1=\mathbf{b}_1/2$, $\mathbf{Q}_2=\mathbf{b}_2/2$ with $N_F=4$, and for the $2\times3$ GWC state, $\mathbf{Q}_1=\mathbf{b}_1/2$, $\mathbf{Q}_2=\mathbf{b}_2/3$ with $N_F=6$. Using this decomposition, the interacting Hamiltonian becomes
\begin{widetext}
\begin{equation}
H_{\mathrm{I}}=\frac{1}{2}
\sum_{\mathbf{k},\mathbf{k}^{\prime},\mathbf{q}}
\sum_{\mathbf{Q},\mathbf{Q}^{\prime},\mathbf{Q}^{\prime\prime}}
\sum_{\mathbf{G},\mathbf{G}^{\prime},\mathbf{G}^{\prime\prime}}
\sum_{\sigma,\sigma^{\prime}}
V_{\mathbf{q}+\mathbf{Q}^{\prime\prime}+\mathbf{G}^{\prime\prime}}
\hat{c}_{\mathbf{k}+\mathbf{Q}+\mathbf{G}+\mathbf{q}+\mathbf{Q}^{\prime\prime}+\mathbf{G}^{\prime\prime},\sigma}^{\dagger}
\hat{c}_{\mathbf{k}^{\prime}+\mathbf{Q}^{\prime}+\mathbf{G}^{\prime}-\mathbf{q}-\mathbf{Q}^{\prime\prime}-\mathbf{G}^{\prime\prime},\sigma^{\prime}}^{\dagger}
\hat{c}_{\mathbf{k}^{\prime}+\mathbf{Q}^\prime+\mathbf{G}^{\prime},\sigma^{\prime}}
\hat{c}_{\mathbf{k}+\mathbf{Q}+\mathbf{G},\sigma}.
\end{equation}

We transform into the band basis by diagonalizing the single-particle Hamiltonian
\begin{equation}\label{eq:nibasis}
\hat{c}_{\mathbf{k},\mathbf{Q},n,\sigma}=\sum_{\mathbf{G}}u_{n}(\mathbf{k}+\mathbf{Q}+\mathbf{G};\sigma)\hat{c}_{\mathbf{k}+\mathbf{Q}+\mathbf{G},\sigma},
\end{equation}
where $n$ indexes energy bands and $\hat{c}_{\mathbf{k}, \mathbf{Q}, \sigma}$ is shorthand for $\hat{c}_{\mathbf{k}+\mathbf{Q}, \sigma}$. In this new basis, the non-interacting Hamiltonian becomes diagonal
\begin{equation}
\begin{aligned}
H_{0}
&=\sum_{\mathbf{k},\mathbf{Q}} \sum_{n,\sigma}  E^{\sigma}_{n}(\mathbf{k},\mathbf{Q})\hat{c}^{\dagger}_{\mathbf{k},\mathbf{Q},n,\sigma} \hat{c}_{\mathbf{k},\mathbf{Q},n,\sigma},\\
\end{aligned}
\end{equation}
where $E^{\sigma}_{n}(\mathbf{k},\mathbf{Q})$ are the eigenvalues (band energy) of the single-particle Hamiltonian for each $\sigma$, and $H_{\mathrm{I}}$ is recasted as
\begin{equation}
\begin{aligned}
H_{\mathrm{I}}&=\frac{1}{2}
\sum_{\{n_i\}}
\sum_{\mathbf{k},\mathbf{k}^{\prime},\mathbf{q}}
\sum_{\mathbf{Q},\mathbf{Q}^{\prime},\mathbf{Q}^{\prime\prime}}
\sum_{\mathbf{G},\mathbf{G}^{\prime},\mathbf{G}^{\prime\prime}}
\sum_{\sigma,\sigma^{\prime}}
V_{\mathbf{q}+\mathbf{Q}^{\prime\prime}+\mathbf{G}^{\prime\prime}}
u^{*}_{n_1}(\mathbf{k}+\mathbf{q}+\mathbf{Q}+\mathbf{Q}^{\prime\prime}+\mathbf{G}+\mathbf{G}^{\prime\prime};\sigma)
u^{*}_{n_2}(\mathbf{k}^{\prime}-\mathbf{q}+\mathbf{Q}^{\prime}-\mathbf{Q}^{\prime\prime}+\mathbf{G}^{\prime}-\mathbf{G}^{\prime\prime};\sigma^\prime)\times\\
&\quad\quad
u_{n_3}(\mathbf{k}^{\prime}+\mathbf{Q}^{\prime},\mathbf{G}^{\prime};\sigma^\prime)
u_{n_4}(\mathbf{k}+\mathbf{Q},\mathbf{G};\sigma)
\hat{c}_{\mathbf{k}+\mathbf{q},\mathbf{Q}^{\prime}+\mathbf{Q}^{\prime\prime},n_1,\sigma}^{\dagger}
\hat{c}_{\mathbf{k}^{\prime}-\mathbf{q},\mathbf{Q}^{\prime}-\mathbf{Q}^{\prime\prime},n_2,\sigma^{\prime}}^{\dagger}
\hat{c}_{\mathbf{k}^{\prime},\mathbf{Q}^{\prime},n_3,\sigma^{\prime}}
\hat{c}_{\mathbf{k},\mathbf{Q},n_4,\sigma}\\
&=\frac{1}{2}
\sum_{\{n_i\}}
\sum_{\mathbf{k},\mathbf{k}^{\prime},\mathbf{q}}
\sum_{\mathbf{Q},\mathbf{Q}^{\prime},\mathbf{Q}^{\prime\prime}}
\sum_{\mathbf{G}^{\prime\prime}}
\sum_{\sigma,\sigma^{\prime}}
V_{\mathbf{q}+\mathbf{Q}^{\prime\prime}+\mathbf{G}^{\prime\prime}}
\left[\sum_{\mathbf{G}}
u^{*}_{n_1}(\mathbf{k}+\mathbf{q}+\mathbf{Q}+\mathbf{Q}^{\prime\prime}+\mathbf{G}+\mathbf{G}^{\prime\prime};\sigma)
u_{n_4}(\mathbf{k}+\mathbf{Q}+\mathbf{G};\sigma)
\right]\times\\
&\quad\quad
\left[\sum_{\mathbf{G}^{\prime}}
u^{*}_{n_2}(\mathbf{k}^{\prime}-\mathbf{q}+\mathbf{Q}^{\prime}-\mathbf{Q}^{\prime\prime}+\mathbf{G}^{\prime}-\mathbf{G}^{\prime\prime};\sigma^\prime)
u_{n_3}(\mathbf{k}^{\prime}+\mathbf{Q}^{\prime},\mathbf{G}^{\prime};\sigma^\prime)
\right]
\hat{c}_{\mathbf{k}+\mathbf{q},\mathbf{Q}+\mathbf{Q}^{\prime\prime},n_1,\sigma}^{\dagger}
\hat{c}_{\mathbf{k}^{\prime}-\mathbf{q},\mathbf{Q}^{\prime}-\mathbf{Q}^{\prime\prime},n_2,\sigma^{\prime}}^{\dagger}
\hat{c}_{\mathbf{k}^{\prime},\mathbf{Q}^{\prime},n_3,\sigma^{\prime}}
\hat{c}_{\mathbf{k},\mathbf{Q},n_4,\sigma}\\
&=\frac{1}{2}
\sum_{\{n_i\}}
\sum_{\mathbf{k},\mathbf{k}^{\prime},\mathbf{q}}
\sum_{\mathbf{Q},\mathbf{Q}^{\prime},\mathbf{Q}^{\prime\prime}}
\sum_{\mathbf{G}^{\prime\prime}}
\sum_{\sigma,\sigma^{\prime}}
V_{\mathbf{q}+\mathbf{Q}^{\prime\prime}+\mathbf{G}^{\prime\prime}}
\left[\Lambda^{*}_{\mathbf{k}+\mathbf{Q},\mathbf{q}+\mathbf{Q}^{\prime\prime}+\mathbf{G}^{\prime\prime}}\right]^{\sigma}_{n_4n_1}
\left[\Lambda^{*}_{\mathbf{k}^{\prime}+\mathbf{Q}^{\prime},-\mathbf{q}-\mathbf{Q}^{\prime\prime}-\mathbf{G}^{\prime\prime}}\right]^{\sigma^\prime}_{n_3n_2}\times\\
&\quad\quad\hat{c}_{\mathbf{k}+\mathbf{q},\mathbf{Q}+\mathbf{Q}^{\prime\prime},n_1,\sigma}^{\dagger}
\hat{c}_{\mathbf{k}^{\prime}-\mathbf{q},\mathbf{Q}^{\prime}-\mathbf{Q}^{\prime\prime},n_2,\sigma^{\prime}}^{\dagger}
\hat{c}_{\mathbf{k}^{\prime},\mathbf{Q}^{\prime},n_3,\sigma^{\prime}}
\hat{c}_{\mathbf{k},\mathbf{Q},n_4,\sigma}.\\
\end{aligned}
\end{equation}
where we have defined the form factor
\begin{equation}
\left[\Lambda_{\mathbf{k}+\mathbf{Q},\mathbf{q}+\mathbf{Q}^{\prime\prime}+\mathbf{G}^{\prime\prime}}\right]^{\sigma}_{m n}=\sum_{\mathbf{G}}
u^*_{m} \left(\mathbf{k}+\mathbf{Q}+\mathbf{G} ; \sigma\right) 
u_{n} \left(\mathbf{k}+\mathbf{Q}+\mathbf{G}    +\mathbf{q}+\mathbf{Q}^{\prime\prime}+\mathbf{G}^{\prime\prime} ; \sigma\right).
\end{equation}
\end{widetext}
We then define the single-particle imaginary-time Green's function in the band basis
\begin{equation}
    \left[G(\mathbf{k},\tau)\right]_{\eta\eta^{\prime}}=-\langle \mathcal{T}_{\tau} \hat{c}_{\mathbf{k},\eta}(\tau)\hat{c}^{\dagger}_{\mathbf{k},\eta^{\prime}}(0) \rangle,
\end{equation}
where $\eta = (n, \sigma)$ combines the band index $n$ and the internal quantum number $\sigma$. Here, $\hat{c}_{\mathbf{k},\eta}(\tau)$ and $\hat{c}^{\dagger}_{\mathbf{k},\eta^{\prime}}(0)$ are the annihilation and creation operators in the Heisenberg picture, $\mathcal{T}_{\tau}$ denotes the time-ordering operator, and $\langle \cdots \rangle$ represents the thermal average over the interacting system. By performing a Fourier transform with respect to imaginary time $\tau$, we obtain the Green's function in frequency space
\begin{equation}
\left[G(\mathbf{k},i\omega_n)\right]_{\eta\eta^{\prime}}=\int_{0}^{\beta}e^{i\omega_n\tau}\left[G(\mathbf{k},\tau)\right]_{\eta\eta^{\prime}}d\tau,
\end{equation}
where $\beta = 1/(k_{\mathrm{B}} T)$ is the inverse temperature, $k_{\mathrm{B}}$ is Boltzmann's constant, and $i\omega_n = (2n + 1)\pi/\beta$ are the fermionic Matsubara frequencies.

In the non-interacting limit, the Green's function is diagonal in $\eta$ and simplifies to
\begin{equation}
\left[G_{0}(\mathbf{k},i\omega_n)\right]_{\eta\eta^{\prime}} = \frac{\delta_{\eta,\eta^{\prime}}}{i\omega_n - E_{n}^{\sigma}(\mathbf{k}) +\mu}, 
\end{equation} 
where $E_{n}^{\sigma}(\mathbf{k})$ are the eigenvalues of the single-particle Hamiltonian, and $\mu$ is the global chemical potential. The interacting Green's function, $\hat{G}(\mathbf{k},i\omega_n)$, is determined by the Dyson equation
\begin{equation}\label{eq:Dyson}
\hat{G}^{-1}(\mathbf{k},i\omega_n) = \hat{G}^{-1}_0(\mathbf{k},i\omega_n) - \hat{\Sigma}(\mathbf{k},i\omega_n),
\end{equation}
where $\hat{\Sigma}(\mathbf{k},i\omega_n)$ represents the electron self-energy, and $\hat{G}_0(\mathbf{k},i\omega_n)$ denotes the non-interacting Green's function. Using Feynman diagram techniques, the Hartree self-energy could be written down as
\begin{widetext}
\begin{equation}
\begin{aligned}
\left[\Sigma_{\mathrm{H}}(\mathbf{k}^{\prime},i\omega_n)\right]_{\mathbf{Q}^{\prime},n_3,\sigma^{\prime}}^{\mathbf{Q}^{\prime}-\mathbf{Q}^{\prime\prime},n_2,\sigma^{\prime}}
&= \frac{1}{\beta}
\sum_{\mathbf{k},\mathbf{G}^{\prime\prime}}
\sum_{m}
\sum_{\mathbf{Q}}
\sum_{n_1,n_4,\sigma}
V_{\mathbf{Q}^{\prime\prime}+\mathbf{G}^{\prime\prime}}
\left[\Lambda^{*}_{\mathbf{k}+\mathbf{Q},\mathbf{Q}^{\prime\prime}+\mathbf{G}^{\prime\prime}}\right]^{\sigma}_{n_4n_1}
\left[\Lambda^*_{\mathbf{k}^{\prime}+\mathbf{Q}^{\prime},-\mathbf{Q}^{\prime\prime}-\mathbf{G}^{\prime\prime}}\right]^{\sigma^\prime}_{n_3n_2}
\left[G(\mathbf{k},i\omega_m)\right]^{\mathbf{Q},n_4,\sigma}_{\mathbf{Q}+\mathbf{Q}^{\prime\prime},n_1,\sigma}\\
&= 
\sum_{\mathbf{G}^{\prime\prime}}
V_{\mathbf{Q}^{\prime\prime}+\mathbf{G}^{\prime\prime}}
\left[\Lambda_{\mathbf{k}^{\prime}+\mathbf{Q}^{\prime},\mathbf{Q}^{\prime\prime}+\mathbf{G}^{\prime\prime}}\right]^{\sigma^\prime}_{n_2n_3}
\sum_{\mathbf{k}}
\frac{1}{\beta}\sum_{m}
\sum_{\mathbf{Q}}
\sum_{n_1,n_4,\sigma}
\left[\Lambda^{*}_{\mathbf{k}+\mathbf{Q},\mathbf{Q}^{\prime\prime}+\mathbf{G}^{\prime\prime}}\right]^{\sigma}_{n_4n_1}
\left[G(\mathbf{k},i\omega_m)\right]^{\mathbf{Q},n_4,\sigma}_{\mathbf{Q}+\mathbf{Q}^{\prime\prime},n_1,\sigma},\\
\end{aligned}
\end{equation}
Similarly, for Fock self-energy
\begin{equation}
\begin{aligned}
&\left[\Sigma_{\mathrm{F}}(\mathbf{k},i\omega_n)\right]^{\mathbf{Q}^{\prime},n_2,\sigma^{\prime}}_{\mathbf{Q},n_4,\sigma}\\
&=-\frac{1}{\beta}\sum_{m}
\sum_{n_1,n_3}
\sum_{\mathbf{q},\mathbf{G}^{\prime\prime},\mathbf{Q}^{\prime\prime}}
V_{\mathbf{q}+\mathbf{Q}^{\prime\prime}+\mathbf{G}^{\prime\prime}}
\left[\Lambda^{*}_{\mathbf{k}+\mathbf{Q},\mathbf{q}+\mathbf{Q}^{\prime\prime}+\mathbf{G}^{\prime\prime}}\right]^{\sigma}_{n_4n_1}
\left[\Lambda^{*}_{\mathbf{k}+\mathbf{Q}^{\prime},-\mathbf{q}-\mathbf{Q}^{\prime\prime}-\mathbf{G}^{\prime\prime}}\right]^{\sigma^\prime}_{n_3n_2}
\left[G(\mathbf{k}+\mathbf{q}+\mathbf{Q}^{\prime\prime},i\omega_n-i\omega_m)\right]^{\mathbf{Q}^{\prime},n_3,\sigma^{\prime}}_{\mathbf{Q},n_1,\sigma}\\
&=-\frac{1}{\beta}\sum_{m}
\sum_{n_1,n_3}
\sum_{\mathbf{q},\mathbf{G}^{\prime\prime},\mathbf{Q}^{\prime\prime}}
V_{\mathbf{q}+\mathbf{Q}^{\prime\prime}+\mathbf{G}^{\prime\prime}}
\left[\Lambda^{*}_{\mathbf{k}+\mathbf{Q},\mathbf{q}+\mathbf{Q}^{\prime\prime}+\mathbf{G}^{\prime\prime}}\right]^{\sigma}_{n_4n_1}
\left[G(\mathbf{k}+\mathbf{q}+\mathbf{Q}^{\prime\prime},i\omega_n-i\omega_m)\right]^{\mathbf{Q}^{\prime},n_3,\sigma^{\prime}}_{\mathbf{Q},n_1,\sigma}
\left[\Lambda_{\mathbf{k}+\mathbf{Q}^{\prime},\mathbf{q}+\mathbf{Q}^{\prime\prime}+\mathbf{G}^{\prime\prime}}\right]^{\sigma^\prime}_{n_2n_3},\\
\end{aligned}
\end{equation}
where we have used the relationship
$\hat{\Lambda}_{\mathbf{k}+\mathbf{q},-\mathbf{q}-\mathbf{G}}=\hat{\Lambda}^{\dagger}_{\mathbf{k},\mathbf{q}+\mathbf{G}}$.

Since we consider only the first-order diagrams, the Matsubara summation can be performed explicitly. The Hartree and Fock self-energy can be rewritten in terms of the single-particle density matrix $\hat{\rho}(\mathbf{k})$, which is related to the Green's function by
\begin{equation}\label{eq:density_matrix}
\hat{\rho}(\mathbf{k})=\hat{G}(\mathbf{k},\tau=0^-)=\frac{1}{\beta}\sum_{n}e^{-i\omega_n0^+}\hat{G}(\mathbf{k},i\omega_n),
\end{equation}
with $0^{\pm}$ ensuring proper analytic continuation. Thus, the Hartree and Fock self-energies become
\begin{equation}\label{eq:hartree}
\begin{aligned}
\left[\Sigma_{\mathrm{H}}(\mathbf{k})\right]_{\mathbf{Q},n_3,\sigma}^{\mathbf{Q}-\mathbf{Q}^{\prime\prime},n_2,\sigma}
&= 
\sum_{\mathbf{G}^{\prime\prime}}
V_{\mathbf{Q}^{\prime\prime}+\mathbf{G}^{\prime\prime}}
\left[\Lambda_{\mathbf{k}+\mathbf{Q},\mathbf{Q}^{\prime\prime}+\mathbf{G}^{\prime\prime}}\right]^{\sigma}_{n_2n_3}
\sum_{\mathbf{k}^{\prime},\mathbf{Q}^{\prime}}
\sum_{n_1,n_4,\sigma^{\prime}}
\left[\Lambda^{*}_{\mathbf{k}^{\prime}+\mathbf{Q}^{\prime},\mathbf{Q}^{\prime\prime}+\mathbf{G}^{\prime\prime}}\right]^{\sigma^{\prime}}_{n_4n_1}
\left[\rho(\mathbf{k}^{\prime})\right]^{\mathbf{Q}^{\prime},n_4,\sigma^{\prime}}_{\mathbf{Q}^{\prime}+\mathbf{Q}^{\prime\prime},n_1,\sigma^{\prime}},\\
\end{aligned}
\end{equation}
and
\begin{equation}\label{eq:fock}
\begin{aligned}
\left[\Sigma_{\mathrm{F}}(\mathbf{k})\right]^{\mathbf{Q}^{\prime},n_2,\sigma^{\prime}}_{\mathbf{Q},n_4,\sigma}
&=\sum_{n_1,n_3}
\sum_{\mathbf{q},\mathbf{G}^{\prime\prime},\mathbf{Q}^{\prime\prime}}
V_{\mathbf{q}+\mathbf{Q}^{\prime\prime}+\mathbf{G}^{\prime\prime}}
\left[\Lambda^{*}_{\mathbf{k}+\mathbf{Q},\mathbf{q}+\mathbf{Q}^{\prime\prime}+\mathbf{G}^{\prime\prime}}\right]^{\sigma}_{n_4n_1}
\left[\rho(\mathbf{k}+\mathbf{q}+\mathbf{Q}^{\prime\prime})\right]^{\mathbf{Q}^{\prime},n_3,\sigma^{\prime}}_{\mathbf{Q},n_1,\sigma}
\left[\Lambda_{\mathbf{k}+\mathbf{Q}^{\prime},\mathbf{q}+\mathbf{Q}^{\prime\prime}+\mathbf{G}^{\prime\prime}}\right]^{\sigma^\prime}_{n_2n_3}.\\
\end{aligned}
\end{equation}
\end{widetext}
Therefore, the total self-energy is given by
\begin{equation}\label{eq:self_energy}
\hat{\Sigma}(\mathbf{k})=\hat{\Sigma}_{\mathrm{H}}(\mathbf{k})+\hat{\Sigma}_{\mathrm{F}}(\mathbf{k}).
\end{equation}

In practical calculations, we impose a cutoff in the band basis denoted by $N$. Consequently, the Green's function is represented by a $(2N_F N)\times(2N_F N)$ matrix, where the factor of two comes from the spin degree of freedom. We set $N=3$ in this work and verify that increasing this cutoff does not qualitatively change our results.

We solve the Hartree-Fock equations using an iterative method. We begin with an initial ansatz (guess) of the density matrix $\hat{\rho}(\mathbf{k})$. Next, we compute the Hartree and Fock self-energies using Eqs.\eqref{eq:hartree}–\eqref{eq:self_energy}. A new Green's function is obtained from Dyson's equation (Eq.\eqref{eq:Dyson}), from which we then compute the updated density matrix using Eq.\eqref{eq:density_matrix}. We repeat these steps until convergence to a self-consistent solution is achieved. 

For the initial ansatz, we fill the top moir\'e band of the density matrix according to the specific phase under consideration, denoted as $\hat{\rho}_0$. For states that preserve translational symmetry, the density matrix for the top band is a $2\times 2$ matrix in spin space. Specifically, for the symmetry unbroken state, we take $\hat{\rho}_0 = \alpha \sigma_0$; for the FM$_z$ state, $\hat{\rho}_0 = \alpha \sigma_z$; and for the AFM$_{xy}$ state, $\hat{\rho}_0 = \alpha \sigma_x$, where $\sigma_i$ denote the Pauli matrices acting on spin indices, $\sigma_0$ is the identity matrix, and $\alpha$ is a energy constant.

For the $120^{\circ}$ AFM state, the density matrix for the top band is a $6 \times 6$ matrix, constructed as
\begin{equation}
\hat{\rho}_0 =
\alpha \begin{pmatrix}
\hat{0} & \hat{M} \\
\hat{M}^\dagger & \hat{0}
\end{pmatrix},
\end{equation}
where
\begin{equation}
\hat{M} =
\begin{cases}
\begin{pmatrix}
0 & 1 & 0 \\
0 & 0 & 1 \\
1 & 0 & 0
\end{pmatrix} & \text{for } V_z \geq 0 \\
\\
\begin{pmatrix}
0 & 0 & 1 \\
1 & 0 & 0 \\
0 & 1 & 0
\end{pmatrix} & \text{for } V_z < 0
\end{cases}
\end{equation}
in the basis $\{|\bm{k}, \uparrow\rangle,\ |\bm{k}+\bm{Q}_1, \uparrow\rangle,\ |\bm{k}+\bm{Q}_2, \uparrow\rangle,\ |\bm{k}, \downarrow\rangle,\ |\bm{k}+\bm{Q}_1, \downarrow\rangle,\ |\bm{k}+\bm{Q}_2, \downarrow\rangle\}$. These two ansatze correspond to opposite chiralities of the $120^{\circ}$ AFM state.

For the generalized Wigner crystal, which is a valley-polarized spin-density wave, the density matrix in the top band has dimension $2N_F\times 2N_F$ and is written as  
\begin{equation}
\hat{\rho}_0=
\alpha
\begin{pmatrix}
\hat{J} & \hat{0} \\
\hat{0} & \hat{0}
\end{pmatrix},
\end{equation}
where $\hat{J}$ is the all-ones matrix of size $N_F\times N_F$ (that is, $(\hat{J})_{ij}=1$ for all $i,j$) and $\hat{0}$ denotes the zero matrix of the same size.

Once the interacting Green's function is determined, the total energy of the system can be calculated using the Galitskii-Migdal formula \cite{galitskii_application_1958}
\begin{equation}
E_{\mathrm{tot}}
=\frac{1}{\beta}\sum_{\mathbf{k},n}e^{-i\omega_n0^+}\mathrm{Tr}\left[\left(\hat{H}_0(\mathbf{k})
+\frac{1}{2}\hat{\Sigma}(\mathbf{k},i\omega_n)\right)\hat{G}(\mathbf{k},i\omega_n)\right].
\end{equation}
If the self-energy is frequency-independent, this expression simplifies to
\begin{equation}\label{eq:total_energy}
E_{\mathrm{tot}}
=\sum_{\mathbf{k}}\mathrm{Tr}\left[\left(\hat{H}_0(\mathbf{k})
+\frac{1}{2}\hat{\Sigma}(\mathbf{k})\right)\hat{\rho}(\mathbf{k})\right].
\end{equation}

\begin{figure}
    \includegraphics[trim={0cm 0 0cm 0},clip,width=0.9\columnwidth]{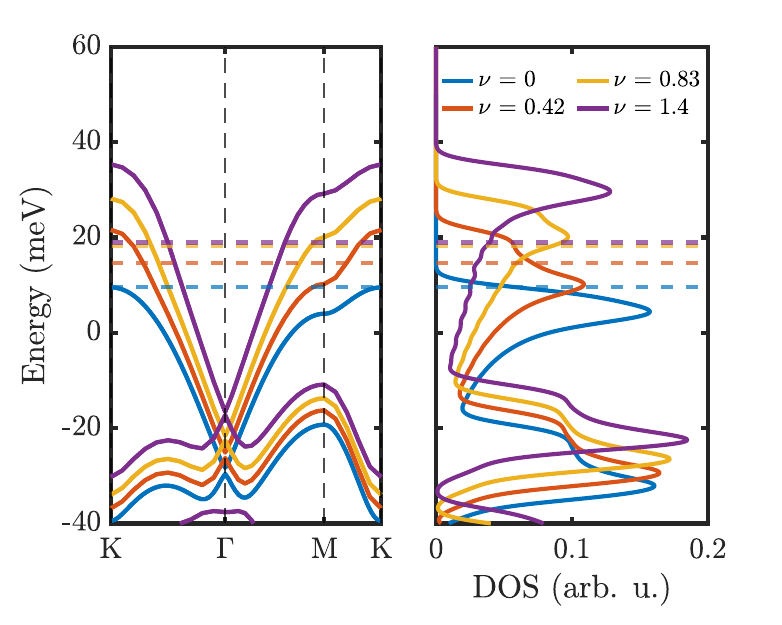}
    \caption{Symmetry-unbroken Hartree--Fock band structure (left) and density of states (right) for several fillings. Curves correspond to $\nu = 0$, $0.42$, $0.83$, and $1.4$. The HF self-energy shifts the top valence bands and moves the Van Hove saddle near the moiré $M$ point as filling increases. The corresponding DOS shows the sharpening and upward shift of the Van Hove peak. The horizontal dashed lines mark the Fermi levels for each filling. This illustrates the interaction-driven pinning of the saddle point close to $\nu \approx 0.8$, which precedes the onset of the Stoner instability.}
    \label{fig:unbroken_HF_bands}
\end{figure}

\begin{figure}
    \includegraphics[trim={0cm 0 0cm 0},clip,width=0.9\columnwidth]{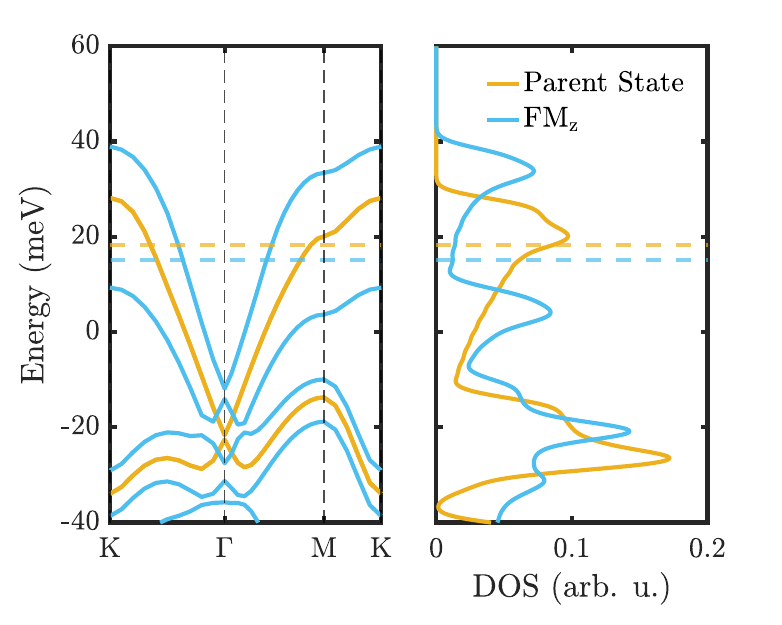}
    \caption{Comparison between the symmetry-unbroken HF parent state and the FM$_z$ phase at the same filling. The FM$_z$ band structure (left) shows a clear splitting of the top valence band relative to the parent state. In the DOS (right), the single Van Hove peak of the parent state splits into two peaks in the FM$_z$ phase. The horizontal dashed lines indicate the Fermi energies. This splitting reflects the spontaneous polarization of the FM$_z$ state and is the signature of the Stoner mechanism described in the main text.}
    \label{fig:broken_HF_bands}
\end{figure}
\section{Additional Hartree--Fock Results for twisted $\mathrm{WSe}_2$}
\label{sec:HF_additional}

This section summarizes the full set of HF results for twisted WSe$_2$, including real-space spin textures of all symmetry-breaking phases, the HF-renormalized band structure, and the HF ground-state energy comparison among competing orders.

In Figs.~\ref{fig:spin_FMz}–\ref{fig:spin_GWC}, we show the spin density wave for different symmetry-breaking phases, separately for the top layer (left), bottom layer (middle), and the total density (right). The color represents the magnitude of the spin component, and arrows indicate the direction of the in-plane spin components. The gray solid lines mark the moir\'e unit cells. For the FM$_z$ state (Fig.~\ref{fig:spin_FMz}), $\sigma_z$ is a good quantum number, and all spins are aligned along the $z$-direction. In the AFM$_{xy}$ state (Fig.~\ref{fig:spin_AFMxy}), all spins lie within the plane, with the maximum spin amplitude found in the MM regions and a smaller spin oriented oppositely in the XM and MX regions. Both of these states do not break the moir\'e translational symmetry. For the $120^\circ$ AFM state (Fig.~\ref{fig:spin_120AFM}), a $\sqrt{3}\times\sqrt{3}$ reconstruction occurs, with spins rotating clockwise for $V_z >0$ and anticlockwise for $V_z <0$. These spin rotations are illustrated by red solid lines. Finally, in Fig.~\ref{fig:spin_GWC} we show the generalized Wigner crystal states corresponding to commensurate reconstructions at fractional fillings. These states exhibit strong real-space modulation of the spin density with one hole per reconstructed unit cell. The patterns shown include the $\sqrt{3}\times\sqrt{3}$ GWC at $\nu=1/3$ and $2/3$, the $2\times2$ GWC at $\nu=1/4$, $1/2$, and $3/4$, and the $2\times3$ GWC at $\nu=1/6$. The spin density peaks on the sites of the enlarged unit cell, reflecting the valley-polarized spin-density-wave character of these phases.

\begin{figure}
    \includegraphics[trim={0cm 0 0cm 0},clip,width=1\columnwidth]{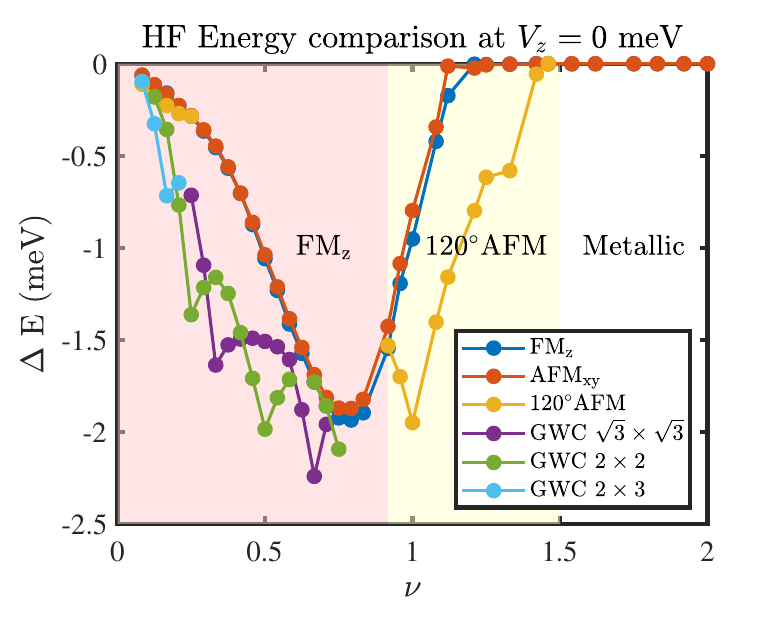}
    \caption{Hartree--Fock ground-state energy comparison at $V_z = 0~\text{meV}$. The quantity $\Delta E_{\alpha}(\nu) = E_{\alpha}(\nu) - E_{\text{parent}}(\nu)$ is shown for all ordered phases: FM$_z$, AFM$_{xy}$, $120^\circ$ AFM, and the generalized Wigner crystal states with $\sqrt{3}\!\times\!\sqrt{3}$, $2\!\times\!2$, and $2\!\times\!3$ reconstruction. Negative values of $\Delta E_{\alpha}$ indicate that the ordered state is favorable relative to the symmetry-unbroken parent state. The shaded regions mark the dominant ground state in each filling window: the FM$_z$ region at low filling, the $120^\circ$ AFM region near half filling, and the metallic regime at larger filling.}
    \label{fig:ground_state_energy}
\end{figure}

\begin{figure}
    \includegraphics[trim={0cm 0 0cm 0},clip,width=1\columnwidth]{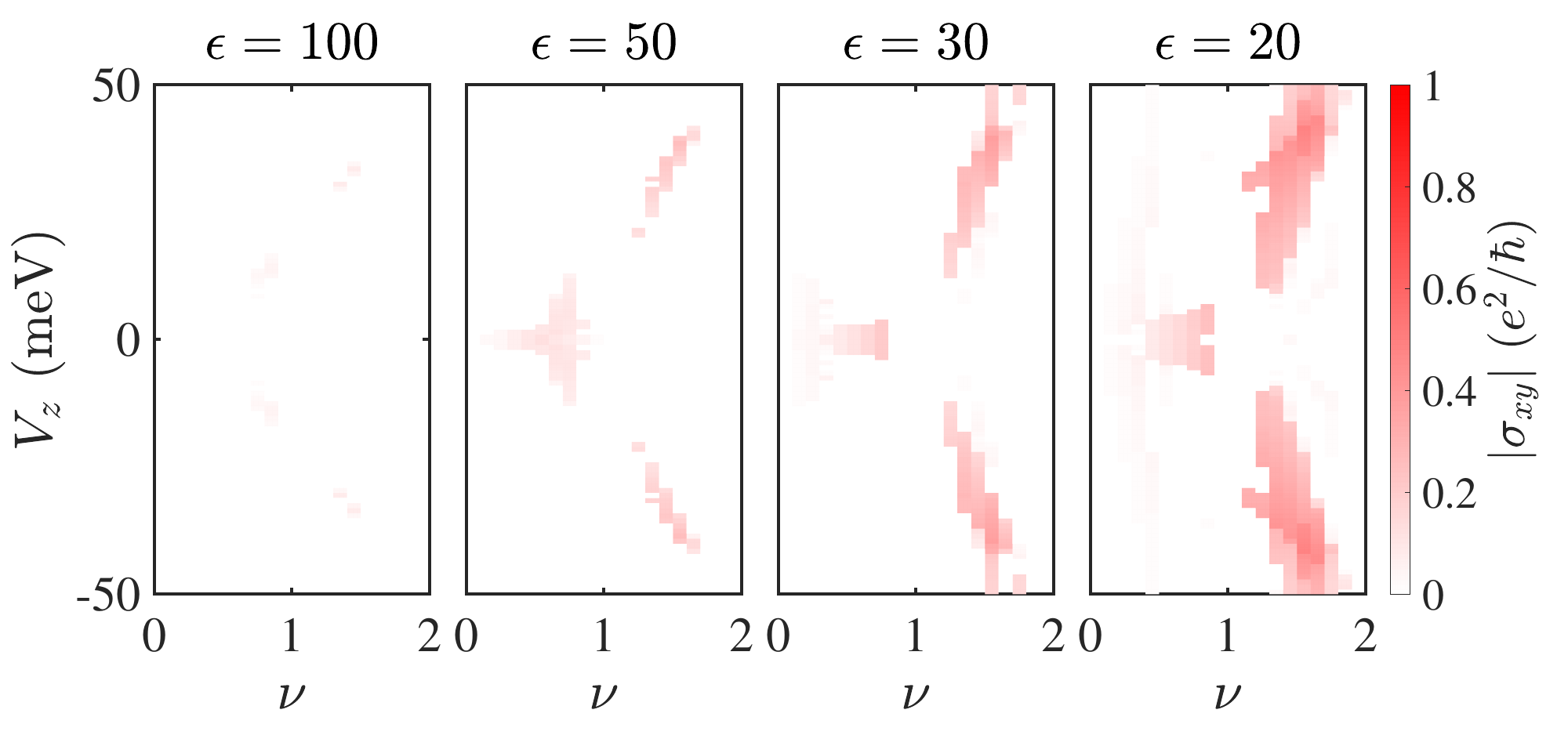}
    \caption{Evolution of DC Hall conductivity $\sigma_{xy}$, which characterizes the degree of $\mathcal T$ breaking, as a function of interlayer bias $V_z$ and hole filling factor $\nu$ for different values of the dielectric constant $\epsilon$.  The ``wing" feature emerges as a consequence of strong correlations.  
    Contrary to $\left<S_z\right>$, we find that $\sigma_{xy}$ is a good probe for the layer-polarized FM$_z$ phase.}
    \label{fig:DC_conductivity}
\end{figure}

To clarify how interactions reshape the moir\'e bands, we present symmetry-unbroken HF band structures and DOS at several fillings in Fig.~\ref{fig:unbroken_HF_bands}. As the hole density increases, exchange shifts the topmost valence bands upward, enhances their bandwidth, and moves the Van Hove saddle near the moir\'e $M$ point. When the filling approaches $\nu \approx 0.8$, the HF solution lowers its energy by shifting the saddle to the Fermi level. This pinning enhances the DOS and drives the Stoner instability. We compare the HF parent state and the FM$_z$ state at the same filling, as shown in Fig.~\ref{fig:broken_HF_bands}. The FM$_z$ phase shows a clear splitting of the top band and a corresponding splitting of the VHS peak in the DOS, which marks the spontaneous polarization of the FM$_z$ phase.

To quantify the competition between ordered phases, we compute the HF ground-state energy per moir\'e cell for each phase relative to the symmetry-unbroken HF parent state,
\begin{equation}
    \Delta E_{\alpha}(\nu) = E_{\alpha}(\nu) - E_{\text{parent}}(\nu),
\end{equation}
as shown in Fig.~\ref{fig:ground_state_energy}. We include FM$_z$, AFM$_{xy}$, $120^\circ$ AFM, and the GWC states ($3\times3$, $2\times2$, and $2\times3$). For $\nu \lesssim 0.9$, the system is generically ferromagnetic with generalized Wigner crystal states having the lowest energy at commensurate fractional fillings like $1/6$, $1/4$ and $1/3$. Throughout this window, the FM$_z$ and AFM$_{xy}$ states are very close in energy, but with the FMz slightly favored. For $0.9\lesssim\nu\lesssim1.5$, the 120$^{\circ}$  AFM becomes the ground state and shows a clear energy gain over any of the competing ferromagnetic states matching the phase boundary found in Fig.~\ref{fig:fig4}(a) of the main text. At higher filling, $\nu\gtrsim1.5$, the energy gain of all ordered phases collapses to zero consistent with a metallic regime. 

\section{Additional Hartree--Fock results for twisted $\mathrm{MoTe}_2$}

For completeness, we also compute the phase diagram for tMoTe$_2$. Figure~\ref{fig:OP_MoTe2} summarizes the symmetry-breaking order parameters obtained for twisted MoTe$_2$ at $\theta = 3.5^\circ$ by scanning the filling $\nu$ and interlayer potential $V_z$. The same Hartree--Fock procedure used for twisted WSe$_2$ is applied here, but MoTe$_2$ shows notable differences due to its large effective mass and narrower moir\'e bands.

The FM$_z$ order (top left) is the dominant phase across a wide region of the $(\nu,V_z)$ plane. In contrast to twisted WSe$_2$, the FM$_z$ state remains strong even at half-filling, consistent with recent experimental observations in twisted MoTe$_2$ \cite{li_universal_2025}. The AFM$_{xy}$ order (top middle) is much weaker and only appears in thin slivers of parameter space. The $120^\circ$ AFM state (top right) also occurs near half-filling but occupies a narrower range compared to WSe$_2$, which reflects the stronger tendency toward spin–valley polarization in MoTe$_2$. The bottom row shows the generalized Wigner crystal states. Their order parameters show strong real-space modulation and appear as narrow vertical bands in the $(\nu,V_z)$ plane. The presence of several GWC phases highlights the enhanced role of exchange in MoTe$_2$, where the reduced bandwidth favors real-space charge and spin order.

Overall, these results indicate that twisted MoTe$_2$ exhibits a stronger FM$_z$ response and more robust Wigner-crystal tendencies than twisted WSe$_2$, reflecting its deeper atomic spin–valley locking and narrower moir\'e bands.

\section{DC Hall conductivity}

The Hall conductivity $\sigma_{xy}$ serves as an important experimental observable in transport that can probe the $\mathcal T$ breaking of the FM$_z$ phase \cite{xia_superconductivity_2025}. We anticipate that $\sigma_{xy}$ will soon be measured in this system both as a function of filling and interlayer bias. The Hall conductivity can be expressed as
\begin{equation}
    \sigma_{xy} = \frac{e^2}{\hbar} \sum_n \int_\text{MBZ} \frac{d^2\bm q}{(2\pi)^2} f_{n,\bm q} \Omega_{n,\bm q},
\end{equation}
where the sum runs over HF bands, $f_{n,\bm q}$ is the occupation and $\Omega_{n,\bm q}$ is the Berry curvature. In Fig.\ \ref{fig:DC_conductivity}, we show $\sigma_{xy}$ in the $\nu$--$V_z$ plane for different $\epsilon$. The symmetry-unbroken and IVC states have vanishing $\sigma_{xy}$ due to $\mathcal T'$. As a result, $\sigma_{xy}$ only probes the FM$_z$ phase. Note that while we find large $\left<\sigma_z \right>$ in Fig.\ \ref{fig:fig2}(b) near charge neutrality, the Hall conductivity is absent because both the density of holes and Berry curvature is small. Moreover, for $V_z=0$ meV, regardless of the interaction strength, we find finite $\sigma_{xy}$ only for $\nu<1$, consistent with experiment \cite{knuppel_correlated_2025}. Note that $\sigma_{xy}$ is absent at half-filling because the gap is trivial and the $120^\circ$ AFM has $\mathcal T'$ symmetry. Contrary to the out-of-plane magnetization, we find that $\sigma_{xy}$ is a good probe for the layer-polarized FM$_z$.

\section{Spin and orbital magnetization}

Here we summarize the formulas used to compute the spin and orbital magnetization from the self-consistent Hartree--Fock bands. All quantities are evaluated per moir\'e unit cell with area $\Omega_\mathrm{M}$, and we use the cell-periodic Bloch states $|u_{n\mathbf{k}}\rangle$ and band energies $E_{n\mathbf{k}}$ of the Hartree--Fock Hamiltonian $H_{\mathbf{k}}$.

The spin magnetization follows directly from the expectation value of the spin operator. For electrons,
\begin{equation}
    \mathbf{M}_{\text{spin}}
    = -\frac{g_s\mu_B}{\hbar\,\Omega_\mathrm{M}}
      \sum_n \int_{\mathrm{BZ}} \frac{d^2\mathbf{k}}{(2\pi)^2}\,
      f(E_{n\mathbf{k}}-\mu)
      \langle u_{n\mathbf{k}} | \mathbf{S} | u_{n\mathbf{k}} \rangle ,
\end{equation}
where $g_s\simeq2$ is the spin $g$-factor, $\mu_B=e\hbar/2m_e$, and $\mathbf{S}=\hbar\boldsymbol{\sigma}/2$. For the $z$ component used in the main text,
\begin{equation}
    M_{\text{spin}}^z
    = -\frac{g_s \mu_B}{2\,\Omega_\mathrm{M}}
      \sum_n \int_{\mathrm{BZ}} \frac{d^2\mathbf{k}}{(2\pi)^2}\,
      f(E_{n\mathbf{k}}-\mu)
      \langle u_{n\mathbf{k}} | \sigma_z | u_{n\mathbf{k}} \rangle .
\end{equation}

The orbital magnetization requires the modern theory of orbital magnetization, which expresses it in terms of Berry curvature and the orbital moment of the Bloch states. At finite temperature, the result is \cite{xiao_berry_2005,shi_quantum_2007}
\begin{equation}
\begin{aligned}
    \mathbf{M}_{\text{orb}}
    = \sum_n \int_{\mathrm{BZ}} \frac{d^2\mathbf{k}}{(2\pi)^2}
      \left[
          f(E_{n\mathbf{k}}-\mu)\,\mathbf{m}_n(\mathbf{k}) \right. \\
          + \left. \frac{e}{\hbar}\,\boldsymbol{\Omega}_n(\mathbf{k})
            \frac{1}{\beta}\ln\!\big(1+e^{-\beta(E_{n\mathbf{k}}-\mu)}\big)
      \right],
\end{aligned}
\end{equation}
where $\beta=1/k_B T$, $\mu$ is the chemical potential, $\mathbf{m}_n(\mathbf{k})$ is the orbital moment, and $\boldsymbol{\Omega}_n(\mathbf{k})$ is the Berry curvature. These are
\begin{align}
    \mathbf{m}_n(\mathbf{k})
    &= \frac{e}{2\hbar}\,\mathrm{Im}
       \langle \partial_{\mathbf{k}} u_{n\mathbf{k}} |
       \times (H_{\mathbf{k}}-E_{n\mathbf{k}})
       | \partial_{\mathbf{k}} u_{n\mathbf{k}} \rangle ,\\
    \boldsymbol{\Omega}_n(\mathbf{k})
    &= i\langle \partial_{\mathbf{k}} u_{n\mathbf{k}} |
       \times
       | \partial_{\mathbf{k}} u_{n\mathbf{k}} \rangle .
\end{align}
Only the $z$ component appears in two dimensions. Taking the limit $\beta\to\infty$ gives the compact zero-temperature expression
\begin{equation}
\begin{aligned}
    \mathbf{M}_{\text{orb}}(T=0)
    = \frac{e}{2\hbar}\,\mathrm{Im}
      \sum_n \int_{\mathrm{BZ}} \frac{d^2\mathbf{k}}{(2\pi)^2}\,
      \Theta(\mu-E_{n\mathbf{k}}) \times \\
      \langle \partial_{\mathbf{k}} u_{n\mathbf{k}} |
      \times \big(H_{\mathbf{k}} + E_{n\mathbf{k}} - 2\mu\big)
      | \partial_{\mathbf{k}} u_{n\mathbf{k}} \rangle .
\end{aligned}
\end{equation}

For numerical calculations, it is convenient to express the orbital magnetization in terms of velocity matrix elements. We define
\begin{align}
    v_\alpha(\mathbf{k})
    &= \frac{1}{\hbar}\,\frac{\partial H_{\mathbf{k}}}{\partial k_\alpha},
    \\
    v^{\alpha}_{nm}(\mathbf{k})
    &= \langle u_{n\mathbf{k}}|v_\alpha(\mathbf{k})|u_{m\mathbf{k}}\rangle
    = \frac{1}{\hbar}\,
      \langle u_{n\mathbf{k}}|\partial_{k_\alpha}H_{\mathbf{k}}|u_{m\mathbf{k}}\rangle .
\end{align}
The $z$ component of the orbital magnetization can then be written as
\begin{equation}
    M_{\text{orb}}^{z}(\mu)
    = \sum_{n} \int_{\mathrm{BZ}} \frac{d^{2}\mathbf{k}}{(2\pi)^{2}}\,
      \mathcal{M}_{n}(\mathbf{k},\mu)\,
      f(E_{n\mathbf{k}}-\mu),
\end{equation}
with the kernel
\begin{equation}
    \mathcal{M}_{n}(\mathbf{k},\mu)
    = \frac{e}{\hbar}\,
      \mathrm{Im}
      \sum_{m\neq n}
      \frac{
        v^{x}_{nm}(\mathbf{k})\,v^{y}_{mn}(\mathbf{k})
      }{\big(E_{n\mathbf{k}} - E_{m\mathbf{k}}\big)^{2}}\,
      \big( E_{n\mathbf{k}} + E_{m\mathbf{k}} - 2\mu \big).
\end{equation}
All quantities are computed from the Hartree--Fock eigenvalues and eigenvectors at each momentum point, and the Brillouin-zone integral is performed on the same $\mathbf{k}$ mesh used for the self-consistent Hartree--Fock calculation.

\begin{figure*}[h]
    \centering
    \includegraphics[width=1\linewidth]{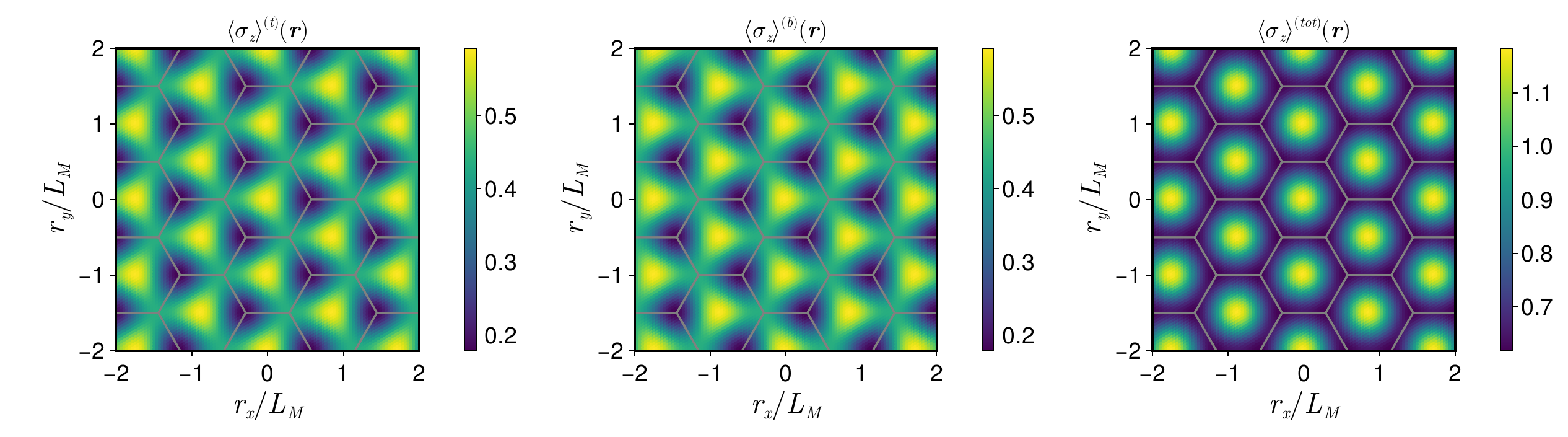}
    \caption{Spin density for $z$ direction of FM$_z$ order at $\nu=0.8$ and $V_z=0$ meV for top layer (left), bottom layer (middle) and total spin density (right). Gray lines represent moir\'e periodicity.}
    \label{fig:spin_FMz}
\end{figure*}

\begin{figure*}[h]
    \centering
    \includegraphics[width=1\linewidth]{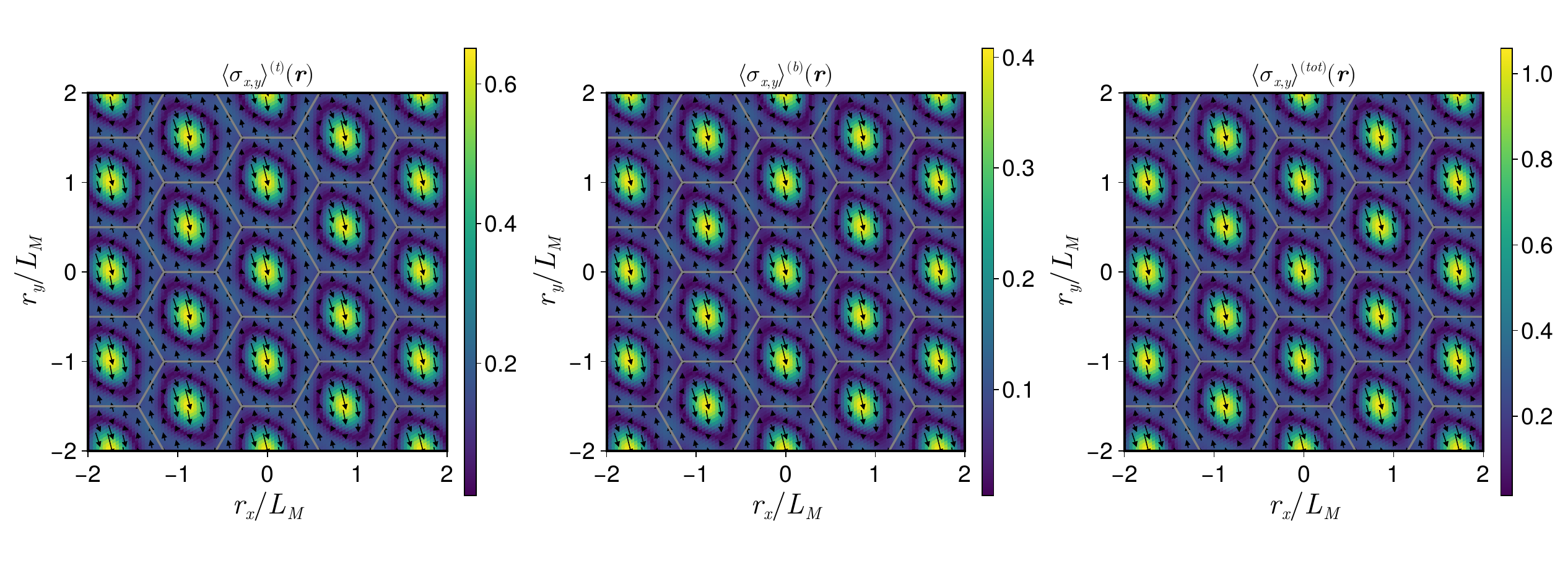}
    \caption{In-plane spin density for AFM$_{xy}$ at $\nu=0.8$ and $V_z=10$ meV for top layer (left), bottom layer (middle) and total spin density (right). Gray lines represent moir\'e periodicity.}
    \label{fig:spin_AFMxy}
\end{figure*}

\begin{figure*}[h]
    \centering
    \includegraphics[width=1\linewidth]{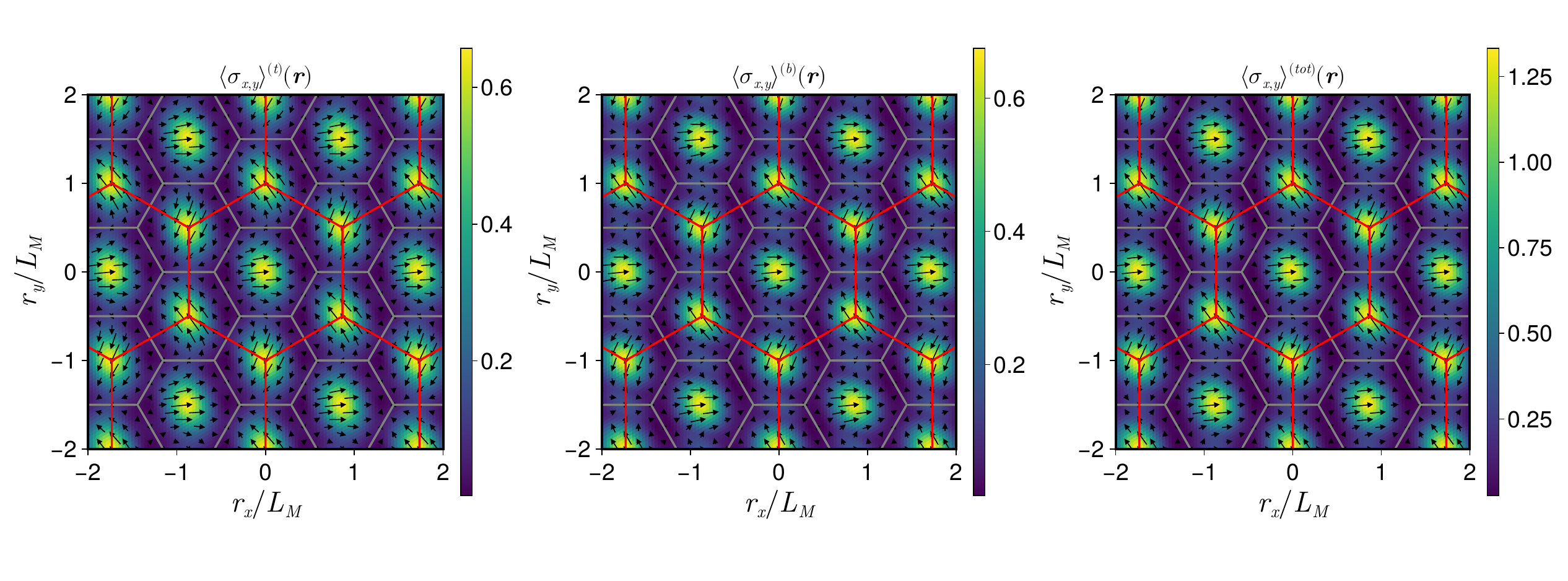}
    \caption{In-plane spin density for $120^\circ$ AFM at $\nu=1$ and $V_z=0$ meV for top layer (left), bottom layer (middle) and total spin density (right). Gray and red lines represent moir\'e and enlarged moir\'e periodicity, respectively.}
    \label{fig:spin_120AFM}
\end{figure*}

\begin{figure*}[h]
    \centering
    \includegraphics[width=1\linewidth]{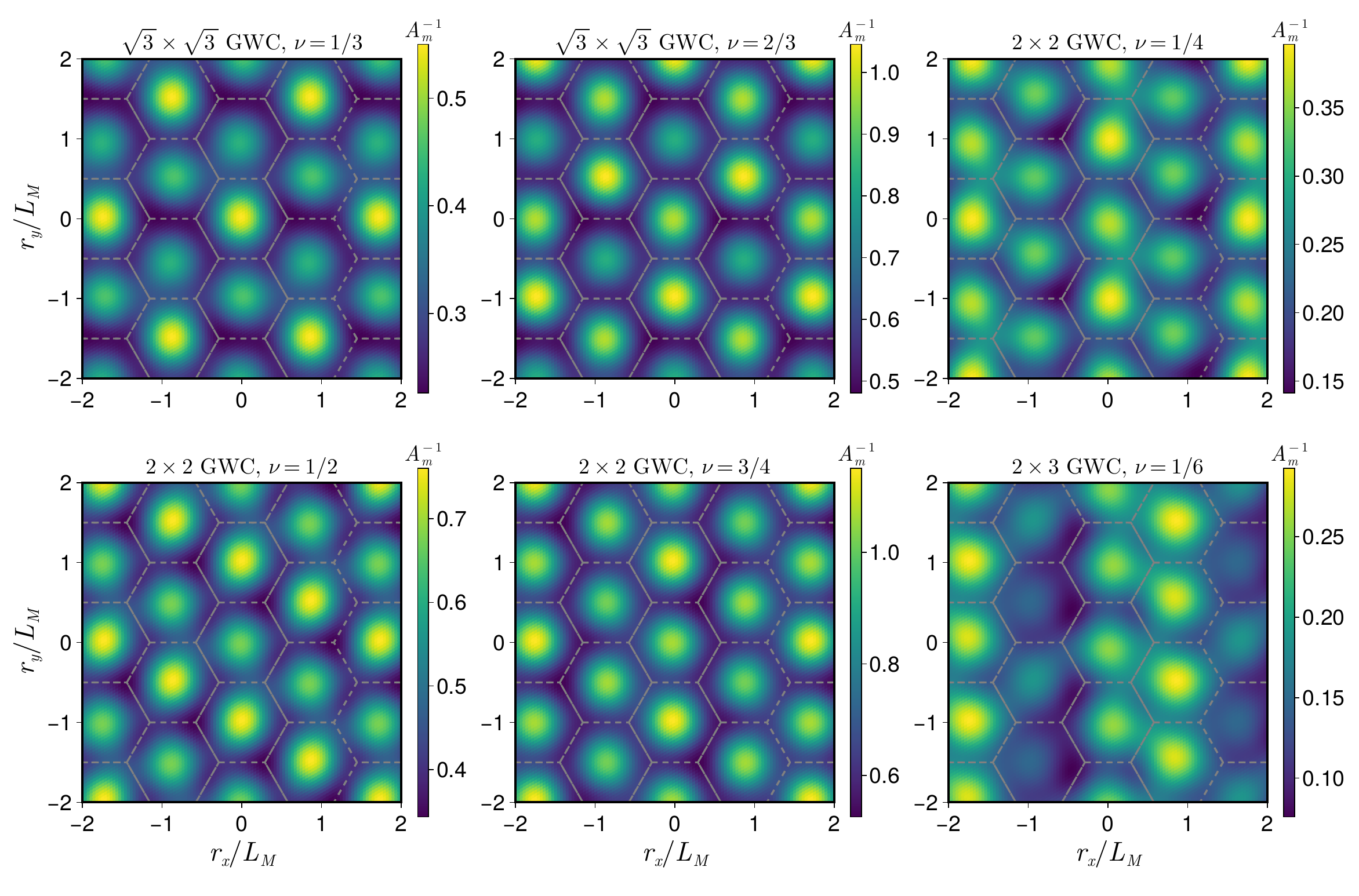}
    \caption{Real-space spin–density pattern of the generalized Wigner crystal states obtained from the self-consistent Hartree–Fock calculation at $V_z=0$ meV. Each panel shows the local spin amplitude $A_m^{-1}$ in the moir\'e unit cell for a different commensurate reconstruction and filling factor $\nu$: (top row) $\sqrt{3}\times\sqrt{3}$ GWC at $\nu=1/3$ and $\nu=2/3$, and $2\times2$ GWC at $\nu=1/4$; (bottom row) $2\times2$ GWC at $\nu=1/2$ and $\nu=3/4$, and $2\times3$ GWC at $\nu=1/6$. The dashed hexagons indicate the underlying moir\'e lattice, and the color scale shows the spatial modulation of the spin density. These states correspond to one hole per reconstructed unit cell and arise from the valley-polarized spin-density-wave instability.}
    \label{fig:spin_GWC}
\end{figure*}

\begin{figure*}[h]
    \centering
    \includegraphics[width=0.9\linewidth]{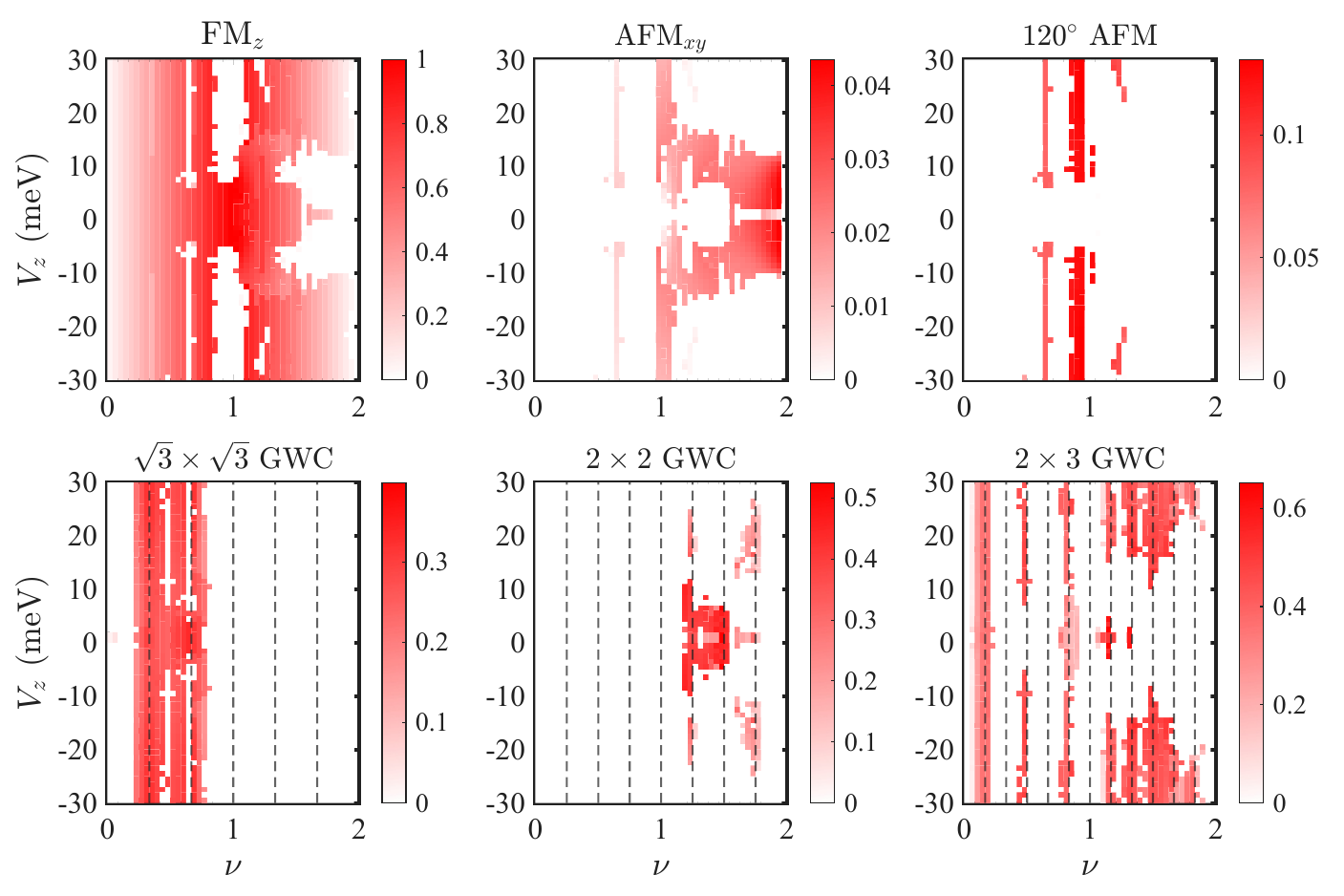}
    \caption{Order parameters of twisted MoTe$_2$ at $\theta = 3.5^\circ$ as functions of filling $\nu$ and interlayer potential $V_z$. Each panel shows the Hartree--Fock expectation value of the corresponding order: (top row) FM$_z$, AFM$_{xy}$, and $120^\circ$ AFM; (bottom row) $\sqrt{3}\!\times\!\sqrt{3}$ GWC, $2\!\times\!2$ GWC, and $2\!\times\!3$ GWC. The color scale indicates the magnitude of the order parameter, with white regions denoting either vanishing order or competing phases with lower energy. The dashed vertical lines mark rational fillings that allow commensurate real-space reconstructions. These results show that MoTe$_2$ favors a strong FM$_z$ response over a wide range of $(\nu,V_z)$, while the GWC phases appear near fractional fillings with one hole per reconstructed unit cell.}
    \label{fig:OP_MoTe2}
\end{figure*}


%% file: main.bbl
\begin{thebibliography}{55}%
\makeatletter
\providecommand \@ifxundefined [1]{%
 \@ifx{#1\undefined}
}%
\providecommand \@ifnum [1]{%
 \ifnum #1\expandafter \@firstoftwo
 \else \expandafter \@secondoftwo
 \fi
}%
\providecommand \@ifx [1]{%
 \ifx #1\expandafter \@firstoftwo
 \else \expandafter \@secondoftwo
 \fi
}%
\providecommand \natexlab [1]{#1}%
\providecommand \enquote  [1]{``#1''}%
\providecommand \bibnamefont  [1]{#1}%
\providecommand \bibfnamefont [1]{#1}%
\providecommand \citenamefont [1]{#1}%
\providecommand \href@noop [0]{\@secondoftwo}%
\providecommand \href [0]{\begingroup \@sanitize@url \@href}%
\providecommand \@href[1]{\@@startlink{#1}\@@href}%
\providecommand \@@href[1]{\endgroup#1\@@endlink}%
\providecommand \@sanitize@url [0]{\catcode `\\12\catcode `\$12\catcode `\&12\catcode `\#12\catcode `\^12\catcode `\_12\catcode `\%12\relax}%
\providecommand \@@startlink[1]{}%
\providecommand \@@endlink[0]{}%
\providecommand \url  [0]{\begingroup\@sanitize@url \@url }%
\providecommand \@url [1]{\endgroup\@href {#1}{\urlprefix }}%
\providecommand \urlprefix  [0]{URL }%
\providecommand \Eprint [0]{\href }%
\providecommand \doibase [0]{https://doi.org/}%
\providecommand \selectlanguage [0]{\@gobble}%
\providecommand \bibinfo  [0]{\@secondoftwo}%
\providecommand \bibfield  [0]{\@secondoftwo}%
\providecommand \translation [1]{[#1]}%
\providecommand \BibitemOpen [0]{}%
\providecommand \bibitemStop [0]{}%
\providecommand \bibitemNoStop [0]{.\EOS\space}%
\providecommand \EOS [0]{\spacefactor3000\relax}%
\providecommand \BibitemShut  [1]{\csname bibitem#1\endcsname}%
\let\auto@bib@innerbib\@empty
\bibitem [{\citenamefont {Cao}\ \emph {et~al.}(2018{\natexlab{a}})\citenamefont {Cao}, \citenamefont {Fatemi}, \citenamefont {Demir}, \citenamefont {Fang}, \citenamefont {Tomarken}, \citenamefont {Luo}, \citenamefont {Sanchez-Yamagishi}, \citenamefont {Watanabe}, \citenamefont {Taniguchi}, \citenamefont {Kaxiras}, \citenamefont {Ashoori},\ and\ \citenamefont {Jarillo-Herrero}}]{cao_correlated_2018}%
  \BibitemOpen
  \bibfield  {author} {\bibinfo {author} {\bibfnamefont {Y.}~\bibnamefont {Cao}}, \bibinfo {author} {\bibfnamefont {V.}~\bibnamefont {Fatemi}}, \bibinfo {author} {\bibfnamefont {A.}~\bibnamefont {Demir}}, \bibinfo {author} {\bibfnamefont {S.}~\bibnamefont {Fang}}, \bibinfo {author} {\bibfnamefont {S.~L.}\ \bibnamefont {Tomarken}}, \bibinfo {author} {\bibfnamefont {J.~Y.}\ \bibnamefont {Luo}}, \bibinfo {author} {\bibfnamefont {J.~D.}\ \bibnamefont {Sanchez-Yamagishi}}, \bibinfo {author} {\bibfnamefont {K.}~\bibnamefont {Watanabe}}, \bibinfo {author} {\bibfnamefont {T.}~\bibnamefont {Taniguchi}}, \bibinfo {author} {\bibfnamefont {E.}~\bibnamefont {Kaxiras}}, \bibinfo {author} {\bibfnamefont {R.~C.}\ \bibnamefont {Ashoori}},\ and\ \bibinfo {author} {\bibfnamefont {P.}~\bibnamefont {Jarillo-Herrero}},\ }\bibfield  {title} {\bibinfo {title} {Correlated insulator behaviour at half-filling in magic-angle graphene superlattices},\ }\href {https://doi.org/10.1038/nature26154} {\bibfield  {journal} {\bibinfo  {journal}
  {Nature}\ }\textbf {\bibinfo {volume} {556}},\ \bibinfo {pages} {80} (\bibinfo {year} {2018}{\natexlab{a}})}\BibitemShut {NoStop}%
\bibitem [{\citenamefont {Cao}\ \emph {et~al.}(2018{\natexlab{b}})\citenamefont {Cao}, \citenamefont {Fatemi}, \citenamefont {Fang}, \citenamefont {Watanabe}, \citenamefont {Taniguchi}, \citenamefont {Kaxiras},\ and\ \citenamefont {Jarillo-Herrero}}]{cao_unconventional_2018}%
  \BibitemOpen
  \bibfield  {author} {\bibinfo {author} {\bibfnamefont {Y.}~\bibnamefont {Cao}}, \bibinfo {author} {\bibfnamefont {V.}~\bibnamefont {Fatemi}}, \bibinfo {author} {\bibfnamefont {S.}~\bibnamefont {Fang}}, \bibinfo {author} {\bibfnamefont {K.}~\bibnamefont {Watanabe}}, \bibinfo {author} {\bibfnamefont {T.}~\bibnamefont {Taniguchi}}, \bibinfo {author} {\bibfnamefont {E.}~\bibnamefont {Kaxiras}},\ and\ \bibinfo {author} {\bibfnamefont {P.}~\bibnamefont {Jarillo-Herrero}},\ }\bibfield  {title} {\bibinfo {title} {Unconventional superconductivity in magic-angle graphene superlattices},\ }\href {https://doi.org/10.1038/nature26160} {\bibfield  {journal} {\bibinfo  {journal} {Nature}\ }\textbf {\bibinfo {volume} {556}},\ \bibinfo {pages} {43} (\bibinfo {year} {2018}{\natexlab{b}})}\BibitemShut {NoStop}%
\bibitem [{\citenamefont {Anderson}\ \emph {et~al.}(2023)\citenamefont {Anderson}, \citenamefont {Fan}, \citenamefont {Cai}, \citenamefont {Holtzmann}, \citenamefont {Taniguchi}, \citenamefont {Watanabe}, \citenamefont {Xiao}, \citenamefont {Yao},\ and\ \citenamefont {Xu}}]{anderson_programming_2023}%
  \BibitemOpen
  \bibfield  {author} {\bibinfo {author} {\bibfnamefont {E.}~\bibnamefont {Anderson}}, \bibinfo {author} {\bibfnamefont {F.-R.}\ \bibnamefont {Fan}}, \bibinfo {author} {\bibfnamefont {J.}~\bibnamefont {Cai}}, \bibinfo {author} {\bibfnamefont {W.}~\bibnamefont {Holtzmann}}, \bibinfo {author} {\bibfnamefont {T.}~\bibnamefont {Taniguchi}}, \bibinfo {author} {\bibfnamefont {K.}~\bibnamefont {Watanabe}}, \bibinfo {author} {\bibfnamefont {D.}~\bibnamefont {Xiao}}, \bibinfo {author} {\bibfnamefont {W.}~\bibnamefont {Yao}},\ and\ \bibinfo {author} {\bibfnamefont {X.}~\bibnamefont {Xu}},\ }\bibfield  {title} {\bibinfo {title} {Programming correlated magnetic states with gate-controlled moiré geometry},\ }\href {https://doi.org/10.1126/science.adg4268} {\bibfield  {journal} {\bibinfo  {journal} {Science}\ }\textbf {\bibinfo {volume} {381}},\ \bibinfo {pages} {325} (\bibinfo {year} {2023})}\BibitemShut {NoStop}%
\bibitem [{\citenamefont {Cai}\ \emph {et~al.}(2023)\citenamefont {Cai}, \citenamefont {Anderson}, \citenamefont {Wang}, \citenamefont {Zhang}, \citenamefont {Liu}, \citenamefont {Holtzmann}, \citenamefont {Zhang}, \citenamefont {Fan}, \citenamefont {Taniguchi}, \citenamefont {Watanabe}, \citenamefont {Ran}, \citenamefont {Cao}, \citenamefont {Fu}, \citenamefont {Xiao}, \citenamefont {Yao},\ and\ \citenamefont {Xu}}]{cai_signatures_2023}%
  \BibitemOpen
  \bibfield  {author} {\bibinfo {author} {\bibfnamefont {J.}~\bibnamefont {Cai}}, \bibinfo {author} {\bibfnamefont {E.}~\bibnamefont {Anderson}}, \bibinfo {author} {\bibfnamefont {C.}~\bibnamefont {Wang}}, \bibinfo {author} {\bibfnamefont {X.}~\bibnamefont {Zhang}}, \bibinfo {author} {\bibfnamefont {X.}~\bibnamefont {Liu}}, \bibinfo {author} {\bibfnamefont {W.}~\bibnamefont {Holtzmann}}, \bibinfo {author} {\bibfnamefont {Y.}~\bibnamefont {Zhang}}, \bibinfo {author} {\bibfnamefont {F.}~\bibnamefont {Fan}}, \bibinfo {author} {\bibfnamefont {T.}~\bibnamefont {Taniguchi}}, \bibinfo {author} {\bibfnamefont {K.}~\bibnamefont {Watanabe}}, \bibinfo {author} {\bibfnamefont {Y.}~\bibnamefont {Ran}}, \bibinfo {author} {\bibfnamefont {T.}~\bibnamefont {Cao}}, \bibinfo {author} {\bibfnamefont {L.}~\bibnamefont {Fu}}, \bibinfo {author} {\bibfnamefont {D.}~\bibnamefont {Xiao}}, \bibinfo {author} {\bibfnamefont {W.}~\bibnamefont {Yao}},\ and\ \bibinfo {author} {\bibfnamefont {X.}~\bibnamefont {Xu}},\ }\bibfield  {title}
  {\bibinfo {title} {Signatures of fractional quantum anomalous {Hall} states in twisted {MoTe2}},\ }\href {https://doi.org/10.1038/s41586-023-06289-w} {\bibfield  {journal} {\bibinfo  {journal} {Nature}\ }\textbf {\bibinfo {volume} {622}},\ \bibinfo {pages} {63} (\bibinfo {year} {2023})}\BibitemShut {NoStop}%
\bibitem [{\citenamefont {Kang}\ \emph {et~al.}(2024)\citenamefont {Kang}, \citenamefont {Shen}, \citenamefont {Qiu}, \citenamefont {Zeng}, \citenamefont {Xia}, \citenamefont {Watanabe}, \citenamefont {Taniguchi}, \citenamefont {Shan},\ and\ \citenamefont {Mak}}]{kang_evidence_2024}%
  \BibitemOpen
  \bibfield  {author} {\bibinfo {author} {\bibfnamefont {K.}~\bibnamefont {Kang}}, \bibinfo {author} {\bibfnamefont {B.}~\bibnamefont {Shen}}, \bibinfo {author} {\bibfnamefont {Y.}~\bibnamefont {Qiu}}, \bibinfo {author} {\bibfnamefont {Y.}~\bibnamefont {Zeng}}, \bibinfo {author} {\bibfnamefont {Z.}~\bibnamefont {Xia}}, \bibinfo {author} {\bibfnamefont {K.}~\bibnamefont {Watanabe}}, \bibinfo {author} {\bibfnamefont {T.}~\bibnamefont {Taniguchi}}, \bibinfo {author} {\bibfnamefont {J.}~\bibnamefont {Shan}},\ and\ \bibinfo {author} {\bibfnamefont {K.~F.}\ \bibnamefont {Mak}},\ }\bibfield  {title} {\bibinfo {title} {Evidence of the fractional quantum spin {Hall} effect in moiré {MoTe2}},\ }\href {https://doi.org/10.1038/s41586-024-07214-5} {\bibfield  {journal} {\bibinfo  {journal} {Nature}\ }\textbf {\bibinfo {volume} {628}},\ \bibinfo {pages} {522} (\bibinfo {year} {2024})}\BibitemShut {NoStop}%
\bibitem [{\citenamefont {Ghiotto}\ \emph {et~al.}(2024)\citenamefont {Ghiotto}, \citenamefont {Wei}, \citenamefont {Song}, \citenamefont {Zang}, \citenamefont {Tazi}, \citenamefont {Ostrom}, \citenamefont {Watanabe}, \citenamefont {Taniguchi}, \citenamefont {Hone}, \citenamefont {Rhodes}, \citenamefont {Millis}, \citenamefont {Dean}, \citenamefont {Wang},\ and\ \citenamefont {Pasupathy}}]{ghiotto_stoner_2024}%
  \BibitemOpen
  \bibfield  {author} {\bibinfo {author} {\bibfnamefont {A.}~\bibnamefont {Ghiotto}}, \bibinfo {author} {\bibfnamefont {L.}~\bibnamefont {Wei}}, \bibinfo {author} {\bibfnamefont {L.}~\bibnamefont {Song}}, \bibinfo {author} {\bibfnamefont {J.}~\bibnamefont {Zang}}, \bibinfo {author} {\bibfnamefont {A.~B.}\ \bibnamefont {Tazi}}, \bibinfo {author} {\bibfnamefont {D.}~\bibnamefont {Ostrom}}, \bibinfo {author} {\bibfnamefont {K.}~\bibnamefont {Watanabe}}, \bibinfo {author} {\bibfnamefont {T.}~\bibnamefont {Taniguchi}}, \bibinfo {author} {\bibfnamefont {J.~C.}\ \bibnamefont {Hone}}, \bibinfo {author} {\bibfnamefont {D.~A.}\ \bibnamefont {Rhodes}}, \bibinfo {author} {\bibfnamefont {A.~J.}\ \bibnamefont {Millis}}, \bibinfo {author} {\bibfnamefont {C.~R.}\ \bibnamefont {Dean}}, \bibinfo {author} {\bibfnamefont {L.}~\bibnamefont {Wang}},\ and\ \bibinfo {author} {\bibfnamefont {A.~N.}\ \bibnamefont {Pasupathy}},\ }\href {https://doi.org/10.48550/arXiv.2405.17316} {\bibinfo {title} {Stoner instabilities and {Ising}
  excitonic states in twisted transition metal dichalcogenides}} (\bibinfo {year} {2024}),\ \bibinfo {note} {arXiv:2405.17316}\BibitemShut {NoStop}%
\bibitem [{\citenamefont {Knüppel}\ \emph {et~al.}(2025)\citenamefont {Knüppel}, \citenamefont {Zhu}, \citenamefont {Xia}, \citenamefont {Xia}, \citenamefont {Han}, \citenamefont {Zeng}, \citenamefont {Watanabe}, \citenamefont {Taniguchi}, \citenamefont {Shan},\ and\ \citenamefont {Mak}}]{knuppel_correlated_2025}%
  \BibitemOpen
  \bibfield  {author} {\bibinfo {author} {\bibfnamefont {P.}~\bibnamefont {Knüppel}}, \bibinfo {author} {\bibfnamefont {J.}~\bibnamefont {Zhu}}, \bibinfo {author} {\bibfnamefont {Y.}~\bibnamefont {Xia}}, \bibinfo {author} {\bibfnamefont {Z.}~\bibnamefont {Xia}}, \bibinfo {author} {\bibfnamefont {Z.}~\bibnamefont {Han}}, \bibinfo {author} {\bibfnamefont {Y.}~\bibnamefont {Zeng}}, \bibinfo {author} {\bibfnamefont {K.}~\bibnamefont {Watanabe}}, \bibinfo {author} {\bibfnamefont {T.}~\bibnamefont {Taniguchi}}, \bibinfo {author} {\bibfnamefont {J.}~\bibnamefont {Shan}},\ and\ \bibinfo {author} {\bibfnamefont {K.~F.}\ \bibnamefont {Mak}},\ }\bibfield  {title} {\bibinfo {title} {Correlated states controlled by a tunable van {Hove} singularity in moiré {WSe2} bilayers},\ }\href {https://doi.org/10.1038/s41467-025-57235-5} {\bibfield  {journal} {\bibinfo  {journal} {Nature Communications}\ }\textbf {\bibinfo {volume} {16}},\ \bibinfo {pages} {1959} (\bibinfo {year} {2025})}\BibitemShut {NoStop}%
\bibitem [{\citenamefont {Xia}\ \emph {et~al.}(2025)\citenamefont {Xia}, \citenamefont {Han}, \citenamefont {Watanabe}, \citenamefont {Taniguchi}, \citenamefont {Shan},\ and\ \citenamefont {Mak}}]{xia_superconductivity_2025}%
  \BibitemOpen
  \bibfield  {author} {\bibinfo {author} {\bibfnamefont {Y.}~\bibnamefont {Xia}}, \bibinfo {author} {\bibfnamefont {Z.}~\bibnamefont {Han}}, \bibinfo {author} {\bibfnamefont {K.}~\bibnamefont {Watanabe}}, \bibinfo {author} {\bibfnamefont {T.}~\bibnamefont {Taniguchi}}, \bibinfo {author} {\bibfnamefont {J.}~\bibnamefont {Shan}},\ and\ \bibinfo {author} {\bibfnamefont {K.~F.}\ \bibnamefont {Mak}},\ }\bibfield  {title} {\bibinfo {title} {Superconductivity in twisted bilayer {WSe2}},\ }\href {https://doi.org/10.1038/s41586-024-08116-2} {\bibfield  {journal} {\bibinfo  {journal} {Nature}\ }\textbf {\bibinfo {volume} {637}},\ \bibinfo {pages} {833} (\bibinfo {year} {2025})}\BibitemShut {NoStop}%
\bibitem [{\citenamefont {Guo}\ \emph {et~al.}(2025)\citenamefont {Guo}, \citenamefont {Pack}, \citenamefont {Swann}, \citenamefont {Holtzman}, \citenamefont {Cothrine}, \citenamefont {Watanabe}, \citenamefont {Taniguchi}, \citenamefont {Mandrus}, \citenamefont {Barmak}, \citenamefont {Hone}, \citenamefont {Millis}, \citenamefont {Pasupathy},\ and\ \citenamefont {Dean}}]{guo_superconductivity_2025}%
  \BibitemOpen
  \bibfield  {author} {\bibinfo {author} {\bibfnamefont {Y.}~\bibnamefont {Guo}}, \bibinfo {author} {\bibfnamefont {J.}~\bibnamefont {Pack}}, \bibinfo {author} {\bibfnamefont {J.}~\bibnamefont {Swann}}, \bibinfo {author} {\bibfnamefont {L.}~\bibnamefont {Holtzman}}, \bibinfo {author} {\bibfnamefont {M.}~\bibnamefont {Cothrine}}, \bibinfo {author} {\bibfnamefont {K.}~\bibnamefont {Watanabe}}, \bibinfo {author} {\bibfnamefont {T.}~\bibnamefont {Taniguchi}}, \bibinfo {author} {\bibfnamefont {D.~G.}\ \bibnamefont {Mandrus}}, \bibinfo {author} {\bibfnamefont {K.}~\bibnamefont {Barmak}}, \bibinfo {author} {\bibfnamefont {J.}~\bibnamefont {Hone}}, \bibinfo {author} {\bibfnamefont {A.~J.}\ \bibnamefont {Millis}}, \bibinfo {author} {\bibfnamefont {A.}~\bibnamefont {Pasupathy}},\ and\ \bibinfo {author} {\bibfnamefont {C.~R.}\ \bibnamefont {Dean}},\ }\bibfield  {title} {\bibinfo {title} {Superconductivity in 5.0° twisted bilayer {WSe2}},\ }\href {https://doi.org/10.1038/s41586-024-08381-1} {\bibfield  {journal}
  {\bibinfo  {journal} {Nature}\ }\textbf {\bibinfo {volume} {637}},\ \bibinfo {pages} {839} (\bibinfo {year} {2025})}\BibitemShut {NoStop}%
\bibitem [{\citenamefont {Devakul}\ \emph {et~al.}(2021)\citenamefont {Devakul}, \citenamefont {Crépel}, \citenamefont {Zhang},\ and\ \citenamefont {Fu}}]{devakul_magic_2021}%
  \BibitemOpen
  \bibfield  {author} {\bibinfo {author} {\bibfnamefont {T.}~\bibnamefont {Devakul}}, \bibinfo {author} {\bibfnamefont {V.}~\bibnamefont {Crépel}}, \bibinfo {author} {\bibfnamefont {Y.}~\bibnamefont {Zhang}},\ and\ \bibinfo {author} {\bibfnamefont {L.}~\bibnamefont {Fu}},\ }\bibfield  {title} {\bibinfo {title} {Magic in twisted transition metal dichalcogenide bilayers},\ }\href {https://doi.org/10.1038/s41467-021-27042-9} {\bibfield  {journal} {\bibinfo  {journal} {Nature Communications}\ }\textbf {\bibinfo {volume} {12}},\ \bibinfo {pages} {6730} (\bibinfo {year} {2021})}\BibitemShut {NoStop}%
\bibitem [{\citenamefont {Kim}\ \emph {et~al.}(2024)\citenamefont {Kim}, \citenamefont {Mendez-Valderrama}, \citenamefont {Wang},\ and\ \citenamefont {Chowdhury}}]{kim_theory_2024}%
  \BibitemOpen
  \bibfield  {author} {\bibinfo {author} {\bibfnamefont {S.}~\bibnamefont {Kim}}, \bibinfo {author} {\bibfnamefont {J.~F.}\ \bibnamefont {Mendez-Valderrama}}, \bibinfo {author} {\bibfnamefont {X.}~\bibnamefont {Wang}},\ and\ \bibinfo {author} {\bibfnamefont {D.}~\bibnamefont {Chowdhury}},\ }\href {https://doi.org/10.48550/arXiv.2406.03525} {\bibinfo {title} {Theory of {Correlated} {Insulator}(s) and {Superconductor} at {\textbackslash}nu=1 in {Twisted} {WSe2}}} (\bibinfo {year} {2024}),\ \bibinfo {note} {[arXiv:2406.03525]}\BibitemShut {NoStop}%
\bibitem [{\citenamefont {Zhu}\ \emph {et~al.}(2024)\citenamefont {Zhu}, \citenamefont {Chou}, \citenamefont {Xie},\ and\ \citenamefont {Sarma}}]{zhu_theory_2024}%
  \BibitemOpen
  \bibfield  {author} {\bibinfo {author} {\bibfnamefont {J.}~\bibnamefont {Zhu}}, \bibinfo {author} {\bibfnamefont {Y.-Z.}\ \bibnamefont {Chou}}, \bibinfo {author} {\bibfnamefont {M.}~\bibnamefont {Xie}},\ and\ \bibinfo {author} {\bibfnamefont {S.~D.}\ \bibnamefont {Sarma}},\ }\href {https://doi.org/10.48550/arXiv.2406.19348} {\bibinfo {title} {Theory of superconductivity in twisted transition metal dichalcogenide homobilayers}} (\bibinfo {year} {2024}),\ \bibinfo {note} {[arXiv:2406.19348]}\BibitemShut {NoStop}%
\bibitem [{\citenamefont {Guerci}\ \emph {et~al.}(2024)\citenamefont {Guerci}, \citenamefont {Kaplan}, \citenamefont {Ingham}, \citenamefont {Pixley},\ and\ \citenamefont {Millis}}]{guerci_topological_2024}%
  \BibitemOpen
  \bibfield  {author} {\bibinfo {author} {\bibfnamefont {D.}~\bibnamefont {Guerci}}, \bibinfo {author} {\bibfnamefont {D.}~\bibnamefont {Kaplan}}, \bibinfo {author} {\bibfnamefont {J.}~\bibnamefont {Ingham}}, \bibinfo {author} {\bibfnamefont {J.~H.}\ \bibnamefont {Pixley}},\ and\ \bibinfo {author} {\bibfnamefont {A.~J.}\ \bibnamefont {Millis}},\ }\href {https://doi.org/10.48550/arXiv.2408.16075} {\bibinfo {title} {Topological superconductivity from repulsive interactions in twisted {WSe2}}} (\bibinfo {year} {2024}),\ \bibinfo {note} {[arXiv:2408.16075]}\BibitemShut {NoStop}%
\bibitem [{\citenamefont {Schrade}\ and\ \citenamefont {Fu}(2024)}]{schrade_nematic_2024}%
  \BibitemOpen
  \bibfield  {author} {\bibinfo {author} {\bibfnamefont {C.}~\bibnamefont {Schrade}}\ and\ \bibinfo {author} {\bibfnamefont {L.}~\bibnamefont {Fu}},\ }\bibfield  {title} {\bibinfo {title} {Nematic, chiral, and topological superconductivity in twisted transition metal dichalcogenides},\ }\href {https://doi.org/10.1103/PhysRevB.110.035143} {\bibfield  {journal} {\bibinfo  {journal} {Physical Review B}\ }\textbf {\bibinfo {volume} {110}},\ \bibinfo {pages} {035143} (\bibinfo {year} {2024})}\BibitemShut {NoStop}%
\bibitem [{\citenamefont {Christos}\ \emph {et~al.}(2024)\citenamefont {Christos}, \citenamefont {Bonetti},\ and\ \citenamefont {Scheurer}}]{christos_approximate_2024}%
  \BibitemOpen
  \bibfield  {author} {\bibinfo {author} {\bibfnamefont {M.}~\bibnamefont {Christos}}, \bibinfo {author} {\bibfnamefont {P.~M.}\ \bibnamefont {Bonetti}},\ and\ \bibinfo {author} {\bibfnamefont {M.~S.}\ \bibnamefont {Scheurer}},\ }\href {https://arxiv.org/abs/2407.02393v2} {\bibinfo {title} {Approximate symmetries, insulators, and superconductivity in continuum-model description of twisted {WSe2}}} (\bibinfo {year} {2024}),\ \bibinfo {note} {[arXiv:2407.02393]}\BibitemShut {NoStop}%
\bibitem [{\citenamefont {Tuo}\ \emph {et~al.}(2024)\citenamefont {Tuo}, \citenamefont {Li}, \citenamefont {Wu}, \citenamefont {Sun},\ and\ \citenamefont {Yao}}]{tuo_theory_2024}%
  \BibitemOpen
  \bibfield  {author} {\bibinfo {author} {\bibfnamefont {C.}~\bibnamefont {Tuo}}, \bibinfo {author} {\bibfnamefont {M.-R.}\ \bibnamefont {Li}}, \bibinfo {author} {\bibfnamefont {Z.}~\bibnamefont {Wu}}, \bibinfo {author} {\bibfnamefont {W.}~\bibnamefont {Sun}},\ and\ \bibinfo {author} {\bibfnamefont {H.}~\bibnamefont {Yao}},\ }\href {https://doi.org/10.48550/arXiv.2409.06779} {\bibinfo {title} {Theory of {Topological} {Superconductivity} and {Antiferromagnetic} {Correlated} {Insulators} in {Twisted} {Bilayer} {WSe2}}} (\bibinfo {year} {2024}),\ \bibinfo {note} {[arXiv:2409.06779]}\BibitemShut {NoStop}%
\bibitem [{\citenamefont {Qin}\ \emph {et~al.}(2024)\citenamefont {Qin}, \citenamefont {Qiu},\ and\ \citenamefont {Wu}}]{qin_kohn-luttinger_2024}%
  \BibitemOpen
  \bibfield  {author} {\bibinfo {author} {\bibfnamefont {W.}~\bibnamefont {Qin}}, \bibinfo {author} {\bibfnamefont {W.-X.}\ \bibnamefont {Qiu}},\ and\ \bibinfo {author} {\bibfnamefont {F.}~\bibnamefont {Wu}},\ }\href {https://doi.org/10.48550/arXiv.2409.16114} {\bibinfo {title} {Kohn-{Luttinger} {Mechanism} of {Superconductivity} in {Twisted} {Bilayer} {WSe2}: {Gate}-{Tunable} {Unconventional} {Pairing} {Symmetry}}} (\bibinfo {year} {2024}),\ \bibinfo {note} {[arXiv:2409.16114]}\BibitemShut {NoStop}%
\bibitem [{\citenamefont {Xie}\ \emph {et~al.}(2024)\citenamefont {Xie}, \citenamefont {Chen}, \citenamefont {Sur}, \citenamefont {Fang}, \citenamefont {Cano},\ and\ \citenamefont {Si}}]{xie_superconductivity_2024}%
  \BibitemOpen
  \bibfield  {author} {\bibinfo {author} {\bibfnamefont {F.}~\bibnamefont {Xie}}, \bibinfo {author} {\bibfnamefont {L.}~\bibnamefont {Chen}}, \bibinfo {author} {\bibfnamefont {S.}~\bibnamefont {Sur}}, \bibinfo {author} {\bibfnamefont {Y.}~\bibnamefont {Fang}}, \bibinfo {author} {\bibfnamefont {J.}~\bibnamefont {Cano}},\ and\ \bibinfo {author} {\bibfnamefont {Q.}~\bibnamefont {Si}},\ }\href {https://doi.org/10.48550/arXiv.2408.10185} {\bibinfo {title} {Superconductivity in twisted {WSe2} from topology-induced quantum fluctuations}} (\bibinfo {year} {2024}),\ \bibinfo {note} {[arXiv:2408.10185]}\BibitemShut {NoStop}%
\bibitem [{\citenamefont {Fischer}\ \emph {et~al.}(2024)\citenamefont {Fischer}, \citenamefont {Klebl}, \citenamefont {Crépel}, \citenamefont {Ryee}, \citenamefont {Rubio}, \citenamefont {Xian}, \citenamefont {Wehling}, \citenamefont {Georges}, \citenamefont {Kennes},\ and\ \citenamefont {Millis}}]{fischer_theory_2024}%
  \BibitemOpen
  \bibfield  {author} {\bibinfo {author} {\bibfnamefont {A.}~\bibnamefont {Fischer}}, \bibinfo {author} {\bibfnamefont {L.}~\bibnamefont {Klebl}}, \bibinfo {author} {\bibfnamefont {V.}~\bibnamefont {Crépel}}, \bibinfo {author} {\bibfnamefont {S.}~\bibnamefont {Ryee}}, \bibinfo {author} {\bibfnamefont {A.}~\bibnamefont {Rubio}}, \bibinfo {author} {\bibfnamefont {L.}~\bibnamefont {Xian}}, \bibinfo {author} {\bibfnamefont {T.~O.}\ \bibnamefont {Wehling}}, \bibinfo {author} {\bibfnamefont {A.}~\bibnamefont {Georges}}, \bibinfo {author} {\bibfnamefont {D.~M.}\ \bibnamefont {Kennes}},\ and\ \bibinfo {author} {\bibfnamefont {A.~J.}\ \bibnamefont {Millis}},\ }\href {https://doi.org/10.48550/arXiv.2412.14296} {\bibinfo {title} {Theory of intervalley-coherent {AFM} order and topological superconductivity in {tWSe2}}} (\bibinfo {year} {2024}),\ \bibinfo {note} {arXiv:2412.14296}\BibitemShut {NoStop}%
\bibitem [{\citenamefont {Jung}\ \emph {et~al.}(2014)\citenamefont {Jung}, \citenamefont {Raoux}, \citenamefont {Qiao},\ and\ \citenamefont {MacDonald}}]{jung_ab_2014}%
  \BibitemOpen
  \bibfield  {author} {\bibinfo {author} {\bibfnamefont {J.}~\bibnamefont {Jung}}, \bibinfo {author} {\bibfnamefont {A.}~\bibnamefont {Raoux}}, \bibinfo {author} {\bibfnamefont {Z.}~\bibnamefont {Qiao}},\ and\ \bibinfo {author} {\bibfnamefont {A.~H.}\ \bibnamefont {MacDonald}},\ }\bibfield  {title} {\bibinfo {title} {Ab initio theory of moiré superlattice bands in layered two-dimensional materials},\ }\href {https://doi.org/10.1103/PhysRevB.89.205414} {\bibfield  {journal} {\bibinfo  {journal} {Physical Review B}\ }\textbf {\bibinfo {volume} {89}},\ \bibinfo {pages} {205414} (\bibinfo {year} {2014})}\BibitemShut {NoStop}%
\bibitem [{\citenamefont {Jung}\ \emph {et~al.}(2015)\citenamefont {Jung}, \citenamefont {DaSilva}, \citenamefont {MacDonald},\ and\ \citenamefont {Adam}}]{jung_origin_2015}%
  \BibitemOpen
  \bibfield  {author} {\bibinfo {author} {\bibfnamefont {J.}~\bibnamefont {Jung}}, \bibinfo {author} {\bibfnamefont {A.~M.}\ \bibnamefont {DaSilva}}, \bibinfo {author} {\bibfnamefont {A.~H.}\ \bibnamefont {MacDonald}},\ and\ \bibinfo {author} {\bibfnamefont {S.}~\bibnamefont {Adam}},\ }\bibfield  {title} {\bibinfo {title} {Origin of band gaps in graphene on hexagonal boron nitride},\ }\href {https://doi.org/10.1038/ncomms7308} {\bibfield  {journal} {\bibinfo  {journal} {Nature Communications}\ }\textbf {\bibinfo {volume} {6}},\ \bibinfo {pages} {6308} (\bibinfo {year} {2015})}\BibitemShut {NoStop}%
\bibitem [{\citenamefont {Carr}\ \emph {et~al.}(2018)\citenamefont {Carr}, \citenamefont {Massatt}, \citenamefont {Torrisi}, \citenamefont {Cazeaux}, \citenamefont {Luskin},\ and\ \citenamefont {Kaxiras}}]{carr_relaxation_2018}%
  \BibitemOpen
  \bibfield  {author} {\bibinfo {author} {\bibfnamefont {S.}~\bibnamefont {Carr}}, \bibinfo {author} {\bibfnamefont {D.}~\bibnamefont {Massatt}}, \bibinfo {author} {\bibfnamefont {S.~B.}\ \bibnamefont {Torrisi}}, \bibinfo {author} {\bibfnamefont {P.}~\bibnamefont {Cazeaux}}, \bibinfo {author} {\bibfnamefont {M.}~\bibnamefont {Luskin}},\ and\ \bibinfo {author} {\bibfnamefont {E.}~\bibnamefont {Kaxiras}},\ }\bibfield  {title} {\bibinfo {title} {Relaxation and domain formation in incommensurate two-dimensional heterostructures},\ }\href {https://doi.org/10.1103/PhysRevB.98.224102} {\bibfield  {journal} {\bibinfo  {journal} {Physical Review B}\ }\textbf {\bibinfo {volume} {98}},\ \bibinfo {pages} {224102} (\bibinfo {year} {2018})}\BibitemShut {NoStop}%
\bibitem [{\citenamefont {Ezzi}\ \emph {et~al.}(2024{\natexlab{a}})\citenamefont {Ezzi}, \citenamefont {Pallewela}, \citenamefont {De~Beule}, \citenamefont {Mele},\ and\ \citenamefont {Adam}}]{ezzi_analytical_2024}%
  \BibitemOpen
  \bibfield  {author} {\bibinfo {author} {\bibfnamefont {M.~M.~A.}\ \bibnamefont {Ezzi}}, \bibinfo {author} {\bibfnamefont {G.~N.}\ \bibnamefont {Pallewela}}, \bibinfo {author} {\bibfnamefont {C.}~\bibnamefont {De~Beule}}, \bibinfo {author} {\bibfnamefont {E.}~\bibnamefont {Mele}},\ and\ \bibinfo {author} {\bibfnamefont {S.}~\bibnamefont {Adam}},\ }\bibfield  {title} {\bibinfo {title} {Analytical {Model} for {Atomic} {Relaxation} in {Twisted} {Moiré} {Materials}},\ }\href {https://doi.org/10.1103/PhysRevLett.133.266201} {\bibfield  {journal} {\bibinfo  {journal} {Physical Review Letters}\ }\textbf {\bibinfo {volume} {133}},\ \bibinfo {pages} {266201} (\bibinfo {year} {2024}{\natexlab{a}})}\BibitemShut {NoStop}%
\bibitem [{Note1()}]{Note1}%
  \BibitemOpen
  \bibinfo {note} {See Supplemental Material [url] for \protect \ldots}\BibitemShut {NoStop}%
\bibitem [{Note2()}]{Note2}%
  \BibitemOpen
  \bibinfo {note} {In this case $t^*(\protect \bm {r}) \protect \neq t(-\protect \bm {r})$ and we only have $\varepsilon _2(x,y) = \varepsilon _1(-x,y)$ by $\protect \mathcal C_{2y}\protect \mathcal T$ symmetry. For example, $\protect \mathcal P$ symmetry is broken by $\arg (w_1/w_2) \protect \neq 0,\pi $ and $\psi _2 \protect \neq 0,\pi $ which is the phase of the second star that is equal for both layers.}\BibitemShut {Stop}%
\bibitem [{\citenamefont {Zhang}\ \emph {et~al.}(2024)\citenamefont {Zhang}, \citenamefont {Wang}, \citenamefont {Liu}, \citenamefont {Fan}, \citenamefont {Cao},\ and\ \citenamefont {Xiao}}]{zhang_polarization-driven_2024}%
  \BibitemOpen
  \bibfield  {author} {\bibinfo {author} {\bibfnamefont {X.-W.}\ \bibnamefont {Zhang}}, \bibinfo {author} {\bibfnamefont {C.}~\bibnamefont {Wang}}, \bibinfo {author} {\bibfnamefont {X.}~\bibnamefont {Liu}}, \bibinfo {author} {\bibfnamefont {Y.}~\bibnamefont {Fan}}, \bibinfo {author} {\bibfnamefont {T.}~\bibnamefont {Cao}},\ and\ \bibinfo {author} {\bibfnamefont {D.}~\bibnamefont {Xiao}},\ }\bibfield  {title} {\bibinfo {title} {Polarization-driven band topology evolution in twisted {MoTe2} and {WSe2}},\ }\href {https://doi.org/10.1038/s41467-024-48511-x} {\bibfield  {journal} {\bibinfo  {journal} {Nature Communications}\ }\textbf {\bibinfo {volume} {15}},\ \bibinfo {pages} {4223} (\bibinfo {year} {2024})}\BibitemShut {NoStop}%
\bibitem [{\citenamefont {Jia}\ \emph {et~al.}(2024)\citenamefont {Jia}, \citenamefont {Yu}, \citenamefont {Liu}, \citenamefont {Herzog-Arbeitman}, \citenamefont {Qi}, \citenamefont {Pi}, \citenamefont {Regnault}, \citenamefont {Weng}, \citenamefont {Bernevig},\ and\ \citenamefont {Wu}}]{jia_moire_2024}%
  \BibitemOpen
  \bibfield  {author} {\bibinfo {author} {\bibfnamefont {Y.}~\bibnamefont {Jia}}, \bibinfo {author} {\bibfnamefont {J.}~\bibnamefont {Yu}}, \bibinfo {author} {\bibfnamefont {J.}~\bibnamefont {Liu}}, \bibinfo {author} {\bibfnamefont {J.}~\bibnamefont {Herzog-Arbeitman}}, \bibinfo {author} {\bibfnamefont {Z.}~\bibnamefont {Qi}}, \bibinfo {author} {\bibfnamefont {H.}~\bibnamefont {Pi}}, \bibinfo {author} {\bibfnamefont {N.}~\bibnamefont {Regnault}}, \bibinfo {author} {\bibfnamefont {H.}~\bibnamefont {Weng}}, \bibinfo {author} {\bibfnamefont {B.~A.}\ \bibnamefont {Bernevig}},\ and\ \bibinfo {author} {\bibfnamefont {Q.}~\bibnamefont {Wu}},\ }\bibfield  {title} {\bibinfo {title} {Moiré fractional {Chern} insulators. {I}. {First}-principles calculations and continuum models of twisted bilayer {MoTe2}},\ }\href {https://doi.org/10.1103/PhysRevB.109.205121} {\bibfield  {journal} {\bibinfo  {journal} {Physical Review B}\ }\textbf {\bibinfo {volume} {109}},\ \bibinfo {pages} {205121} (\bibinfo {year}
  {2024})}\BibitemShut {NoStop}%
\bibitem [{\citenamefont {Peng}\ \emph {et~al.}(2025)\citenamefont {Peng}, \citenamefont {Vignale},\ and\ \citenamefont {Adam}}]{peng_many-body_2025}%
  \BibitemOpen
  \bibfield  {author} {\bibinfo {author} {\bibfnamefont {L.}~\bibnamefont {Peng}}, \bibinfo {author} {\bibfnamefont {G.}~\bibnamefont {Vignale}},\ and\ \bibinfo {author} {\bibfnamefont {S.}~\bibnamefont {Adam}},\ }\href {https://doi.org/10.48550/arXiv.2502.06968} {\bibinfo {title} {Many-body perturbation theory for moiré systems}} (\bibinfo {year} {2025}),\ \bibinfo {note} {[arXiv:2502.06968]}\BibitemShut {NoStop}%
\bibitem [{\citenamefont {Ezzi}\ \emph {et~al.}(2024{\natexlab{b}})\citenamefont {Ezzi}, \citenamefont {Peng}, \citenamefont {Liu}, \citenamefont {Chao}, \citenamefont {Pallewela}, \citenamefont {Foo},\ and\ \citenamefont {Adam}}]{ezzi_self-consistent_2024}%
  \BibitemOpen
  \bibfield  {author} {\bibinfo {author} {\bibfnamefont {M.~M.~A.}\ \bibnamefont {Ezzi}}, \bibinfo {author} {\bibfnamefont {L.}~\bibnamefont {Peng}}, \bibinfo {author} {\bibfnamefont {Z.}~\bibnamefont {Liu}}, \bibinfo {author} {\bibfnamefont {J.~H.~Z.}\ \bibnamefont {Chao}}, \bibinfo {author} {\bibfnamefont {G.~N.}\ \bibnamefont {Pallewela}}, \bibinfo {author} {\bibfnamefont {D.}~\bibnamefont {Foo}},\ and\ \bibinfo {author} {\bibfnamefont {S.}~\bibnamefont {Adam}},\ }\href {https://doi.org/10.48550/arXiv.2404.17638} {\bibinfo {title} {A self-consistent {Hartree} theory for lattice-relaxed magic-angle twisted bilayer graphene}} (\bibinfo {year} {2024}{\natexlab{b}}),\ \bibinfo {note} {[arXiv:2404.17638]}\BibitemShut {NoStop}%
\bibitem [{\citenamefont {Lewandowski}\ \emph {et~al.}(2021)\citenamefont {Lewandowski}, \citenamefont {Nadj-Perge},\ and\ \citenamefont {Chowdhury}}]{lewandowski_does_2021}%
  \BibitemOpen
  \bibfield  {author} {\bibinfo {author} {\bibfnamefont {C.}~\bibnamefont {Lewandowski}}, \bibinfo {author} {\bibfnamefont {S.}~\bibnamefont {Nadj-Perge}},\ and\ \bibinfo {author} {\bibfnamefont {D.}~\bibnamefont {Chowdhury}},\ }\bibfield  {title} {\bibinfo {title} {Does filling-dependent band renormalization aid pairing in twisted bilayer graphene?},\ }\href {https://doi.org/10.1038/s41535-021-00379-6} {\bibfield  {journal} {\bibinfo  {journal} {npj Quantum Materials}\ }\textbf {\bibinfo {volume} {6}},\ \bibinfo {pages} {1} (\bibinfo {year} {2021})}\BibitemShut {NoStop}%
\bibitem [{\citenamefont {Klein}\ \emph {et~al.}(2024)\citenamefont {Klein}, \citenamefont {Zondiner}, \citenamefont {Keren}, \citenamefont {Birkbeck}, \citenamefont {Inbar}, \citenamefont {Xiao}, \citenamefont {Sidorova}, \citenamefont {Ezzi}, \citenamefont {Peng}, \citenamefont {Watanabe}, \citenamefont {Taniguchi}, \citenamefont {Adam},\ and\ \citenamefont {Ilani}}]{klein_imaging_2024}%
  \BibitemOpen
  \bibfield  {author} {\bibinfo {author} {\bibfnamefont {D.~R.}\ \bibnamefont {Klein}}, \bibinfo {author} {\bibfnamefont {U.}~\bibnamefont {Zondiner}}, \bibinfo {author} {\bibfnamefont {A.}~\bibnamefont {Keren}}, \bibinfo {author} {\bibfnamefont {J.}~\bibnamefont {Birkbeck}}, \bibinfo {author} {\bibfnamefont {A.}~\bibnamefont {Inbar}}, \bibinfo {author} {\bibfnamefont {J.}~\bibnamefont {Xiao}}, \bibinfo {author} {\bibfnamefont {M.}~\bibnamefont {Sidorova}}, \bibinfo {author} {\bibfnamefont {M.~M.~A.}\ \bibnamefont {Ezzi}}, \bibinfo {author} {\bibfnamefont {L.}~\bibnamefont {Peng}}, \bibinfo {author} {\bibfnamefont {K.}~\bibnamefont {Watanabe}}, \bibinfo {author} {\bibfnamefont {T.}~\bibnamefont {Taniguchi}}, \bibinfo {author} {\bibfnamefont {S.}~\bibnamefont {Adam}},\ and\ \bibinfo {author} {\bibfnamefont {S.}~\bibnamefont {Ilani}},\ }\href {https://doi.org/10.48550/arXiv.2410.22277} {\bibinfo {title} {Imaging the {Sub}-{Moiré} {Potential} {Landscape} using an {Atomic} {Single} {Electron} {Transistor}}}
  (\bibinfo {year} {2024}),\ \bibinfo {note} {[arXiv:2410.22277]}\BibitemShut {NoStop}%
\bibitem [{\citenamefont {Pan}\ \emph {et~al.}(2022)\citenamefont {Pan}, \citenamefont {Xie}, \citenamefont {Wu},\ and\ \citenamefont {Das~Sarma}}]{pan_topological_2022}%
  \BibitemOpen
  \bibfield  {author} {\bibinfo {author} {\bibfnamefont {H.}~\bibnamefont {Pan}}, \bibinfo {author} {\bibfnamefont {M.}~\bibnamefont {Xie}}, \bibinfo {author} {\bibfnamefont {F.}~\bibnamefont {Wu}},\ and\ \bibinfo {author} {\bibfnamefont {S.}~\bibnamefont {Das~Sarma}},\ }\bibfield  {title} {\bibinfo {title} {Topological {Phases} in {AB}-{Stacked} {MoTe2}/{WSe2}: {Z2} {Topological} {Insulators}, {Chern} {Insulators}, and {Topological} {Charge} {Density} {Waves}},\ }\href {https://doi.org/10.1103/PhysRevLett.129.056804} {\bibfield  {journal} {\bibinfo  {journal} {Physical Review Letters}\ }\textbf {\bibinfo {volume} {129}},\ \bibinfo {pages} {056804} (\bibinfo {year} {2022})}\BibitemShut {NoStop}%
\bibitem [{\citenamefont {Qiu}\ \emph {et~al.}(2023)\citenamefont {Qiu}, \citenamefont {Li}, \citenamefont {Luo},\ and\ \citenamefont {Wu}}]{qiu_interaction-driven_2023}%
  \BibitemOpen
  \bibfield  {author} {\bibinfo {author} {\bibfnamefont {W.-X.}\ \bibnamefont {Qiu}}, \bibinfo {author} {\bibfnamefont {B.}~\bibnamefont {Li}}, \bibinfo {author} {\bibfnamefont {X.-J.}\ \bibnamefont {Luo}},\ and\ \bibinfo {author} {\bibfnamefont {F.}~\bibnamefont {Wu}},\ }\bibfield  {title} {\bibinfo {title} {Interaction-{Driven} {Topological} {Phase} {Diagram} of {Twisted} {Bilayer} {MoTe2}},\ }\href {https://doi.org/10.1103/PhysRevX.13.041026} {\bibfield  {journal} {\bibinfo  {journal} {Physical Review X}\ }\textbf {\bibinfo {volume} {13}},\ \bibinfo {pages} {041026} (\bibinfo {year} {2023})}\BibitemShut {NoStop}%
\bibitem [{\citenamefont {Li}\ \emph {et~al.}(2024)\citenamefont {Li}, \citenamefont {Qiu},\ and\ \citenamefont {Wu}}]{li_electrically_2024}%
  \BibitemOpen
  \bibfield  {author} {\bibinfo {author} {\bibfnamefont {B.}~\bibnamefont {Li}}, \bibinfo {author} {\bibfnamefont {W.-X.}\ \bibnamefont {Qiu}},\ and\ \bibinfo {author} {\bibfnamefont {F.}~\bibnamefont {Wu}},\ }\bibfield  {title} {\bibinfo {title} {Electrically tuned topology and magnetism in twisted bilayer {MoTe2} at {\textbackslash}nu\_h=1},\ }\href {https://doi.org/10.1103/PhysRevB.109.L041106} {\bibfield  {journal} {\bibinfo  {journal} {Physical Review B}\ }\textbf {\bibinfo {volume} {109}},\ \bibinfo {pages} {L041106} (\bibinfo {year} {2024})}\BibitemShut {NoStop}%
\bibitem [{Note3()}]{Note3}%
  \BibitemOpen
  \bibinfo {note} {For a uniform electron gas the Hartree term is canceled by the jellium background}\BibitemShut {NoStop}%
\bibitem [{\citenamefont {Bultinck}\ \emph {et~al.}(2020)\citenamefont {Bultinck}, \citenamefont {Khalaf}, \citenamefont {Liu}, \citenamefont {Chatterjee}, \citenamefont {Vishwanath},\ and\ \citenamefont {Zaletel}}]{bultinck_ground_2020}%
  \BibitemOpen
  \bibfield  {author} {\bibinfo {author} {\bibfnamefont {N.}~\bibnamefont {Bultinck}}, \bibinfo {author} {\bibfnamefont {E.}~\bibnamefont {Khalaf}}, \bibinfo {author} {\bibfnamefont {S.}~\bibnamefont {Liu}}, \bibinfo {author} {\bibfnamefont {S.}~\bibnamefont {Chatterjee}}, \bibinfo {author} {\bibfnamefont {A.}~\bibnamefont {Vishwanath}},\ and\ \bibinfo {author} {\bibfnamefont {M.~P.}\ \bibnamefont {Zaletel}},\ }\bibfield  {title} {\bibinfo {title} {Ground {State} and {Hidden} {Symmetry} of {Magic}-{Angle} {Graphene} at {Even} {Integer} {Filling}},\ }\href {https://doi.org/10.1103/PhysRevX.10.031034} {\bibfield  {journal} {\bibinfo  {journal} {Physical Review X}\ }\textbf {\bibinfo {volume} {10}},\ \bibinfo {pages} {031034} (\bibinfo {year} {2020})}\BibitemShut {NoStop}%
\bibitem [{\citenamefont {Wang}\ \emph {et~al.}(2023)\citenamefont {Wang}, \citenamefont {Wang}, \citenamefont {Kim}, \citenamefont {Louie}, \citenamefont {Fu},\ and\ \citenamefont {Zaletel}}]{wang_topology_2023}%
  \BibitemOpen
  \bibfield  {author} {\bibinfo {author} {\bibfnamefont {T.}~\bibnamefont {Wang}}, \bibinfo {author} {\bibfnamefont {M.}~\bibnamefont {Wang}}, \bibinfo {author} {\bibfnamefont {W.}~\bibnamefont {Kim}}, \bibinfo {author} {\bibfnamefont {S.~G.}\ \bibnamefont {Louie}}, \bibinfo {author} {\bibfnamefont {L.}~\bibnamefont {Fu}},\ and\ \bibinfo {author} {\bibfnamefont {M.~P.}\ \bibnamefont {Zaletel}},\ }\href {https://doi.org/10.48550/arXiv.2312.12531} {\bibinfo {title} {Topology, magnetism and charge order in twisted {MoTe2} at higher integer hole fillings}} (\bibinfo {year} {2023}),\ \bibinfo {note} {[arXiv:2312.12531]}\BibitemShut {NoStop}%
\bibitem [{\citenamefont {Stoner}(1997)}]{stoner_collective_1997}%
  \BibitemOpen
  \bibfield  {author} {\bibinfo {author} {\bibfnamefont {E.~C.}\ \bibnamefont {Stoner}},\ }\bibfield  {title} {\bibinfo {title} {Collective electron ferromagnetism},\ }\href {https://doi.org/10.1098/rspa.1938.0066} {\bibfield  {journal} {\bibinfo  {journal} {Proceedings of the Royal Society of London. Series A. Mathematical and Physical Sciences}\ }\textbf {\bibinfo {volume} {165}},\ \bibinfo {pages} {372} (\bibinfo {year} {1997})}\BibitemShut {NoStop}%
\bibitem [{\citenamefont {Li}\ \emph {et~al.}(2025)\citenamefont {Li}, \citenamefont {Redekop}, \citenamefont {Beach}, \citenamefont {Zhang}, \citenamefont {Zhang}, \citenamefont {Liu}, \citenamefont {Holtzmann}, \citenamefont {Hu}, \citenamefont {Anderson}, \citenamefont {Park}, \citenamefont {Taniguchi}, \citenamefont {Watanabe}, \citenamefont {Chu}, \citenamefont {Fu}, \citenamefont {Cao}, \citenamefont {Xiao}, \citenamefont {Young},\ and\ \citenamefont {Xu}}]{li_universal_2025}%
  \BibitemOpen
  \bibfield  {author} {\bibinfo {author} {\bibfnamefont {W.}~\bibnamefont {Li}}, \bibinfo {author} {\bibfnamefont {E.}~\bibnamefont {Redekop}}, \bibinfo {author} {\bibfnamefont {C.~W.}\ \bibnamefont {Beach}}, \bibinfo {author} {\bibfnamefont {C.}~\bibnamefont {Zhang}}, \bibinfo {author} {\bibfnamefont {X.}~\bibnamefont {Zhang}}, \bibinfo {author} {\bibfnamefont {X.}~\bibnamefont {Liu}}, \bibinfo {author} {\bibfnamefont {W.}~\bibnamefont {Holtzmann}}, \bibinfo {author} {\bibfnamefont {C.}~\bibnamefont {Hu}}, \bibinfo {author} {\bibfnamefont {E.}~\bibnamefont {Anderson}}, \bibinfo {author} {\bibfnamefont {H.}~\bibnamefont {Park}}, \bibinfo {author} {\bibfnamefont {T.}~\bibnamefont {Taniguchi}}, \bibinfo {author} {\bibfnamefont {K.}~\bibnamefont {Watanabe}}, \bibinfo {author} {\bibfnamefont {J.-h.}\ \bibnamefont {Chu}}, \bibinfo {author} {\bibfnamefont {L.}~\bibnamefont {Fu}}, \bibinfo {author} {\bibfnamefont {T.}~\bibnamefont {Cao}}, \bibinfo {author} {\bibfnamefont {D.}~\bibnamefont {Xiao}}, \bibinfo {author}
  {\bibfnamefont {A.~F.}\ \bibnamefont {Young}},\ and\ \bibinfo {author} {\bibfnamefont {X.}~\bibnamefont {Xu}},\ }\href {https://doi.org/10.48550/arXiv.2507.22354} {\bibinfo {title} {Universal {Magnetic} {Phases} in {Twisted} {Bilayer} {MoTe}\$\_2\$}} (\bibinfo {year} {2025}),\ \bibinfo {note} {arXiv:2507.22354}\BibitemShut {NoStop}%
\bibitem [{\citenamefont {Xiao}\ \emph {et~al.}(2005)\citenamefont {Xiao}, \citenamefont {Shi},\ and\ \citenamefont {Niu}}]{xiao_berry_2005}%
  \BibitemOpen
  \bibfield  {author} {\bibinfo {author} {\bibfnamefont {D.}~\bibnamefont {Xiao}}, \bibinfo {author} {\bibfnamefont {J.}~\bibnamefont {Shi}},\ and\ \bibinfo {author} {\bibfnamefont {Q.}~\bibnamefont {Niu}},\ }\bibfield  {title} {\bibinfo {title} {Berry {Phase} {Correction} to {Electron} {Density} of {States} in {Solids}},\ }\href {https://doi.org/10.1103/PhysRevLett.95.137204} {\bibfield  {journal} {\bibinfo  {journal} {Physical Review Letters}\ }\textbf {\bibinfo {volume} {95}},\ \bibinfo {pages} {137204} (\bibinfo {year} {2005})}\BibitemShut {NoStop}%
\bibitem [{\citenamefont {Shi}\ \emph {et~al.}(2007)\citenamefont {Shi}, \citenamefont {Vignale}, \citenamefont {Xiao},\ and\ \citenamefont {Niu}}]{shi_quantum_2007}%
  \BibitemOpen
  \bibfield  {author} {\bibinfo {author} {\bibfnamefont {J.}~\bibnamefont {Shi}}, \bibinfo {author} {\bibfnamefont {G.}~\bibnamefont {Vignale}}, \bibinfo {author} {\bibfnamefont {D.}~\bibnamefont {Xiao}},\ and\ \bibinfo {author} {\bibfnamefont {Q.}~\bibnamefont {Niu}},\ }\bibfield  {title} {\bibinfo {title} {Quantum {Theory} of {Orbital} {Magnetization} and {Its} {Generalization} to {Interacting} {Systems}},\ }\href {https://doi.org/10.1103/PhysRevLett.99.197202} {\bibfield  {journal} {\bibinfo  {journal} {Physical Review Letters}\ }\textbf {\bibinfo {volume} {99}},\ \bibinfo {pages} {197202} (\bibinfo {year} {2007})}\BibitemShut {NoStop}%
\bibitem [{\citenamefont {Tschirhart}\ \emph {et~al.}(2021)\citenamefont {Tschirhart}, \citenamefont {Serlin}, \citenamefont {Polshyn}, \citenamefont {Shragai}, \citenamefont {Xia}, \citenamefont {Zhu}, \citenamefont {Zhang}, \citenamefont {Watanabe}, \citenamefont {Taniguchi}, \citenamefont {Huber},\ and\ \citenamefont {Young}}]{tschirhart_imaging_2021}%
  \BibitemOpen
  \bibfield  {author} {\bibinfo {author} {\bibfnamefont {C.~L.}\ \bibnamefont {Tschirhart}}, \bibinfo {author} {\bibfnamefont {M.}~\bibnamefont {Serlin}}, \bibinfo {author} {\bibfnamefont {H.}~\bibnamefont {Polshyn}}, \bibinfo {author} {\bibfnamefont {A.}~\bibnamefont {Shragai}}, \bibinfo {author} {\bibfnamefont {Z.}~\bibnamefont {Xia}}, \bibinfo {author} {\bibfnamefont {J.}~\bibnamefont {Zhu}}, \bibinfo {author} {\bibfnamefont {Y.}~\bibnamefont {Zhang}}, \bibinfo {author} {\bibfnamefont {K.}~\bibnamefont {Watanabe}}, \bibinfo {author} {\bibfnamefont {T.}~\bibnamefont {Taniguchi}}, \bibinfo {author} {\bibfnamefont {M.~E.}\ \bibnamefont {Huber}},\ and\ \bibinfo {author} {\bibfnamefont {A.~F.}\ \bibnamefont {Young}},\ }\bibfield  {title} {\bibinfo {title} {Imaging orbital ferromagnetism in a moiré {Chern} insulator},\ }\href {https://doi.org/10.1126/science.abd3190} {\bibfield  {journal} {\bibinfo  {journal} {Science}\ }\textbf {\bibinfo {volume} {372}},\ \bibinfo {pages} {1323} (\bibinfo {year}
  {2021})}\BibitemShut {NoStop}%
\bibitem [{\citenamefont {Song}\ \emph {et~al.}(2024)\citenamefont {Song}, \citenamefont {Qi}, \citenamefont {Liebman},\ and\ \citenamefont {Narang}}]{song_collective_2024}%
  \BibitemOpen
  \bibfield  {author} {\bibinfo {author} {\bibfnamefont {Z.}~\bibnamefont {Song}}, \bibinfo {author} {\bibfnamefont {J.}~\bibnamefont {Qi}}, \bibinfo {author} {\bibfnamefont {O.}~\bibnamefont {Liebman}},\ and\ \bibinfo {author} {\bibfnamefont {P.}~\bibnamefont {Narang}},\ }\bibfield  {title} {\bibinfo {title} {Collective spin in twisted bilayer materials},\ }\href {https://doi.org/10.1103/PhysRevB.110.024401} {\bibfield  {journal} {\bibinfo  {journal} {Physical Review B}\ }\textbf {\bibinfo {volume} {110}},\ \bibinfo {pages} {024401} (\bibinfo {year} {2024})}\BibitemShut {NoStop}%
\bibitem [{\citenamefont {Redekop}\ \emph {et~al.}(2024)\citenamefont {Redekop}, \citenamefont {Zhang}, \citenamefont {Park}, \citenamefont {Cai}, \citenamefont {Anderson}, \citenamefont {Sheekey}, \citenamefont {Arp}, \citenamefont {Babikyan}, \citenamefont {Salters}, \citenamefont {Watanabe}, \citenamefont {Taniguchi}, \citenamefont {Huber}, \citenamefont {Xu},\ and\ \citenamefont {Young}}]{redekop_direct_2024}%
  \BibitemOpen
  \bibfield  {author} {\bibinfo {author} {\bibfnamefont {E.}~\bibnamefont {Redekop}}, \bibinfo {author} {\bibfnamefont {C.}~\bibnamefont {Zhang}}, \bibinfo {author} {\bibfnamefont {H.}~\bibnamefont {Park}}, \bibinfo {author} {\bibfnamefont {J.}~\bibnamefont {Cai}}, \bibinfo {author} {\bibfnamefont {E.}~\bibnamefont {Anderson}}, \bibinfo {author} {\bibfnamefont {O.}~\bibnamefont {Sheekey}}, \bibinfo {author} {\bibfnamefont {T.}~\bibnamefont {Arp}}, \bibinfo {author} {\bibfnamefont {G.}~\bibnamefont {Babikyan}}, \bibinfo {author} {\bibfnamefont {S.}~\bibnamefont {Salters}}, \bibinfo {author} {\bibfnamefont {K.}~\bibnamefont {Watanabe}}, \bibinfo {author} {\bibfnamefont {T.}~\bibnamefont {Taniguchi}}, \bibinfo {author} {\bibfnamefont {M.~E.}\ \bibnamefont {Huber}}, \bibinfo {author} {\bibfnamefont {X.}~\bibnamefont {Xu}},\ and\ \bibinfo {author} {\bibfnamefont {A.~F.}\ \bibnamefont {Young}},\ }\bibfield  {title} {\bibinfo {title} {Direct magnetic imaging of fractional {Chern} insulators in twisted {MoTe2}},\
  }\href {https://doi.org/10.1038/s41586-024-08153-x} {\bibfield  {journal} {\bibinfo  {journal} {Nature}\ }\textbf {\bibinfo {volume} {635}},\ \bibinfo {pages} {584} (\bibinfo {year} {2024})}\BibitemShut {NoStop}%
\bibitem [{\citenamefont {Deilmann}\ \emph {et~al.}(2020)\citenamefont {Deilmann}, \citenamefont {Krüger},\ and\ \citenamefont {Rohlfing}}]{deilmann_ab_2020}%
  \BibitemOpen
  \bibfield  {author} {\bibinfo {author} {\bibfnamefont {T.}~\bibnamefont {Deilmann}}, \bibinfo {author} {\bibfnamefont {P.}~\bibnamefont {Krüger}},\ and\ \bibinfo {author} {\bibfnamefont {M.}~\bibnamefont {Rohlfing}},\ }\bibfield  {title} {\bibinfo {title} {Ab {Initio} {Studies} of {Exciton} \$g\$ {Factors}: {Monolayer} {Transition} {Metal} {Dichalcogenides} in {Magnetic} {Fields}},\ }\href {https://doi.org/10.1103/PhysRevLett.124.226402} {\bibfield  {journal} {\bibinfo  {journal} {Physical Review Letters}\ }\textbf {\bibinfo {volume} {124}},\ \bibinfo {pages} {226402} (\bibinfo {year} {2020})}\BibitemShut {NoStop}%
\bibitem [{\citenamefont {Woźniak}\ \emph {et~al.}(2020)\citenamefont {Woźniak}, \citenamefont {Faria~Junior}, \citenamefont {Seifert}, \citenamefont {Chaves},\ and\ \citenamefont {Kunstmann}}]{wozniak_exciton_2020}%
  \BibitemOpen
  \bibfield  {author} {\bibinfo {author} {\bibfnamefont {T.}~\bibnamefont {Woźniak}}, \bibinfo {author} {\bibfnamefont {P.~E.}\ \bibnamefont {Faria~Junior}}, \bibinfo {author} {\bibfnamefont {G.}~\bibnamefont {Seifert}}, \bibinfo {author} {\bibfnamefont {A.}~\bibnamefont {Chaves}},\ and\ \bibinfo {author} {\bibfnamefont {J.}~\bibnamefont {Kunstmann}},\ }\bibfield  {title} {\bibinfo {title} {Exciton \$g\$ factors of van der {Waals} heterostructures from first-principles calculations},\ }\href {https://doi.org/10.1103/PhysRevB.101.235408} {\bibfield  {journal} {\bibinfo  {journal} {Physical Review B}\ }\textbf {\bibinfo {volume} {101}},\ \bibinfo {pages} {235408} (\bibinfo {year} {2020})}\BibitemShut {NoStop}%
\bibitem [{\citenamefont {Xie}\ \emph {et~al.}(2023{\natexlab{a}})\citenamefont {Xie}, \citenamefont {Pan}, \citenamefont {Wu},\ and\ \citenamefont {Das~Sarma}}]{xie_nematic_2023}%
  \BibitemOpen
  \bibfield  {author} {\bibinfo {author} {\bibfnamefont {M.}~\bibnamefont {Xie}}, \bibinfo {author} {\bibfnamefont {H.}~\bibnamefont {Pan}}, \bibinfo {author} {\bibfnamefont {F.}~\bibnamefont {Wu}},\ and\ \bibinfo {author} {\bibfnamefont {S.}~\bibnamefont {Das~Sarma}},\ }\bibfield  {title} {\bibinfo {title} {Nematic {Excitonic} {Insulator} in {Transition} {Metal} {Dichalcogenide} {Moiré} {Heterobilayers}},\ }\href {https://doi.org/10.1103/PhysRevLett.131.046402} {\bibfield  {journal} {\bibinfo  {journal} {Physical Review Letters}\ }\textbf {\bibinfo {volume} {131}},\ \bibinfo {pages} {046402} (\bibinfo {year} {2023}{\natexlab{a}})}\BibitemShut {NoStop}%
\bibitem [{\citenamefont {Wang}\ \emph {et~al.}(2024)\citenamefont {Wang}, \citenamefont {Devakul}, \citenamefont {Zaletel},\ and\ \citenamefont {Fu}}]{wang_diverse_2024}%
  \BibitemOpen
  \bibfield  {author} {\bibinfo {author} {\bibfnamefont {T.}~\bibnamefont {Wang}}, \bibinfo {author} {\bibfnamefont {T.}~\bibnamefont {Devakul}}, \bibinfo {author} {\bibfnamefont {M.~P.}\ \bibnamefont {Zaletel}},\ and\ \bibinfo {author} {\bibfnamefont {L.}~\bibnamefont {Fu}},\ }\href {https://doi.org/10.48550/arXiv.2306.02501} {\bibinfo {title} {Diverse magnetic orders and quantum anomalous {Hall} effect in twisted bilayer {MoTe2} and {WSe2}}} (\bibinfo {year} {2024}),\ \bibinfo {note} {[arXiv:2306.02501]}\BibitemShut {NoStop}%
\bibitem [{\citenamefont {Nambu}(1960)}]{nambu_quasi-particles_1960}%
  \BibitemOpen
  \bibfield  {author} {\bibinfo {author} {\bibfnamefont {Y.}~\bibnamefont {Nambu}},\ }\bibfield  {title} {\bibinfo {title} {Quasi-{Particles} and {Gauge} {Invariance} in the {Theory} of {Superconductivity}},\ }\href {https://doi.org/10.1103/PhysRev.117.648} {\bibfield  {journal} {\bibinfo  {journal} {Physical Review}\ }\textbf {\bibinfo {volume} {117}},\ \bibinfo {pages} {648} (\bibinfo {year} {1960})}\BibitemShut {NoStop}%
\bibitem [{\citenamefont {Muñoz-Segovia}\ \emph {et~al.}(2025)\citenamefont {Muñoz-Segovia}, \citenamefont {Crépel}, \citenamefont {Queiroz},\ and\ \citenamefont {Millis}}]{munoz-segovia_twist-angle_2025}%
  \BibitemOpen
  \bibfield  {author} {\bibinfo {author} {\bibfnamefont {D.}~\bibnamefont {Muñoz-Segovia}}, \bibinfo {author} {\bibfnamefont {V.}~\bibnamefont {Crépel}}, \bibinfo {author} {\bibfnamefont {R.}~\bibnamefont {Queiroz}},\ and\ \bibinfo {author} {\bibfnamefont {A.~J.}\ \bibnamefont {Millis}},\ }\href {https://doi.org/10.48550/arXiv.2503.11763} {\bibinfo {title} {Twist-angle evolution of the intervalley-coherent antiferromagnet in twisted {WSe2}}} (\bibinfo {year} {2025}),\ \bibinfo {note} {arXiv:2503.11763}\BibitemShut {NoStop}%
\bibitem [{\citenamefont {Wu}\ \emph {et~al.}(2019)\citenamefont {Wu}, \citenamefont {Lovorn}, \citenamefont {Tutuc}, \citenamefont {Martin},\ and\ \citenamefont {MacDonald}}]{wu_topological_2019}%
  \BibitemOpen
  \bibfield  {author} {\bibinfo {author} {\bibfnamefont {F.}~\bibnamefont {Wu}}, \bibinfo {author} {\bibfnamefont {T.}~\bibnamefont {Lovorn}}, \bibinfo {author} {\bibfnamefont {E.}~\bibnamefont {Tutuc}}, \bibinfo {author} {\bibfnamefont {I.}~\bibnamefont {Martin}},\ and\ \bibinfo {author} {\bibfnamefont {A.}~\bibnamefont {MacDonald}},\ }\bibfield  {title} {\bibinfo {title} {Topological {Insulators} in {Twisted} {Transition} {Metal} {Dichalcogenide} {Homobilayers}},\ }\href {https://doi.org/10.1103/PhysRevLett.122.086402} {\bibfield  {journal} {\bibinfo  {journal} {Physical Review Letters}\ }\textbf {\bibinfo {volume} {122}},\ \bibinfo {pages} {086402} (\bibinfo {year} {2019})}\BibitemShut {NoStop}%
\bibitem [{\citenamefont {Troullier}\ and\ \citenamefont {Martins}(1991)}]{troullier_efficient_1991}%
  \BibitemOpen
  \bibfield  {author} {\bibinfo {author} {\bibfnamefont {N.}~\bibnamefont {Troullier}}\ and\ \bibinfo {author} {\bibfnamefont {J.~L.}\ \bibnamefont {Martins}},\ }\bibfield  {title} {\bibinfo {title} {Efficient pseudopotentials for plane-wave calculations},\ }\href {https://doi.org/10.1103/PhysRevB.43.1993} {\bibfield  {journal} {\bibinfo  {journal} {Physical Review B}\ }\textbf {\bibinfo {volume} {43}},\ \bibinfo {pages} {1993} (\bibinfo {year} {1991})}\BibitemShut {NoStop}%
\bibitem [{\citenamefont {Giannozzi}\ \emph {et~al.}(2017)\citenamefont {Giannozzi}, \citenamefont {Andreussi}, \citenamefont {Brumme}, \citenamefont {Bunau}, \citenamefont {Buongiorno~Nardelli}, \citenamefont {Calandra}, \citenamefont {Car}, \citenamefont {Cavazzoni}, \citenamefont {Ceresoli}, \citenamefont {Cococcioni}, \citenamefont {Colonna}, \citenamefont {Carnimeo}, \citenamefont {Dal~Corso}, \citenamefont {de~Gironcoli}, \citenamefont {Delugas}, \citenamefont {DiStasio}, \citenamefont {Ferretti}, \citenamefont {Floris}, \citenamefont {Fratesi}, \citenamefont {Fugallo}, \citenamefont {Gebauer}, \citenamefont {Gerstmann}, \citenamefont {Giustino}, \citenamefont {Gorni}, \citenamefont {Jia}, \citenamefont {Kawamura}, \citenamefont {Ko}, \citenamefont {Kokalj}, \citenamefont {Küçükbenli}, \citenamefont {Lazzeri}, \citenamefont {Marsili}, \citenamefont {Marzari}, \citenamefont {Mauri}, \citenamefont {Nguyen}, \citenamefont {Nguyen}, \citenamefont {Otero-de-la Roza}, \citenamefont {Paulatto},
  \citenamefont {Poncé}, \citenamefont {Rocca}, \citenamefont {Sabatini}, \citenamefont {Santra}, \citenamefont {Schlipf}, \citenamefont {Seitsonen}, \citenamefont {Smogunov}, \citenamefont {Timrov}, \citenamefont {Thonhauser}, \citenamefont {Umari}, \citenamefont {Vast}, \citenamefont {Wu},\ and\ \citenamefont {Baroni}}]{giannozzi_advanced_2017}%
  \BibitemOpen
  \bibfield  {author} {\bibinfo {author} {\bibfnamefont {P.}~\bibnamefont {Giannozzi}}, \bibinfo {author} {\bibfnamefont {O.}~\bibnamefont {Andreussi}}, \bibinfo {author} {\bibfnamefont {T.}~\bibnamefont {Brumme}}, \bibinfo {author} {\bibfnamefont {O.}~\bibnamefont {Bunau}}, \bibinfo {author} {\bibfnamefont {M.}~\bibnamefont {Buongiorno~Nardelli}}, \bibinfo {author} {\bibfnamefont {M.}~\bibnamefont {Calandra}}, \bibinfo {author} {\bibfnamefont {R.}~\bibnamefont {Car}}, \bibinfo {author} {\bibfnamefont {C.}~\bibnamefont {Cavazzoni}}, \bibinfo {author} {\bibfnamefont {D.}~\bibnamefont {Ceresoli}}, \bibinfo {author} {\bibfnamefont {M.}~\bibnamefont {Cococcioni}}, \bibinfo {author} {\bibfnamefont {N.}~\bibnamefont {Colonna}}, \bibinfo {author} {\bibfnamefont {I.}~\bibnamefont {Carnimeo}}, \bibinfo {author} {\bibfnamefont {A.}~\bibnamefont {Dal~Corso}}, \bibinfo {author} {\bibfnamefont {S.}~\bibnamefont {de~Gironcoli}}, \bibinfo {author} {\bibfnamefont {P.}~\bibnamefont {Delugas}}, \bibinfo {author} {\bibfnamefont
  {R.~A.}\ \bibnamefont {DiStasio}}, \bibinfo {author} {\bibfnamefont {A.}~\bibnamefont {Ferretti}}, \bibinfo {author} {\bibfnamefont {A.}~\bibnamefont {Floris}}, \bibinfo {author} {\bibfnamefont {G.}~\bibnamefont {Fratesi}}, \bibinfo {author} {\bibfnamefont {G.}~\bibnamefont {Fugallo}}, \bibinfo {author} {\bibfnamefont {R.}~\bibnamefont {Gebauer}}, \bibinfo {author} {\bibfnamefont {U.}~\bibnamefont {Gerstmann}}, \bibinfo {author} {\bibfnamefont {F.}~\bibnamefont {Giustino}}, \bibinfo {author} {\bibfnamefont {T.}~\bibnamefont {Gorni}}, \bibinfo {author} {\bibfnamefont {J.}~\bibnamefont {Jia}}, \bibinfo {author} {\bibfnamefont {M.}~\bibnamefont {Kawamura}}, \bibinfo {author} {\bibfnamefont {H.-Y.}\ \bibnamefont {Ko}}, \bibinfo {author} {\bibfnamefont {A.}~\bibnamefont {Kokalj}}, \bibinfo {author} {\bibfnamefont {E.}~\bibnamefont {Küçükbenli}}, \bibinfo {author} {\bibfnamefont {M.}~\bibnamefont {Lazzeri}}, \bibinfo {author} {\bibfnamefont {M.}~\bibnamefont {Marsili}}, \bibinfo {author} {\bibfnamefont
  {N.}~\bibnamefont {Marzari}}, \bibinfo {author} {\bibfnamefont {F.}~\bibnamefont {Mauri}}, \bibinfo {author} {\bibfnamefont {N.~L.}\ \bibnamefont {Nguyen}}, \bibinfo {author} {\bibfnamefont {H.-V.}\ \bibnamefont {Nguyen}}, \bibinfo {author} {\bibfnamefont {A.}~\bibnamefont {Otero-de-la Roza}}, \bibinfo {author} {\bibfnamefont {L.}~\bibnamefont {Paulatto}}, \bibinfo {author} {\bibfnamefont {S.}~\bibnamefont {Poncé}}, \bibinfo {author} {\bibfnamefont {D.}~\bibnamefont {Rocca}}, \bibinfo {author} {\bibfnamefont {R.}~\bibnamefont {Sabatini}}, \bibinfo {author} {\bibfnamefont {B.}~\bibnamefont {Santra}}, \bibinfo {author} {\bibfnamefont {M.}~\bibnamefont {Schlipf}}, \bibinfo {author} {\bibfnamefont {A.~P.}\ \bibnamefont {Seitsonen}}, \bibinfo {author} {\bibfnamefont {A.}~\bibnamefont {Smogunov}}, \bibinfo {author} {\bibfnamefont {I.}~\bibnamefont {Timrov}}, \bibinfo {author} {\bibfnamefont {T.}~\bibnamefont {Thonhauser}}, \bibinfo {author} {\bibfnamefont {P.}~\bibnamefont {Umari}}, \bibinfo {author}
  {\bibfnamefont {N.}~\bibnamefont {Vast}}, \bibinfo {author} {\bibfnamefont {X.}~\bibnamefont {Wu}},\ and\ \bibinfo {author} {\bibfnamefont {S.}~\bibnamefont {Baroni}},\ }\bibfield  {title} {\bibinfo {title} {Advanced capabilities for materials modelling with {Quantum} {ESPRESSO}},\ }\href {https://doi.org/10.1088/1361-648X/aa8f79} {\bibfield  {journal} {\bibinfo  {journal} {Journal of Physics: Condensed Matter}\ }\textbf {\bibinfo {volume} {29}},\ \bibinfo {pages} {465901} (\bibinfo {year} {2017})}\BibitemShut {NoStop}%
\bibitem [{\citenamefont {Xie}\ \emph {et~al.}(2023{\natexlab{b}})\citenamefont {Xie}, \citenamefont {Kang}, \citenamefont {Bernevig}, \citenamefont {Vafek},\ and\ \citenamefont {Regnault}}]{xie_phase_2023}%
  \BibitemOpen
  \bibfield  {author} {\bibinfo {author} {\bibfnamefont {F.}~\bibnamefont {Xie}}, \bibinfo {author} {\bibfnamefont {J.}~\bibnamefont {Kang}}, \bibinfo {author} {\bibfnamefont {B.~A.}\ \bibnamefont {Bernevig}}, \bibinfo {author} {\bibfnamefont {O.}~\bibnamefont {Vafek}},\ and\ \bibinfo {author} {\bibfnamefont {N.}~\bibnamefont {Regnault}},\ }\bibfield  {title} {\bibinfo {title} {Phase diagram of twisted bilayer graphene at filling factor {\textbackslash}nu=3},\ }\href {https://doi.org/10.1103/PhysRevB.107.075156} {\bibfield  {journal} {\bibinfo  {journal} {Physical Review B}\ }\textbf {\bibinfo {volume} {107}},\ \bibinfo {pages} {075156} (\bibinfo {year} {2023}{\natexlab{b}})}\BibitemShut {NoStop}%
\bibitem [{\citenamefont {Galitskii}\ and\ \citenamefont {Migdal}(1958)}]{galitskii_application_1958}%
  \BibitemOpen
  \bibfield  {author} {\bibinfo {author} {\bibfnamefont {V.~M.}\ \bibnamefont {Galitskii}}\ and\ \bibinfo {author} {\bibfnamefont {A.~B.}\ \bibnamefont {Migdal}},\ }\bibfield  {title} {\bibinfo {title} {Application of {Quantum} {Field} {Theory} {Methods} to the {Many} {Body} {Problem}},\ }\href {http://jetp.ras.ru/cgi-bin/e/index/e/7/1/p96?a=list} {\bibfield  {journal} {\bibinfo  {journal} {JETP}\ }\textbf {\bibinfo {volume} {7}},\ \bibinfo {pages} {96} (\bibinfo {year} {1958})}\BibitemShut {NoStop}%
\end{thebibliography}%
